\documentclass[10pt]{article}
\usepackage{times}
\usepackage[T1]{fontenc}
\usepackage[latin1]{inputenc}
\usepackage{geometry}
\usepackage{graphicx,color,rotating,tabularx,dcolumn,amsmath}
\usepackage{pdflscape,longtable}
\usepackage{pdfpages}
\usepackage{soul}  %highlight texts by \hl

\usepackage[table]{xcolor}
\usepackage{colortbl}

\usepackage{longtable}
\usepackage{natbib}
\bibpunct{(}{)}{;}{author-year}{}{,}

\usepackage{setspace}
\usepackage{caption}
\makeatletter
\def\@seccntformat#1{\@ifundefined{#1@cntformat}%
   {\csname the#1\endcsname\quad}  % default
   {\csname #1@cntformat\endcsname}% enable individual control
}
\let\oldappendix\appendix %% save current definition of \appendix
\renewcommand\appendix{%
    \oldappendix
    \newcommand{\section@cntformat}{\appendixname~\thesection\quad}
}
\makeatother

\begin{document}

\begin{flushleft}
{\Large
\textbf{Marker-based estimation of heritability in immortal populations}
}
% Insert Author names, affiliations and corresponding author email.
\\
Willem Kruijer$^{1}$, % Corresponding author:
Martin P. Boer$^{1}$,
Marcos Malosetti$^{1}$,
P\'adraic J. Flood$^{2,3}$,
Bas Engel$^{1}$,
Rik Kooke$^{2,4}$,
Joost J.B. Keurentjes$^{2,5}$,
Fred A. van Eeuwijk$^{1}$
\\
\bf{1} Biometris, Wageningen University and Research Centre, Wageningen, Netherlands
\\
\bf{2} Laboratory of Genetics, Wageningen University and Research Centre, Wageningen, Netherlands
\\
\bf{3} Horticulture \& Product Physiology, Wageningen University and Research Centre, Wageningen, Netherlands
\\
\bf{4} Laboratory of Plant Physiology, Wageningen University and Research Centre, Wageningen, Netherlands
\\
\bf{5} Swammerdam Institute for Life Sciences, University of Amsterdam, Netherlands
%\\
%$\ast$ E-mail: willem.kruijer@wur.nl
\end{flushleft}

%\clearpage
\vspace*{0.1in}
\noindent
\textbf{Published in \emph{Genetics} (2015), Vol. 199(2), p. 379-398}; early online December 19, 2014.
The article is available at \verb|www.genetics.org| with DOI \verb|doi:10.1534/genetics.114.167916|
\vspace*{0.1in}

\noindent
\textbf{Running head}: estimation of heritability


% \textit{Arabidopsis}

\vspace*{0.1in}
\noindent
\textbf{*Corresponding author}: \\
Willem Kruijer \\
Biometris \\
Wageningen University and Research Centre \\
PO Box 100, 6700AC Wageningen \\
The Netherlands \\
Phone: +31 317 480806 \\
Email: willem.kruijer@wur.nl

%\clearpage

% Please keep the abstract between 250 and 300 words
\section*{Abstract}

Heritability is a central parameter in quantitative genetics, both from an evolutionary and a
breeding perspective. For plant traits heritability is traditionally estimated by comparing within
and between genotype variability. This approach estimates broad-sense heritability, and does
not account for different genetic relatedness. With the availability of high-density markers there
is growing interest in marker based estimates of narrow-sense heritability, using mixed models
in which genetic relatedness is estimated from genetic markers. Such estimates have received
much attention in human genetics but are rarely reported for plant traits. A major obstacle is
that current methodology and software assume a single phenotypic value per genotype, hence
requiring genotypic means. An alternative that we propose here, is to use mixed models at
individual plant or plot level. Using statistical arguments, simulations and real data we investigate the feasibility of both approaches, and how these affect genomic prediction with G-BLUP
and genome-wide association studies. Heritability estimates obtained from genotypic means
had very large standard errors and were sometimes biologically unrealistic. Mixed models at
individual plant or plot level produced more realistic estimates, and for simulated traits standard errors
were up to 13 times smaller.
%
%Also genomic prediction improved using these mixed models:
%accuracy increased up to 49\%.
Genomic prediction was also improved by using these mixed models, with up to a 49\% increase in accuracy.
For GWAS on simulated traits, the use of individual plant data gave almost no
increase in power. The new methodology is applicable to any complex trait where
multiple replicates of individual genotypes can be scored. This includes important agronomic crops, as well as
bacteria and fungi.

\section*{Introduction}

Narrow-sense heritability is an important parameter in quantitative genetics, determining the response to selection and representing the proportion of phenotypic variance that is due to additive genetic effects
(\cite{jacquard_1983},
\cite{nyquist_baker_1991},
\cite{Holland_Nyquist_CervantesMartinez_2010},
\cite{ritland_1996},
\cite{visscher_etal_2006},
\cite{visscher_hill_wray_2008},
\cite{sillanpaa_2011}).
This definition of heritability goes back to \cite{Fisher_1918} and \cite{Wright_1920} almost a century ago.
In plant species for which replicates of the same genotype are available (inbred lines, doubled haploids, clones), a different form of heritability, broad sense heritability, is traditionally estimated by the intra-class correlation coefficient for genotypic effects, using estimates for within and between genotype variance. Broad sense heritability is also referred to as repeatability and gives the proportion of phenotypic variance explained by heritable (additive) and non-heritable (dominance, epistasis) genetic variance.

With the arrival of high-density genotyping there is growing interest in marker-based estimation of narrow-sense heritability
(\cite{WTCCC_nature_2007}, \cite{yang_etal_2010}, \cite{Yang_HongLee_Goddard_Visscher_2011}, \cite{speed_hemani_johnson_balding_2012}, \cite{vattikuti_etal_2012}, \cite{visscher_goddard_2014}).
These estimates are obtained from mixed models containing random additive genetic effects, whose covariance structure is estimated from genetic markers.
%\cite{henderson_etal_1959}, \cite{hartley_rao_1967}.
While marker-based heritability estimates have received much attention in human genetics,
these are rarely reported for plant traits, despite their relevance in evolutionary genetics, the dissection of complex traits,
and in the ongoing debate on missing heritability
(\cite{manolio_etal_2009},
\cite{eichler_etal_2010},
\cite{lee_etal_2011},
\cite{brachi_etal_2011},
\cite{zuk_etal_2012}).
Heritability estimates are also of great relevance to plant breeders, as they give a measure for the breeding potential of a trait.
In addition, state-of-the-art phenotyping platforms are making experiments more reproducible,
increasing the relevance of marker-based estimation of heritability, as well as comparison of estimates from different experiments.

Although marker-based heritability estimation for plant traits can in principle be achieved within the same analytic framework that has been developed for human traits, there are important differences.
First, heritability of human traits is usually estimated using panels of unrelated individuals,
in order to avoid confounding with geographical or environmental effects (\cite{browning_browning_2011}).
For most plant species panels of unrelated individuals are not available, as plant genotypes will often share ancestry or adaptation inducing dependence between genotypes.
At the same time however, plant genotypes can be evaluated
under the same experimental conditions (e.g. common garden or growth chamber), and the usual assumption is that this eliminates genotype-environment correlations. Second, plant genotypes are often phenotyped in several genetically identical replicates.
This is the case for all so-called immortal populations (\cite{keurentjes_etal_2007}, \cite{wijnen_keurentjes_2014}).
Mixed model analysis can then be performed either on the individual plant (or plot) data or on genotypic means.
In the literature on multi-environment trials
(\cite{smith_cullis_thompson_2001},
\cite{Smith_Cullis_Thompson_2005},
\cite{oakey_etal_2006},
\cite{Piepho_Williams_2006},
\cite{Piepho_Buchse_Truberg_2006}
\cite{verbyla_cullis_thompson_2007},
\cite{boer_etal_2007},
\cite{Stich_etal_2008},
\cite{mohring_piepho_2009},
\cite{vaneeuwijk_etal_2010},
\cite{welham_etal_2010},
\cite{piepho_etal_2012},
\cite{malosetti_etal_2013})
these approaches are referred to as respectively one-stage and two-stage. These works consider mostly populations for which a pedigree is available,
typically experimental populations. In the context of genomic prediction and GWAS for natural populations, mixed model analysis is usually performed
using a two-stage approach. The (usually tacit) assumption is that the genotypic means and kinship coefficients contain all the relevant information
for estimating the genetic and residual variance.
Here we investigate the feasibility of marker-based estimation of heritability
with one- and two-stage approaches, and look at how heritability estimates affect genomic prediction with the best linear unbiased predictor (G-BLUP) and genome-wide association studies (GWAS).
Although our analysis of observed phenotypes focuses on the model plant \emph{A. thaliana},
the asymptotic variances of different heritability estimators were also computed for diverse panels of  \emph{Z. mays} (\cite{vanheerwaarden_etal_2012}, \cite{riedelsheimer_etal_2012}) and \emph{O. sativa} (\cite{zhao_etal_2011}).

In both published data (\cite{atwell_etal_2010}) and new experiments, % on the model plant \emph{A. thaliana},
we found very large standard errors and sometimes unrealistically high estimates of heritability, which could not be explained by varying linkage disequilibrium (\cite{speed_hemani_johnson_balding_2012}).
Much better heritability estimates were obtained when mixed model analysis was performed at individual plant level. These estimates were based on kinship information as well as additional information on within genotype variability, and in simulations they were found to be up to $13$ times more accurate than for heritability estimates based on genotypic means.
In genomic prediction, correlation between simulated and predicted genetic effects increased in some cases by as much as 49\%. This is a substantial improvement which shows the importance of accurate heritability estimates in plant breeding programs.

All reported heritability estimates can be obtained using our R-package \verb|heritability|, which is freely available from CRAN % \newline
(\verb|http://cran.r-project.org/web/packages/heritability/index.html|).
In contrast to existing packages such as \verb|emma| (\cite{kang_etal_2008}), \verb|rrblup| (\cite{endelman_2011}), and \verb|synbreed| (\cite{wimmer_etal_2012}), it provides confidence intervals for heritability estimates.
We also present software for GWAS: our program \verb|scan_GLS| can efficiently perform GWAS directly on the individual plant or plot-level data, as well as on the means, incorporating a non-diagonal error covariance structure.

\section*{Materials and Methods}

We assume a natural population of $n$ genotypes, where for each genotype a quantitative trait is measured on a number of genetically identical replicates of immortal lines.
These replicates can refer to either individual plants  (e.g. Arabidopsis) or plots in a field trial (e.g. maize).
For convenience we will use the terms 'replicate' and 'individual plant data' as synonyms throughout.
In either case, the observations on replicates are not to be confused with having multiple observations on the same individual.
We focus on %\emph{A. thaliana}
inbred lines, but apart from a few different constants, all expressions are valid for any diploid species.
We do not partition the environmental variance into different contributing factors (see e.g. \cite{visscher_hill_wray_2008}), hence the environmental variance just equals the error variance.
For simplicity we first present marker-based  estimation of heritability for a completely randomized design, in absence of additional covariates.
The expressions for more general designs are given in Appendix \ref{App_marker_based_h2}, which also provides a brief overview of the existing methodology for marker-based heritability estimation.
Details on G-BLUP, GWAS and the phenotypic data are given in appendices \ref{App_GP}, \ref{App_GWAS} and \ref{App_pheno}.

\subsection*{Genotypic data}

We analyze simulated and real traits for sub-populations of the RegMap, which contains 1307 world-wide accessions of \emph{A. thaliana} that have been genotyped at $214051$ SNP-markers (\cite{horton_etal_2012}).
We considered four sub-populations: $298$ Swedish accessions (Swedish RegMap, \cite{horton_etal_2012}, \cite{long_etal_2013}),
$204$ French accessions (French RegMap, \cite{horton_etal_2012}, \cite{brachi_etal_2013}),
$350$ accessions from the HapMap-population (\cite{li_huang_etal_2010}), and a subset of $250$
accessions which we will refer to as the \emph{structured RegMap} (accessions ID's are given
in Table S1). The structured RegMap accessions were chosen to have large differences in genetic relatedness (i.e. variation in kinship coefficients, across pairs of accessions).
These differences were hence largest in the structured RegMap, and smallest in the HapMap
(Figures S1 and S2). For the asymptotic distributions of heritability estimators given below, we also considered marker-based kinship matrices for three populations of crop plants: the panel described in \cite{vanheerwaarden_etal_2012} (\emph{Z. mays}, $400$ accessions), the panel described in \cite{riedelsheimer_etal_2012} (\emph{Z. mays}, $280$ accessions) and the panel from \cite{zhao_etal_2011} (\emph{O. sativa}, $413$ accessions).

\subsection*{Phenotypic data}

We estimated heritabilities for two traits from the literature and four traits from new experiments.
Two flowering traits from \cite{atwell_etal_2010} were analyzed: days to flowering time under long day and vernalization (LDV), and days to flowering time under long day (LD).
In two new experiments (Appendix \ref{App_pheno}), leaf area 13 days after sowing (LA) was measured for the Swedish RegMap and the HapMap, using the same automated phenotyping platform.
In the final experiment, the HapMap was phenotyped for bolting time (BT) and leaf width (LW).
Trait descriptions and abbreviations are given in Table \ref{trait_abbreviations}. %, as well as details on the collection of phenotypic data and plant growth conditions.
In all experiments the individual plants were grouped into complete blocks, but due to non-germinating seeds and dead plants, there is some unbalance for all of the traits.

\subsection*{Marker-based  estimation of heritability}

Let $K$ denote the genetic relatedness or kinship matrix of the genotypes, with elements
\begin{equation} \label{kinship}
K_{i,j} = \frac{1}{p} \sum_{l=1}^p \frac{(x_{i,l}-f_l)(x_{j,l}-f_l)}{4 f_l(1-f_l)}
\end{equation}
(see e.g. \cite{patterson_price_reich_2006} and \cite{Goddard_Wray_Verbyla_Visscher_2010}).
The numbers $x_{i,l} \in \{0,2\}$ denote the minor allele count\footnote{Because we focus on inbred lines there is a small difference with the standard expression for outbreeders under Hardy-Weinberg equilibrium, where $x_{i,l} \in \{0,1,2\}$, and the constant $4$ in \eqref{kinship} is to be replaced by $2$.} at marker $l$ for genotype $i$, and $f_l \in [0,1]$ is the minor allele frequency at marker $l$.
The phenotypic response of replicate $j$ of genotype $i$ is modeled as
\begin{equation} \label{infModelR}
Y_{i,j} = \mu + G_i + E_{i,j} \quad (i=1,\ldots,n, \quad j=1,\ldots,r), % x_{i,j} \beta
\end{equation}
where $G = (G_1,\ldots,G_n)$ has a $N(0,\sigma_A^2 K)$ distribution, and the errors $E_{i,j}$ have independent normal distributions with variance $\sigma_E^2$.
We aim to estimate the heritability
\begin{equation*}
h^2 = \frac{\sigma_A^2}{\sigma_A^2 + \sigma_E^2},
\end{equation*}
which has been referred to as the 'chip' heritability (\cite{speed_hemani_johnson_balding_2012}), and under certain assumption equals narrow sense-heritability (Appendix \ref{App_marker_based_h2}).
In contrast to the \emph{line}-heritability $\sigma_A^2 / (\sigma_A^2 + r^{-1} \sigma_E^2)$, the denominator contains the residual variance for a single individual ($\sigma_E^2$).
Given REML estimates $\hat\sigma_{A,r}^2$ and $ \hat\sigma_{E,r}^2$, $h^2$ can be estimated by
\begin{equation} \label{h2repl}
{\hat h}_r^2 = \frac{\hat \sigma_{A,r}^2}{\hat \sigma_{A,r}^2 + \hat \sigma_{E,r}^2},
\end{equation}
where the subscripts $r$ stress the fact that $\hat \sigma_{A,r}^2$ and $\hat \sigma_{E,r}^2$ are estimates directly obtained from the replicates.
Alternatively, $h^2$ can be estimated using the genotypic means % $\bar Y_i$
\begin{equation} \label{infModelM2}
\bar Y_i = \frac{1}{r} \sum_{j=1}^{r} = \mu + G_i + \bar E_i, \qquad G \sim N(0,\sigma_A^2 K), \quad \bar E_i \sim N(0,r^{-1} \sigma_E^2).
\end{equation}
Given REML estimates $\hat \sigma_{A,m}^2$ and $\hat \sigma_{E,m}^2$ for model \eqref{infModelM2}, we have the heritability estimate
\begin{equation} \label{h2means}
{\hat h}_m^2 = \frac{\hat \sigma_{A,m}^2}{\hat \sigma_{A,m}^2 + \hat \sigma_{E,m}^2}.
\end{equation}
where the subscripts $m$ indicate that the estimates are based on genotypic means. We will omit the
letters $r$ and $m$ if either these are clear from the context, or when a statement holds
for both model \eqref{infModelR} and model \eqref{infModelM2}.
In human association studies, $r$ is usually one, and ${\hat h}_m^2$ and ${\hat h}_r^2$ are identical. %reduce to $\hat h^2$ in \eqref{h2estimate}.

\subsection*{Repeatability and broad-sense heritability}

Given a completely randomized design with $r$ replicates, repeatability or intra-class correlation %(\cite{kempthorne_1957})
can be estimated by
% broad-sense heritability ($H^2$) is estimated by
\begin{equation} \label{H2}
{\hat H}^2 = \frac{\hat \sigma_G^2}{\hat \sigma_G^2 + \hat \sigma_{\textrm{Env}}^2}, \textrm{with} \quad \hat \sigma_G^2 = (MS(G) - MS(Env)) / {r }, \quad \hat \sigma_{\textrm{Env}}^2 = MS(Env),
\end{equation}
where $MS(G)$ and $MS(Env)$ are the mean sums of squares for genotype and residual error,
obtained from analysis of variance (ANOVA). In case $MS(G) < MS(Env)$, $\hat \sigma_G^2$ is set to zero. See \cite{Singh_Ceccarelli_Hamblin_1993} or (in the context of sib-analysis) \cite{lynch_walsh_1998} (p.563). Here we stick to the widely used notation ${\hat H}^2$, although repeatability only equals broad-sense heritability ($H^2$) under the assumption that
all differences between genotypes are indeed genetic, and not due to e.g. genotype-environment correlation.
In any case, no additional information on the genetic structure is used here, hence the mean sums of squares for genotype contains all genetic effects, not only the additive ones.
Consequently, ${\hat H}^2$ is an estimator of (an upper-bound on) broad-sense heritability.

It is known to have a small bias (\cite{nyquist_baker_1991}, \cite{Singh_Ceccarelli_Hamblin_1993}), which tends to zero when the number of genotypes increases.
Additional fixed effects can be included in the ANOVA, which may reduce $MS(Env)$ %the environmental mean sum of squares,
and give a higher heritability estimate. In case the genotypes have differing numbers of replicates, genetic variance is estimated with
$\hat \sigma_G^2 = (MS(G) - MS(Env)) / {\bar r }$, where
\begin{equation}\label{effective_r}
\bar r = (n-1)^{-1} \left[\sum_{i=1}^n r_i - \left(\sum_{i=1}^n r_i^2\right) / \left(\sum_{i=1}^n r_i\right)\right]
\end{equation}
is the effective number of replicates (\cite{lynch_walsh_1998} (p.559)). For a balanced design, $\bar r = r$.

\subsection*{Confidence intervals}

Confidence intervals for broad-sense heritability $H^2$ were obtained from classical theory (e.g. \cite{lynch_walsh_1998}, p.563; for easy reference see File S1).
Confidence intervals for $h^2$ were constructed using $\hat \sigma_A^2$ and $\hat \sigma_E^2$ %($\hat \sigma_{E,r}^2$ in case of ${\hat h}_m^2$)
and the inverse average-information (AI) matrix, obtained from the AI-REML algorithm (\cite{gilmour_thompson_cullis_1995}).
This $2 \times 2$ matrix provides an estimate of the covariance matrix of $(\hat \sigma_A^2, \hat \sigma_E^2)$. The delta-method (see e.g. \cite{vandervaart2000}) applied to
the function $(\hat \sigma_A^2, \hat \sigma_E^2) \rightarrow \hat \sigma_A^2 / (\hat \sigma_A^2 + \hat \sigma_E^2)$
then gives the asymptotic distribution of ${\hat h}^2$, from which a confidence interval for $h^2$ is calculated.
In case the heritability is low or high, confidence intervals may be partly outside $[0,1]$.
We set all negative values to zero and all values larger than one equal to one.
An alternative is application of the delta-method to the function $(\hat \sigma_A^2, \hat \sigma_E^2) \rightarrow \log(\hat \sigma_A^2 / \hat \sigma_E^2)$.
This gives confidence intervals for $\log(\sigma_A^2 / \sigma_E^2)$, which are then back-transformed to confidence intervals for heritability.
The latter intervals will be referred to as 'log-transformed', and the other intervals as 'standard'. % obtained by application of the delta-method on the .

When the likelihood as function of $h^2$ is monotonically increasing, the confidence intervals can become
numerically unstable, alternating between $[0,1]$ and $[1,1]$. However we found that using the AI-matrix
obtained by setting $h^2=1-\epsilon$, the interval was always very close to $[0,1]$, for any $\epsilon >0$.
We therefore defined the interval to be $[0,1]$ in case of a monotonically increasing likelihood.

\subsection*{Simulations}

Each simulated trait consists of $r=3$ replicates for a subset of 200 accessions randomly drawn from sub-population of the Arabidopsis RegMap (\cite{horton_etal_2012}).
We simulate according to model \eqref{infModelR}, except that the genetic effects $G_i$ are not purely polygenic, but a mixture of QTL-effects at a small number of markers and a polygenic signal with genetic structure given by the kinship matrix defined in \eqref{kinship}. More details are given in Appendix \ref{App_simulations}.
Different genetic architectures were considered, but all simulated genetic effects were additive. Hence, broad- and narrow-sense heritability were equal, and ${\hat H}^2$, ${\hat h}_r^2$ and ${\hat h}_m^2$ were in this case different estimators of the (same) simulated heritability.

For the comparison of heritability estimates we simulated traits for the structured RegMap, the HapMap and the Swedish- and French RegMap. For comparing one- and two-stage association mapping and genomic prediction only the structured RegMap and the HapMap were considered.
For genomic prediction and the comparison of heritability estimates we simulated 3 levels of heritability, for each sub-population: low ($0.2$), medium ($0.5$) and high ($0.8$). For each heritability level, $5000$ traits were simulated. For the comparison of one- and two-stage association mapping, $1000$ traits were simulated with heritability $0.5$.
In all simulations, a completely randomized design was assumed.

% Results and Discussion can be combined.
\section*{Results}

\subsection*{Asymptotic variance}

Applying general mixed model theory (\cite{searle_casella_mcculloch_2006}, p.387) to model \eqref{infModelR}, it follows that
%From mixed-model theory it follows that
the REML-estimators $(\hat \sigma_{A,r}^2,\hat \sigma_{E,r}^2)$ are asymptotically\footnote{The number of replicates $r$ is considered fixed here. The asymptotic arguments are all with respect to $n$, the number of genotypes. The behavior of $\textrm{Var}({\hat h}_m^2)$ as a function of $n$, $h^2$ and $K$ can also be derived from approximations recently proposed in \cite{visscher_goddard_2014}, independently from the present work.} Gaussian with covariance
\begin{equation} \label{as_var_general}
\textrm{Var}
\left(
\begin{array}{c}
\hat \sigma_{A,r}^2 \\
\hat \sigma_{E,r}^2
\end{array}
\right)
 \simeq
 2
\left[
\begin{array}{cc}
\textrm{tr}(P (Z K Z^t)P (Z K Z^t)) & \textrm{tr}(P (Z K Z^t) P) \\
\textrm{tr}(P (Z K Z^t) P) & \textrm{tr}(P P)
\end{array}
\right]^{-1} =: \Sigma_{\hat \sigma_{A,r}^2,\hat \sigma_{E,r}^2},
\end{equation}
where $P = V^{-1} - V^{-1} X (X^t V^{-1} X)^{-1} X^t V^{-1}$ for $V=\sigma_A^2 Z K Z^t + \sigma_E^2 I_{nr}$ ,
$X=1_{nr}$ and $Z$ is the $nr \times n$ incidence matrix assigning plants to genotypes. Although this is not made explicit in the
notation, $\Sigma_{\hat \sigma_{A,r}^2,\hat \sigma_{E,r}^2}$ depends on the
true (and usually unknown) values $\sigma_A^2$ and $\sigma_E^2$ through $P$ and $V$.
If $(\sigma_A^2,\sigma_E^2)$ are estimated based on genotypic means following model \eqref{infModelM2},
$(\hat \sigma_{A,m}^2,\hat \sigma_{E,m}^2)$ has asymptotic covariance
$\Sigma_{\hat \sigma_{A,m}^2,\hat \sigma_{E,m}^2}$, which is obtained if %on the right hand
 we replace in \eqref{as_var_general} $Z K Z^t$ by $K$ and $X=1_{nr}$ by $X=1_n$, and substitute
$V=\sigma_A^2 K + \sigma_E^2 r^{-1} I_{n}$ in the definition of $P$.
The asymptotic variance of the heritability estimators ${\hat h}_r^2$ and ${\hat h}_m^2$ can now be obtained by application of the delta-method (\cite{vandervaart2000}) to the function $(\sigma_A^2,\sigma_E^2) \rightarrow \sigma_A^2 / (\sigma_A^2 + \sigma_E^2)$, with gradient
\begin{equation*}
b_{\sigma_A^2,\sigma_E^2} = \frac{1}{(\sigma_A^2 + \sigma_E^2)^2} (\sigma_E^2,-\sigma_A^2)^t.
\end{equation*}
It then follows that given the true $\sigma_A^2$ and $\sigma_E^2$,
\begin{equation}\label{as_var_formula}
\textrm{Var}({\hat h}_r^2)  \simeq
b_{\sigma_A^2,\sigma_E^2} \Sigma_{\hat \sigma_{A,r}^2,\hat \sigma_{E,r}^2} b_{\sigma_A^2,\sigma_E^2}^t, \qquad
\textrm{Var}({\hat h}_m^2)  \simeq b_{\sigma_A^2,\sigma_E^2} \Sigma_{\hat \sigma_{A,m}^2,\hat \sigma_{E,m}^2} b_{\sigma_A^2,\sigma_E^2}^t.
\end{equation}
It is easily verified that both variances do not depend on the absolute values of $\sigma_A^2$ and $\sigma_E^2$, but only on the ratio $h^2 = \sigma_A^2 / (\sigma_A^2 + \sigma_E^2)$.
To explore the potential gains in accuracy due to the use of individual plant data,
we computed $\sqrt{\textrm{Var}({\hat h}_r^2)}$ and the ratio $\sqrt{\textrm{Var}({\hat h}_r^2)} / \sqrt{\textrm{Var}({\hat h}_m^2)}$
of the two standard deviations, for several populations in \emph{A. thaliana}, \emph{Z. mays} and \emph{O. sativa} (Table \ref{asymptotic_variance_table}), and three heritability levels ($h^2=0.2,0.5,0.8$).

The use of individual plant data gave a substantial improvement in accuracy for all populations, which was largest for the Arabidopsis HapMap ($81\%$ reduction with respect to $\sqrt{\textrm{Var}({\hat h}_m^2)}$, for $r=4$ and $h^2=0.8$) and smallest for the structured RegMap, the rice population from \cite{zhao_etal_2011} and the maize population from \cite{riedelsheimer_etal_2012} ($13-15\%$ reduction when $r=2$ and $h^2=0.2$). For the maize population from \cite{vanheerwaarden_etal_2012}
and the other Arabidopsis populations the improvements were similar. The standard deviation of ${\hat h}_r^2$ decreased with increasing heritability and increasing numbers of replicates. The ratio of the standard deviations of ${\hat h}_r^2$ over those of ${\hat h}_m^2$ also decreased substantially with increasing heritability, but remained similar when increasing the number of replicates beyond $2$.

It should be emphasized that the ratios in Table \ref{asymptotic_variance_table} are asymptotic: although the expressions in \eqref{as_var_formula} depend on the kinship matrix $K$ of a finite population, they may only be good approximations for (very) large populations with population structure similar
to that defined by $K$.
Such populations could be modeled as a function of population genetic processes such as drift and selection (\cite{rousset_2002}, \cite{fu_etal_2003}). However, for real plant populations marker information is available for at most a few thousands of genotypes, and more often several hundreds.
To assess variance and bias of ${\hat h}_r^2$ and ${\hat h}_m^2$ in such populations we used simulated traits. %in a more realistic scenario,

\subsection*{Heritability estimates for simulated data}

We analyzed simulated traits for the $4$ populations of \emph{A. thaliana}, for different genetic architectures. We also used the simulations to compare these marker-based estimators with the broad-sense heritability estimator ${\hat H}^2$, which ignores genetic relatedness, and to investigate the quality of heritability confidence intervals.

\begin{figure}[!ht]
\includegraphics[width=16cm,height=18cm]{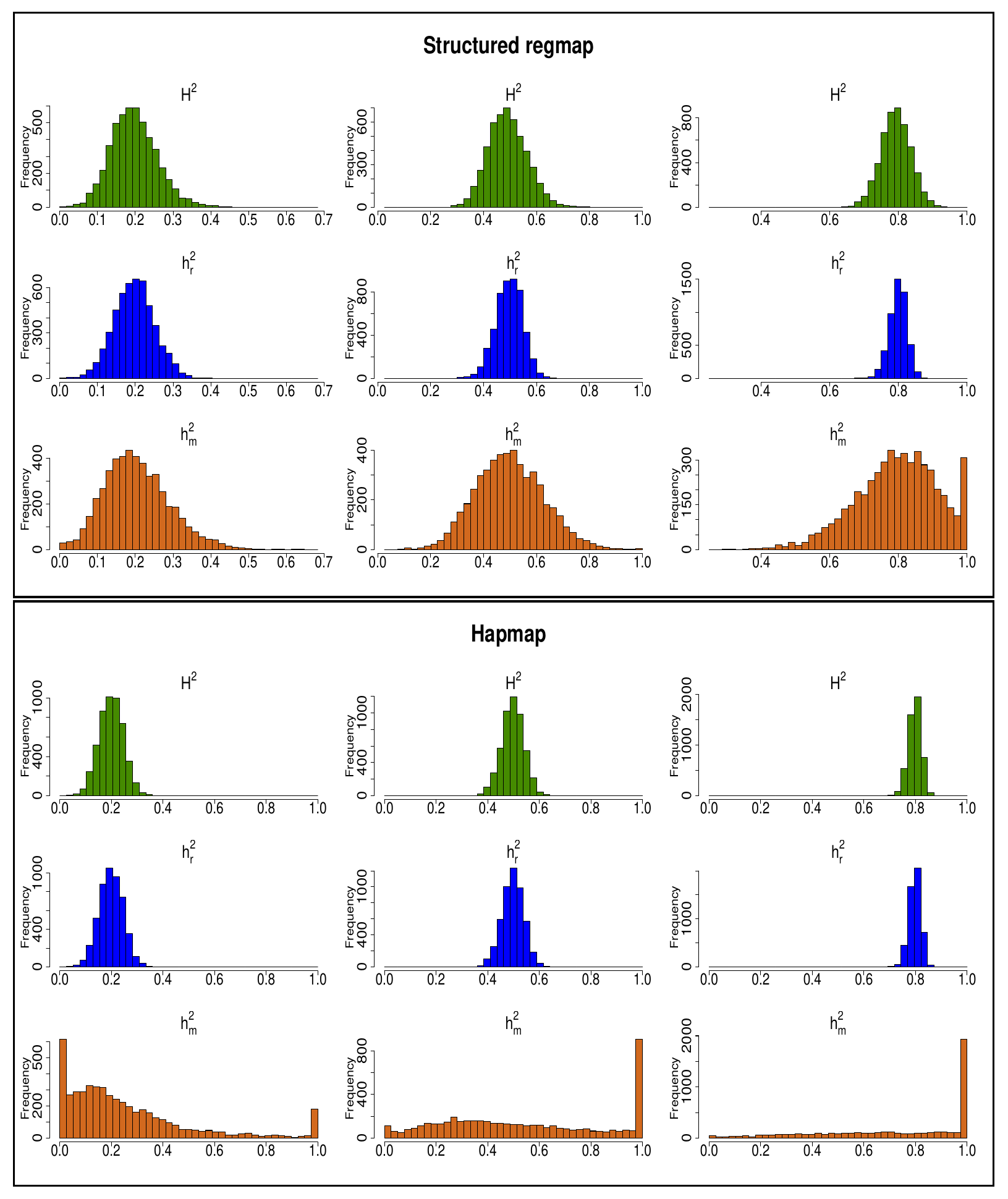}
\caption{{\bf Heritability estimates for 5000 simulated traits for random samples of 200 accessions drawn from the Structured RegMap (top panel) and the HapMap (bottom panel).} 20 QTLs were simulated, which explained half of the genetic variance. The simulated heritability was 0.2 (left column), 0.5 (middle column) and 0.8 (right column). Within each panel, the first row shows the ANOVA-based estimates of broad-sense heritability, the second row the mixed model based estimates based on the individual plant data, and the third row the mixed model based estimates based on genotypic means.}%\label{boxplot3}
\label{h2_sim histograms_structured_and_hapmap}
%\label{h2_sim histograms_structured}
\end{figure}

Heritability estimates for the structured RegMap and HapMap are given in Figure \ref{h2_sim histograms_structured_and_hapmap} and Table \ref{sd_h2_sim_pop1_AND_pop3_regmap_new_gamma5_5000traits_200acc_20QTLs_3rep_ALL}.
Apart from ${\hat h}_m^2$ in the HapMap, none of the estimators showed marked bias. In particular, ${\hat H}^2$ had negligible bias, despite the fact that it does not account for genetic relatedness.
As in the asymptotic results, ${\hat h}_r^2$ had lower variance than ${\hat h}_m^2$, the difference being largest for the HapMap.
The magnitude of the differences was larger than expected from the asymptotic variances in Table \ref{asymptotic_variance_table}, and in the HapMap standard errors were up to $13$ times larger.
As in the asymptotic framework, differences were smallest for the structured RegMap and largest for the HapMap.

Remarkably, also ${\hat H}^2$ was considerably more accurate than ${\hat h}_m^2$. The additional marker information appeared to be less important here than the loss of information on within genotype variability. The estimator ${\hat h}_r^2$, which includes both sources of information, outperformed ${\hat H}^2$.
Differences were largest for the structured RegMap ($12.7\%$ reduction when $h^2=0.2$, $42.6\%$ reduction when $h^2=0.8$),
where the large differences in relatedness provide more information than in case of the HapMap (almost no difference when $h^2=0.2$ and $7\%$ reduction when $h^2=0.8$).
Simulation results for the Swedish and French RegMap are given in Figure S3 and Table S2.
In terms of standard deviations of ${\hat h}_r^2$, ${\hat h}_m^2$ and ${\hat H}^2$, these populations were somewhere in between the structured RegMap and HapMap.

For a substantial proportion of the simulated traits in both the HapMap and structured RegMap, we observed a phenomenon
that was not apparent from the asymptotic arguments above:
${\hat h}_m^2$ was close or equal to $1$, in which case the likelihood as function of $h^2$
was monotonically increasing. This behavior is most likely to occur when the genetic relatedness matrix is close to compound
symmetry (i.e. all genotypes being equally related), in which case the likelihood is constant in $h^2$ (File S3).
The small sample size %(compared to human studies)
can then lead to monotone likelihood profiles (Figure S4).
Indeed ${\hat h}_m^2$ was equal to $1$ most often for the HapMap: even for a simulated heritability of $0.5$, ${\hat h}_m^2$
was larger than $0.99$ for $882$ of the $5000$ traits (Figure \ref{h2_sim histograms_structured_and_hapmap}).
For the HapMap and simulated $h^2=0.2$, we also observed ${\hat h}_m^2$ equal to zero.

The simulations were repeated for a genetic architecture where only ten percent of the
genetic variance consists of polygenic effects, and the remaining $90 \%$ is determined by a single QTL
(File S4). Even in this extreme scenario, ${\hat h}_r^2$ still performed
very well, having almost negligible bias. % (REFER TO SPEED).
Also the improvement in standard error relative to ${\hat h}_m^2$ remained similar.

\subsection*{Confidence intervals}

For all simulated traits we calculated both standard and log-transformed confidence intervals for narrow sense heritability ($h^2$), using genotypic means as well as individual plant data. Also confidence intervals for broad-sense heritability ($H^2$) were calculated (We recall that only additive genetic effects were simulated, hence $h^2=H^2$).
Coverage and width of confidence intervals are given in Table \ref{conf_h2_sim_pop1_and_pop3_regmap_new_gamma5_5000traits_200acc_20QTLs_3rep_ALL}, and Table S3 (Swedish and French RegMap).
The standard $95 \%$ intervals obtained from the individual observations
(i.e. those associated with ${\hat h}_r^2$) had around $95 \%$ coverage in the simulations, while coverage was below $95 \%$ for the intervals based on genotypic means.
The latter intervals also had larger width, which relates to the higher variance of ${\hat h}_m^2$ in simulations. For the HapMap average width was larger than $0.5$, even at a simulated heritability of $0.2$. For the simulated traits with ${\hat h}_m^2=1$, the corresponding confidence interval was $[0,1]$.

Differences between the log-transformed and the standard intervals were small for ${\hat h}_r^2$.
For ${\hat h}_m^2$ the use of the log-transformed intervals gave bigger improvements, although coverage was still below $95 \%$. %Finally,
Also the confidence intervals for broad-sense heritability had insufficient coverage. This is due to model-misspecification: the analysis of variance does not incorporate the genetic structure, and uncertainty in the estimates is therefore underestimated. Coverage was still around $93 \%$ for the HapMap, whereas for the structured RegMap it decreased to $67.6 \%$ for $h^2=0.8$.

\subsection*{Heritability estimates for real data}

The heritability estimates ${\hat h}_r^2$, ${\hat h}_m^2$ and $\hat H^2$ were calculated for the six traits.
Genotypic means were calculated as the Best Linear Unbiased Estimators (BLUEs) for the genetic effects, in a linear model containing fixed effects for both genotype and additional design effects (Appendix \ref{App_marker_based_h2}).
In the calculation of ${\hat h}_m^2$ the non-diagonal covariance structure of the BLUEs was taken into account. % (Appendix \ref{App_marker_based_h2}). % (Materials and Methods).
In the case of ${\hat h}_r^2$ and $\hat H^2$, design effects were directly included in the mixed model.

\begin{figure}[!ht]
\begin{center}
\includegraphics[height=6in,width=16cm]{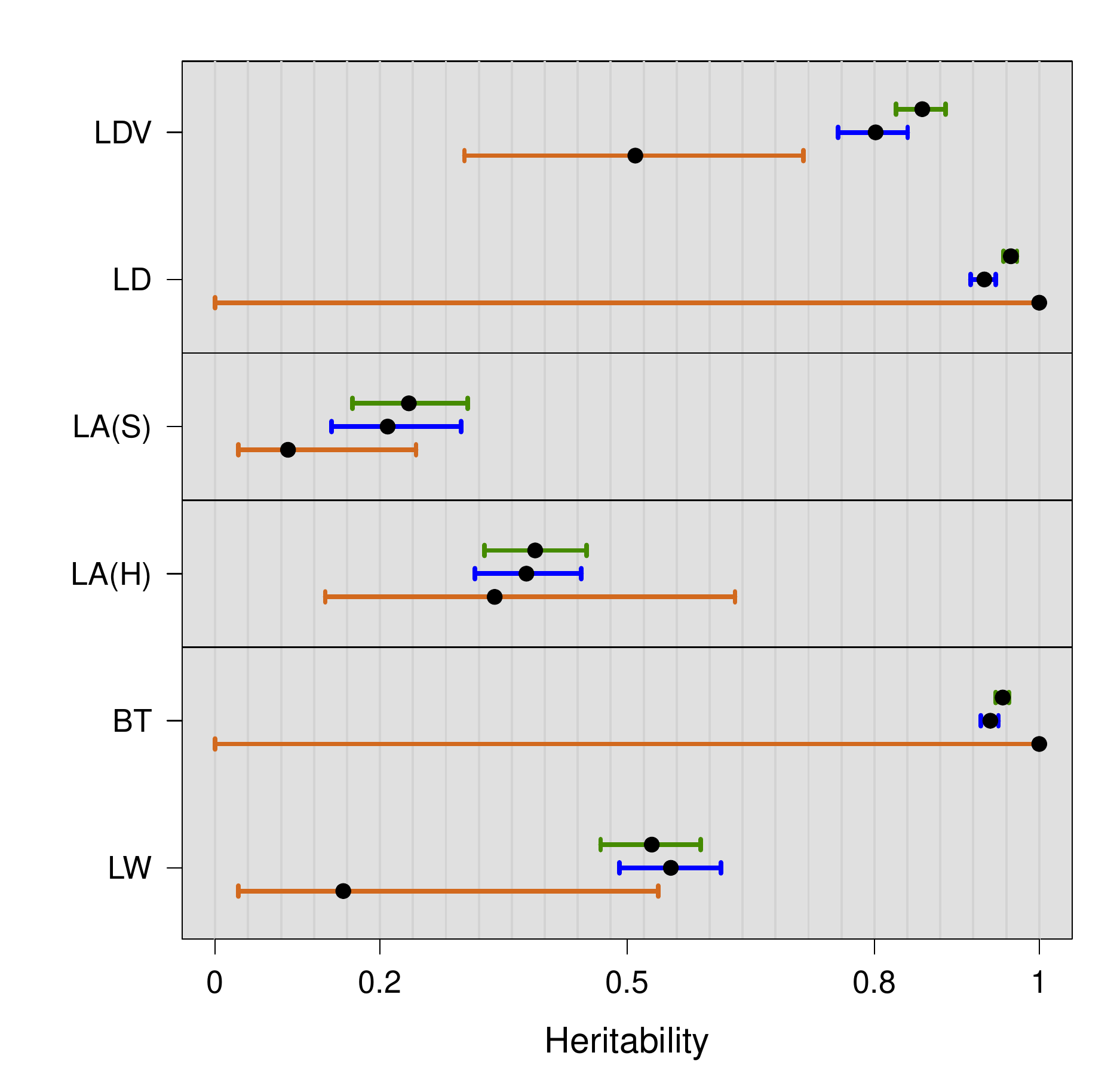}
\end{center}
\caption{
{\bf Heritability estimates and confidence intervals for two flowering traits from Atwell \emph{et al.} 2010 (LDV and LD), and $4$ traits from new experiments.}  Three estimators were used: the ANOVA-based estimator of broad-sense heritability (${\hat H}^2$, green),
the marker-based estimator using individual plant data (${\hat h}_r^2$, blue) and the marker-based estimator using genotypic means (${\hat h}_m^2$, brown). Traits from different experiments are separated by the black horizontal lines. Trait abbreviations are given in Table \ref{trait_abbreviations}. Confidence intervals associated with the marker-based estimates are constructed using the logarithmic transformation described in the Materials and Methods.
}
\label{all_h2}
\end{figure}

For the flowering traits from \cite{atwell_etal_2010}, %, for which the original data were kindly provided by the Gregor Mendel Institute, Vienna.
our estimates of  broad-sense heritability were $0.858$ for LDV and $0.966$ for LD (Table \ref{h2_values}).
For several reasons (File S2), these values were lower than those reported in Supplementary table 7 of \cite{atwell_etal_2010}: $0.94$ for LDV and $0.99$ for LD.
For both traits, marker-based heritability estimates based on individual plant data (${\hat h}_r^2$) were substantially lower, but still high ($0.80$ for LDV and $0.93$ for LD).
Marker-based heritability estimates obtained from genotypic means (${\hat h}_m^2$) were very different: much lower in case of LDV ($0.51$), and equal to $1$ for LD.
(Table \ref{h2_values} and Figure \ref{all_h2}). In the latter case, the likelihood as a function of $h^2$ was monotonically increasing, just as we observed for some of the simulated traits.
The estimate of residual variance (not reported) was virtually zero in this case, and the confidence interval equal to the whole unit interval.
Also in the leaf area experiments %on the Swedish RegMap and the HapMap
there were substantial differences between ${\hat h}_r^2$ and ${\hat h}_m^2$, especially for the Swedish RegMap (${\hat h}_r^2=0.21$
versus ${\hat h}_m^2$=0.09). Heritability estimates were larger for the HapMap (${\hat h}_r^2=0.38$ versus ${\hat h}_m^2$=0.34),
which has greater genetic diversity. Again, confidence intervals associated with ${\hat h}_m^2$
were wider than those associated with ${\hat h}_r^2$. In the final experiment, heritability estimates for BT were very similar to those for LD: ${\hat h}_m^2$ equalled $1$, and ${\hat h}_r^2$ and ${\hat H}^2$ were close to $1$. For LW we found ${\hat h}_m^2=0.16$, much lower than ${\hat h}_r^2=0.55$. The latter was slightly larger than the broad-sense heritability estimate ${\hat H}^2=0.53$, but the confidence intervals largely overlap.

\subsection*{Genomic prediction with G-BLUP}
%{accuracy_table}

The best linear unbiased predictor of the genetic effects (G-BLUP) depends on the estimated genetic- and residual variance, and hence on the estimated heritability (\cite{henderson_1975}, \cite{robinson_1991}).
Therefore genomic prediction with G-BLUP could potentially be improved when individual plant data are used, instead of genotypic means.
We compared G-BLUP based on genotypic means and individual plant data using simulated traits, each including a training set of $n=200$ genotypes and a validation set of $m=50$ genotypes, for which only marker information was available. For both sets of genotypes, we obtained the  G-BLUP ($\hat G$) predicting the true (simulated) genetic effects $G$,
and calculated the prediction accuracy in terms of the correlation ($r$) between $\hat G$ and $G$.

\begin{figure}[!ht] % figure produced with the script genomic_selection_PLOTS_correlations.r
\includegraphics[width=15cm,height=19cm]{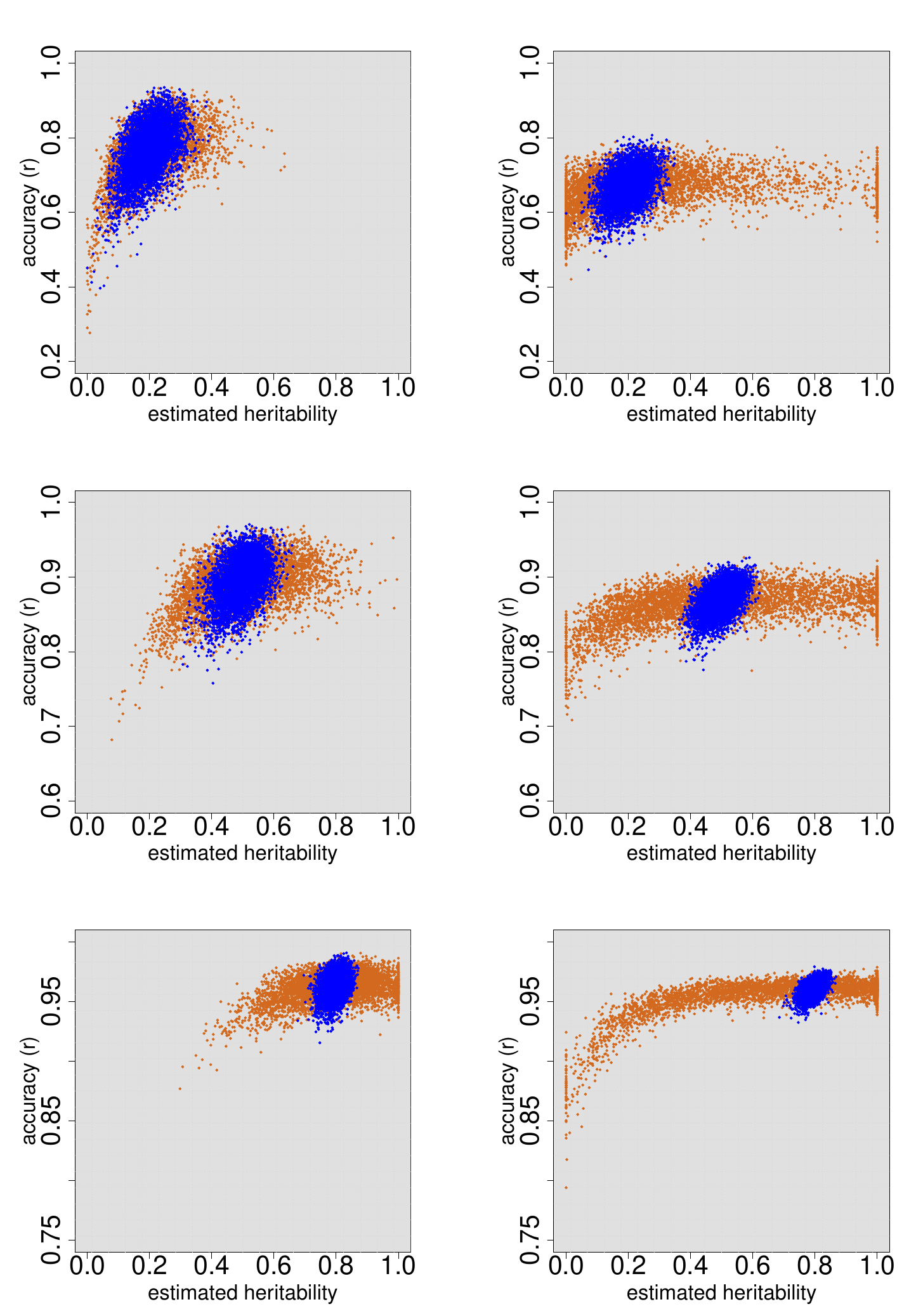}
\caption{{ \bf Prediction accuracy ($r$) of G-BLUP on training sets of $200$ accessions, for $5000$ simulated traits, for the Structured RegMap (left) and HapMap (right).} Within both populations, each trait was simulated for a randomly drawn training set of 200 accessions. Genetic effects were predicted using G-BLUP, based on either a mixed model for the individual plants (blue) or for the genotypic means (orange). 20 QTLs were simulated, which explained 50 percent of the genetic variance. The simulated heritability was $0.2$ (top), $0.5$ (middle) and $0.8$ (bottom).}
\label{gs_h2_all}
%\label{gs_h2_02}
\end{figure}

\begin{figure}[!ht] % figure produced with the script genomic_selection_PLOTS_correlations.r
\includegraphics[width=15cm,height=19cm]{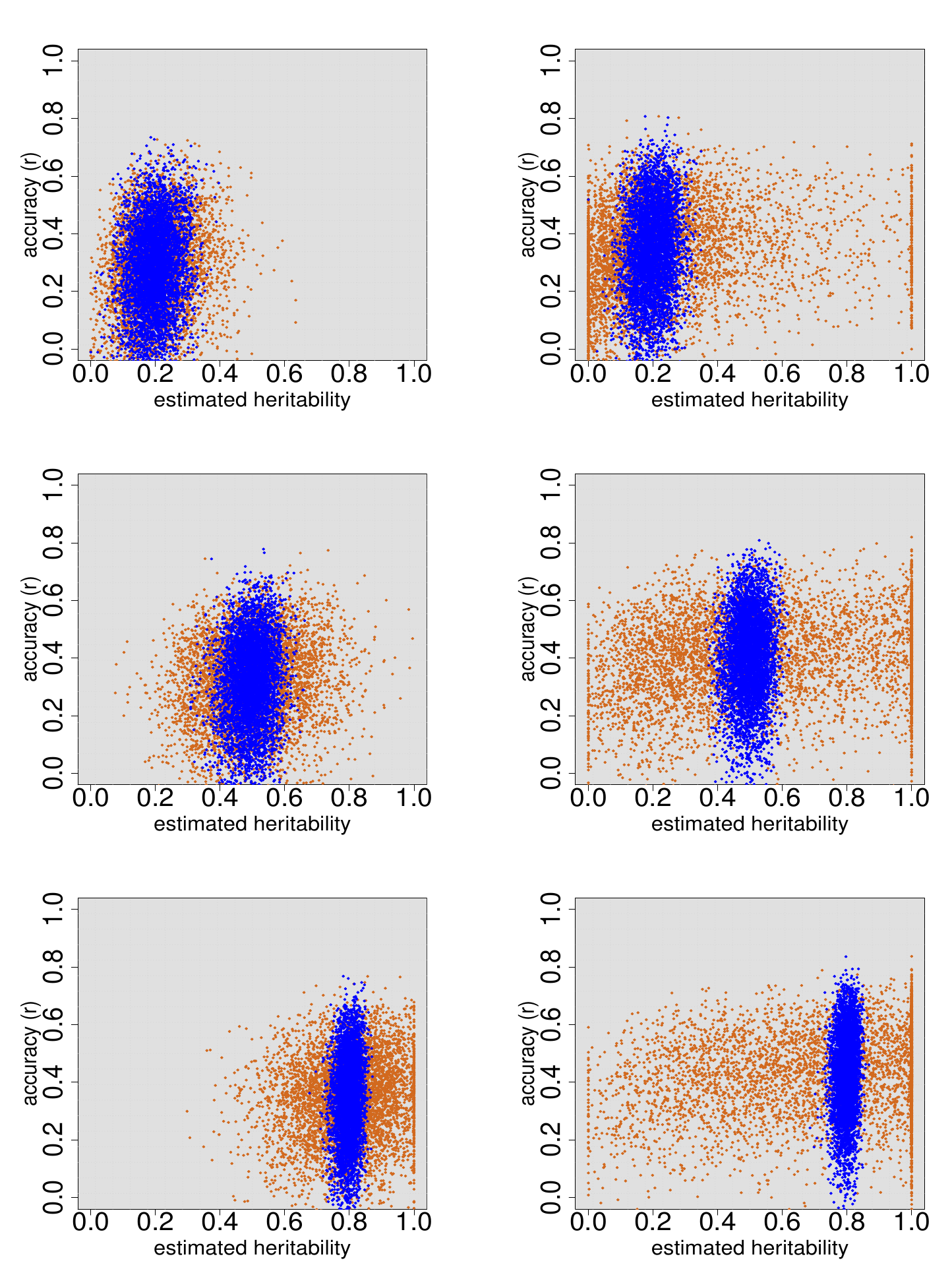}
\caption{{ \bf Prediction accuracy ($r$) of G-BLUP on validation sets of $50$ accessions, for $5000$ simulated traits, for the Structured RegMap (left) and HapMap (right).} Within both populations, each trait was simulated for a randomly drawn training set of 200 accessions. Genetic effects  for a randomly drawn validation set of $50$ accessions were predicted using G-BLUP, based on either a mixed model for the individual plants (blue) or for the genotypic means (orange). 20 QTLs were simulated, which explained 50 percent of the genetic variance. The simulated heritability was $0.2$ (top), $0.5$ (middle) and $0.8$ (bottom).}
\label{gs_h2_all_crit2}
\end{figure}

For the training and validation sets, accuracy decreased when the  estimated heritability was too far from the simulated heritability (Figures \ref{gs_h2_all} and \ref{gs_h2_all_crit2}, Table S5). For the validation sets there were larger differences in accuracy across simulations, because of the additional randomness in the selection of the validation genotypes, creating varying degrees of connectedness with the training-sets. When using individual plant data, heritability estimates were never far from the simulated heritability, and prediction accuracy was close to a constant depending on the simulated heritability.
Using only genotypic means, heritability was often be severely over- or underestimated, in which case accuracy decreased substantially. This decrease was largest when heritability was underestimated, which follows from the mathematical expressions for G-BLUP (Materials and methods and File S5).

In the HapMap population with simulated heritability $0.8$, the estimate ${\hat h}_r^2$ was between $0.7$ and $0.9$ for $4998$ of the $5000$ simulated traits, and accuracy ($r$) averaged over these simulations was $0.431$ (Table \ref{accuracy_table}). The heritability estimate based on means (${\hat h}_m^2$) was however between $0.1$ and $0.3$ for $8.4\%$ of the simulated traits, and for $2.6 \%$ it was even smaller than $0.1$. Averaged over the latter group of traits, accuracy was only $0.289$. Consequently, for these traits an improvement in accuracy of $49\%$ could be realized by genomic prediction based on individual plant data instead of genotypic means.
Averaged over all simulated traits, the prediction accuracy obtained from genotypic means was not much lower, because of the large 'safe zone' where the approaches performed similarly: accuracy was at least $0.4$ once the estimated heritability was above $0.5$.
It is impossible however to determine whether a \emph{given} trait is within this zone using genotypic means only. When individual plant data are available the heritability estimates ${\hat h}_r^2$ and ${\hat h}_m^2$ could be compared, and the G-BLUP based on genotypic means could be used if ${\hat h}_r^2$ and ${\hat h}_m^2$ were similar.

\begin{figure}[!ht] % figure produced with the script genomic_selection_PLOTS_correlations.r
\includegraphics[width=15cm,height=19cm]{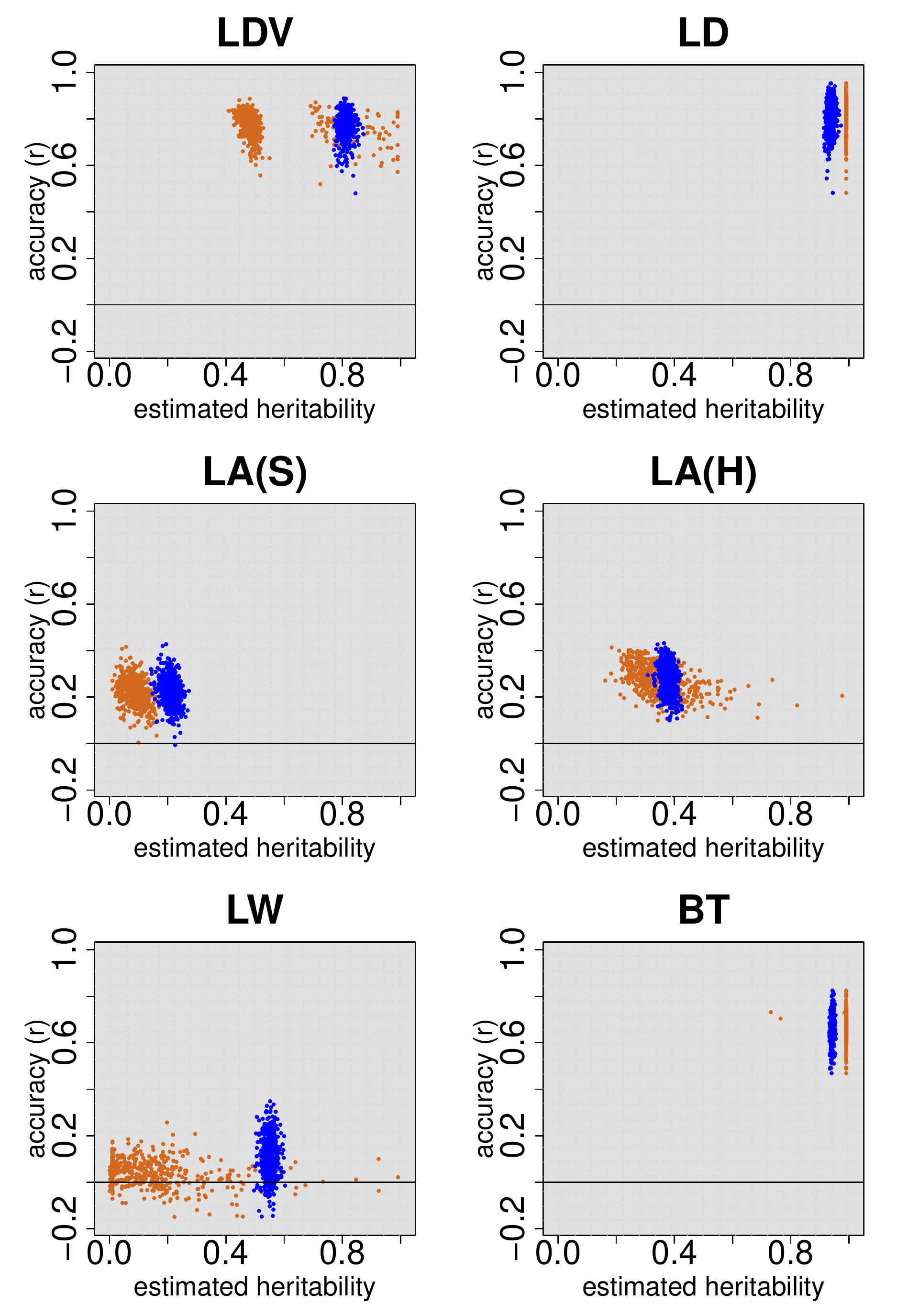}
\caption{{ \bf Prediction accuracy ($r$) of G-BLUP in $500$ cross-validations, for $6$ traits observed on \emph{A. thaliana}.} In each cross-validation, the accessions are randomly partitioned into a training set of $80\%$ and a validation set of $20\%$ of the accessions. Predictions for the individual plant data for for the accessions in the validation set were obtained from equation \eqref{data_prediction} in Appendix \ref{App_GP}. The genetic effects in \eqref{data_prediction} were estimated either using individual plant data (blue) or using genotypic means (orange).
}
\label{gs_h2_real_data}
\end{figure}

Similar results were obtained for prediction accuracy observed in cross-validations on the $6$ observed traits (Figure \ref{gs_h2_real_data}). Averaged over $500$ validation sets, prediction accuracies obtained with individual plant data and means were almost the same for LDV, LD and BT (respectively $0.77$, $0.82$ and $0.67$).
For LA(S), LA(H) and LW, average accuracy was respectively $0.22$, $0.27$ and $0.04$ when using genotypic means, and $0.23$, $0.28$ and $0.11$ when using individual plant data. For LDV, the heritability estimates obtained using genotypic means followed a bimodal pattern, which was largely due to accession 8233, for which only a single observation was available, and which genotypically is an outlier (Figure S6). Despite the fact that the standard errors and correlations among the genotypic means were taken into account, training sets including this accession produced much lower heritability estimates than in cases where it was assigned to the validation set. This however did not lead to lower prediction accuracy. More generally, the relation between low prediction accuracy and underestimating heritability was less clear than in Figure \ref{gs_h2_all_crit2}, which may be due to the additional uncertainty in the estimation of fixed effects, and the fact that this concerns a sub-sampling of observed traits, rather than simulated traits. Nonetheless, genomic prediction based on individual plant data again performed at least as good as the standard approach based on means, and much better in case of LW.

\subsection*{Genome-wide association studies (GWAS)}

\begin{figure}[!ht]
%\renewcommand{\figurename}{Supplementary Figure 6(b)}
%\begin{center}
\includegraphics[height=5in,width=7in,angle=90]{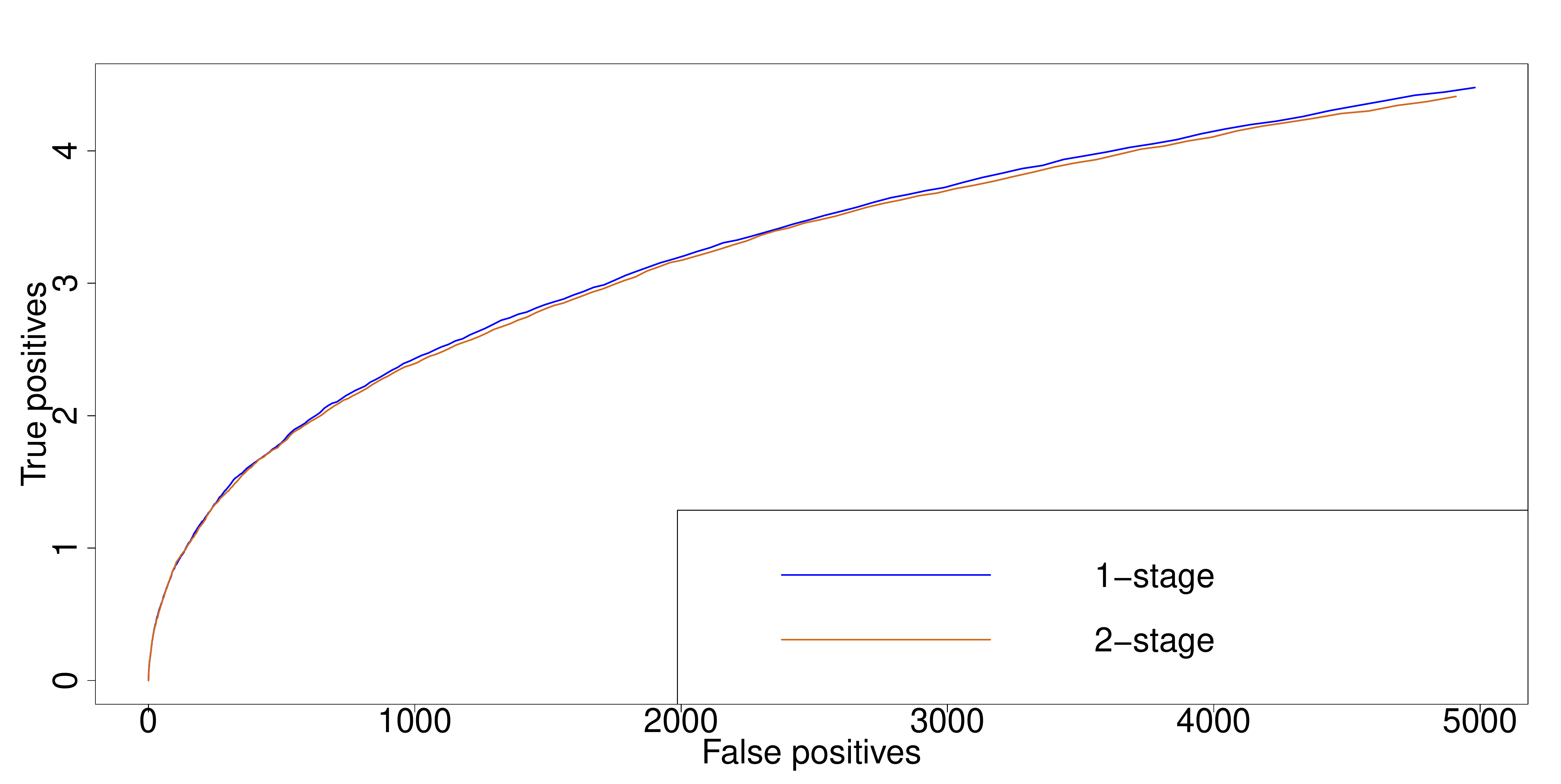}
%\includegraphics[height=6in]{GWAS_ROC_hapmap.pdf}
%\end{center}
\caption{
{\bf Receiver operating characteristic for the GWAS-simulations.}  1000 traits were simulated, for random samples of 200 accessions drawn from the HapMap.  10 QTLs were simulated, at randomly chosen SNP-positions (see the simulation section in the Materials and Methods for more details). The QTLs explained $75 \%$ of the genetic variance. The simulated heritability was $0.5$.
The curves were obtained by moving from a small (conservative) significance threshold to a larger one. Simulated QTLs were counted as true positive if their p-value was below the threshold; all other SNPs below the threshold were counted as false positive. %Similar results are obtained when including LD-windows around the simulated QTLs.
}
\label{roc_hapmap}
\end{figure}

Many state-of-the-art methods for GWAS
(\cite{kang_etal_2010},
\cite{lippert_etal_2011},
\cite{Zhou_Stephens_2012},
\cite{lipka_etal_2012})
use the same mixed model
as in marker-based estimation of heritability, apart from the additional marker effect (Appendix \ref{App_GWAS}).
When testing the significance of this marker effect, the estimated genetic and residual variance (and hence the heritability) determine the correction for population structure or genetic effects elsewhere on the genome.
This suggests that poor estimation of the variance components $\sigma_A^2$ and $\sigma_E^2$ may also affect association mapping, especially when these are estimated in a model without marker effects and then kept fixed when calculating generalized least squares (GLS) estimates of marker effects (\cite{kang_etal_2010}).
The GLS estimate of the fixed effects (including the marker effect) $\hat \beta = (X^t V^{-1} X)^{-1} X^t V^{-1} Y$ is unbiased since the expectation of $Y$ is $X\beta$.
Hence $E(\hat \beta) = \beta$, even conditionally on a poor estimate of $V = \hat \sigma_A^2 Z K Z^t+ \hat  \sigma_E^2 I_N$.
However, p-values for the significance of marker effects will be more sensitive to poor estimates of $(\sigma_A^2,\sigma_E^2)$ than the effect estimates themselves. We investigated this with GWAS on simulated traits for the HapMap and RegMap populations, which was performed on genotypic means (two-stage) and individual plant data (one-stage).
Rank correlations between between one- and two-stage $-{\log}_{10}$ p-values were almost $1$ when ${\hat h}_m^2$ was close to the simulated heritability, and decreased when ${\hat h}_m^2$
under- or over-estimated heritability (Figure S7). However, even then correlations were still high, and the resulting loss in power appeared to be limited (Figure \ref{roc_hapmap}). For the RegMap (results not shown) differences between ROC curves were even smaller. In contrast to the accuracy of genomic prediction, these differences remained small when restricting the set of simulated traits to those for which ${\hat h}_m^2$ underestimated the simulated heritability (curves not shown). This may be explained by the fact that the correlation between $\hat \beta$ and $(\hat \sigma_A^2,\hat \sigma_E^2)$ tends to zero for large numbers of genotypes (\cite{searle_casella_mcculloch_2006}).

\subsection*{Software}

In most of our analyses and simulations we used the commercial R-package \verb|asreml| (\cite{butler_etal_2009}), which contains a fast implementation of the AI-algorithm.
We also made our own implementation of the AI-algorithm, together with functions for estimating heritability. These are contained in the R-package \verb|heritability|, which is freely available online. % \verb|https://sites.google.com/site/wkruijer/|
In contrast to other packages such as \verb|emma| and \verb|synbreed|, it provides confidence intervals for heritability.

For GWAS on simulated traits we developed the command-line program \verb|scan_GLS|, which is available on request.
\verb|scan_GLS| performs generalized least squares calculations conditional on variance components estimated in a model without markers, as proposed by \cite{kang_etal_2010}, and can efficiently handle genetically identical individuals. State-of-the-art association mapping software can only perform association mapping on genetically identical individuals, when these are given different identifiers, i.e. as if they were different genotypes. This leads to large genotypic data files and increases computation time.
The generalized least squares (GLS) calculations in \verb|scan_GLS| are more efficient since they use the fact that $Z K Z^t$ and $K$ have the same rank. This has been proposed in \cite{lippert_etal_2011} (supplement), but to our knowledge this has not been implemented in the Fast-LMM software.
Although we did not find one-stage GWAS to be more powerful in the present study, the ability to perform fast association mapping for genetically identical individuals is useful in the context of a compressed kinship matrix
(\cite{Bradbury_Zhang_Kroon_Casstevens_Ramdoss_Buckler_2007}, \cite{zhang_etal_2010}, \cite{lipka_etal_2012}).
\verb|scan_GLS| also includes a function to perform GLS calculations with non-diagonal residual variance structure, allowing association mapping with extra (possibly non-genetic) random effects.

\section*{Discussion}

We have presented new methodology for marker-based estimation of heritability for plant traits, which accounts for genetic relatedness and includes information on within genotype variability, available from replicates. Our approach offers an alternative to mixed-model analysis based on genotypic means, which is the current practice in GWAS and genomic prediction with G-BLUP.
Although mixed models can indeed estimate heritability from only kinship coefficients and genotypic means, we observed very large standard errors and sometimes unrealistic estimates of heritability, in both published data and new experiments. Using simulations and statistical arguments we showed that marker-based estimation of heritability based on genotypic means has indeed severe limitations when applied to commonly used association panels in \emph{A. thaliana}. The main reason for this is the lack of information on within genotype variability: heritability estimates are exclusively based on the (usually small) differences between genotypes. This is feasible in human cohorts with many thousands of individuals, but gives insufficient information in plant populations with only several hundreds of different genotypes, even if standard errors of genotypic means are taken into account. Much more accurate heritability estimates were obtained %with our new methodology, which is based on
with mixed model analysis at individual plant or plot level.
The resulting heritability estimates had accuracies similar to those reported for human diseases (see e.g. \cite{speed_hemani_johnson_balding_2012}).
In our simulations with $200$ genotypes, the difference in accuracy relative to that of estimates obtained from genotypic means was larger than what was expected from the asymptotic variances. The reason for this larger difference appeared to be the monotone likelihood profile occurring in a substantial part of the simulations, giving heritability estimates of zero or one. However even without this phenomenon,
heritability estimation using individual plant data is more precise, and the asymptotic approximations appear to provide a lower bound on the gain in accuracy.

Mixed model analysis at individual plant level can also improve accuracy of G-BLUP, in particular when the actual heritability is underestimated. Too much shrinkage then leads to lower prediction accuracy.
In GWAS, where the interest is in the estimated marker effects rather than the variance components,
inclusion of individual plant gave almost no increase in power.
However, the possibility to include covariates observed at individual plant level may be an important practical advantage. While two-stage procedures are usually considered preferable in complex multi-experiment settings (\cite{welham_etal_2010}, \cite{piepho_etal_2012}), a one-stage approach may give
a more convenient and less error-prone analysis of a single experiment with a simple design.

State-of-the-art phenotyping platforms can measure plant traits with increasing accuracy and throughput. Compared to human traits, the key advantages are that phenotyping is performed under experimental conditions and can include different individuals of the same genotype. Our findings suggest that in order to fully exploit these advantages, statistical analysis at individual plant level is necessary.
Obviously, this requires the availability of the individual plant data, as well as covariates. For many studies in the literature this information is however not available, since most online resources only store genotypic means. More specifically, our results are relevant in the light of the missing heritability debate. Although the aim here was not to propose specific explanations for missing heritability,
any such explanation clearly requires an accurate estimate of heritability in the first place. Standard errors for heritability are commonly reported for human traits, but usually absent in the \emph{A. thaliana} literature.  We demonstrated that these standard errors can be extremely large when using mixed models at genotypic means level. Explaining missing heritability based on such estimates then becomes an unreasonable goal.

For two flowering traits, from both a published and a new experiment, ${\hat h}_m^2$ was equal to $1$. In these cases, also ${\hat h}_r^2$ and the broad-sense heritability estimates (${\hat H}^2$) were very high. This can partly be explained by the discrete scale (in whole days) on which the traits were measured. Some genotypes therefore had exactly equal phenotypic values in all or many of the replicates. The heritability estimates for LD were also affected by the fact that in the \cite{atwell_etal_2010} data, all non-flowering plants were given a phenotypic value of $200$.  However, the estimates ${\hat h}_m^2$ being equal to $1$ also occurred for some of our simulated (Gaussian) traits, and has a more fundamental reason: the small number of genotypes (compared to human studies) leads to monotone likelihood profiles and very large standard errors, making it hard to make any statement about heritability.

Recently, \cite{speed_hemani_johnson_balding_2012} showed that heritability estimates may become biased when linkage disequilibrium (LD) is not constant over the genome, and proposed LD-adjusted kinship (LDAK) matrices to correct for this.
To assess the effect on ${\hat h}_r^2$ and ${\hat h}_m^2$ we re-calculated these heritability estimates using an LD-adjusted kinship matrix (Figure S5 and Table S4).
For all traits the estimates ${\hat h}_r^2$ were very close to the values obtained using the unadjusted kinship matrix.  The estimates ${\hat h}_m^2$ on the other hand were quite different for several traits. Nevertheless, the same problems occurred: very large confidence intervals, and estimates that appear biologically unrealistic.
We conclude that ${\hat h}_m^2$ is more sensitive to the choice of kinship matrix than ${\hat h}_r^2$, and that the
different behavior of ${\hat h}_r^2$ and ${\hat h}_m^2$ cannot be explained by LD being different over the genome.

The mixed models analysis at individual plant level proposed here
can be extended in several ways, for example by partitioning the genetic variance into different chromosomal contributions, as in \cite{Yang_HongLee_Goddard_Visscher_2011}, or by including epistatic effects. The latter has been proposed for experimental populations with known pedigrees (\cite{oakey_etal_2007}),
but marker-based estimation of epistatic effects for natural populations appears only possible with more advanced statistical methodology,
for example semi-parametric mixed models and reproducing kernel Hilbert spaces (\cite{gianola_vankaam_2008}, \cite{howard_etal_2014}). Another direction for future research is the estimation of heritability in the presence of additional random effects, which would increase the applicability to agricultural field trials (where the raw data are usually at \emph{plot} rather than individual plant level).
Although our approach allows for missing values, it does assume that all design effects can be modeled as fixed. Field trials are often laid out in incomplete blocks, containing only a small number of the genotypes under study. Differences between such blocks are usually best modeled using random block effects.
The definition of heritability in such contexts is however far from obvious: \cite{oakey_etal_2007} proposed generalized heritability,
but for natural populations this definition is not equivalent to the classical definition of heritability (\cite{oakey_etal_2007}, page 813). In this case the ratio of the estimated genetic variance over the total phenotypic variance could be used as a lower bound on heritability.
Nevertheless, we have demonstrated that mixed model analysis at individual plant or plot level offers important advantages over mixed models for genotypic means.

\section*{Supporting information}

\textbf{File S1: Confidence intervals for broad-sense heritability estimates.} %\ref{supmat_H2}

\noindent
\textbf{File S2: Comparison with the broad-sense heritability estimates reported in \cite{atwell_etal_2010}.}
% \ref{supmat_atwell}

%\noindent
%\textbf{File S3: Asymptotic results for marker-based estimation of heritability.}
%%(Supplementary material \ref{theoretical_results}).

\noindent
\textbf{File S3: When all pairs of genotypes are equally related, the likelihood is constant in $h^2$.}%When the population structure is compound symmetry }
%(Supplementary material \ref{supmat_compound_symmetry})

\noindent
\textbf{File S4: Simulation results for a different genetic architecture.}
%Supplementary material \ref{supmat_different_architecture}).

%\noindent
%\textbf{File S5: Genome-wide association mapping for two flowering traits from \cite{atwell_etal_2010} and 4 traits from new experiments.}
%Comparison of one- and two-stage approaches.
%%(full results given in Supplementary material \ref{supmat_gwas}).

\noindent
\textbf{File S5: Expressions for the prediction error variance of G-BLUP for genotypes in the training- and validation set}
%(see Supplementary material \ref{supmat_GS_MSE}),
% and we therefore treat the validation- and training-set separately.

\noindent
\textbf{Figure S1: histograms of the off-diagonal kinship coefficients, for 4 sub-populations of the RegMap.}
%(Supplementary material \ref{supmat_kinship_coefficients}).

\noindent
\textbf{Figure S2: histograms of the off-diagonal identity-by-state coefficients, for 4 sub-populations of the RegMap.}
%(Supplementary material \ref{supmat_kinship_coefficients}).

\noindent
\textbf{Figure S3: Simulation results for the Swedish and French RegMap.}
%supplementary material \ref{swedish_french_simulations}.

\noindent
\textbf{Figure S4: Monotone likelihood.}
%of $0.5$, ${\hat h}_m^2$ was larger than $0.99$ for $882$ of the $5000$ simulated HapMap traits (Figure \ref{h2_sim histograms_hapmap})

\noindent
\textbf{Figure S5: Heritability estimates using an LD-adjusted kinship matrix}
%(Supplementary material \ref{supmat_LDAK}: Table \ref{h2_values_LDAK} and Figure \ref{all_h2_LDAK}.

\noindent
\textbf{Figure S6: Principal component biplot of the genetic markers for the panel of \cite{atwell_etal_2010}.}
%(Supplementary material \ref{supmat_LDAK}: Table \ref{h2_values_LDAK} and Figure \ref{all_h2_LDAK}.

\noindent
\textbf{Figure S7: GWAS on simulated data: rank correlations between effect-size estimates obtained with one- and two-stage approaches.}
%(Supplementary material \ref{supmat_LDAK}: Table \ref{h2_values_LDAK} and Figure \ref{all_h2_LDAK}.

\noindent
\textbf{Table S1: accessions ID's of the structured RegMap.}
%Supplementary material \ref{supmat_structured_regmap}).

\noindent
\textbf{Table S2: Simulation results for the Swedish and French RegMap.}
%supplementary material \ref{swedish_french_simulations}.

\noindent
\textbf{Table S3: Simulations for the Swedish and French RegMap: coverage and width of confidence intervals.}
%Supplementary material \ref{swedish_french_simulations}.

\noindent
\textbf{Table S4: Heritability estimates using an LD-adjusted kinship matrix}
%(Supplementary material \ref{supmat_LDAK}: Table \ref{h2_values_LDAK} and Figure \ref{all_h2_LDAK}.

\noindent
\textbf{Table S5: Correlations between predicted and simulated genetic effects.}
%(Supplementary material \ref{supmat_LDAK}: Table \ref{h2_values_LDAK} and Figure \ref{all_h2_LDAK}.

%\noindent
%\textbf{Supplementary}

\section*{Acknowledgments}

%This work was partially funded by the EU \emph{DROPS} project, STW/NWO (Netherlands Organisation for Scientific Research) \emph{Learning from Nature project} and the CipY project.
This work was partially funded by the STW/NWO (Netherlands Organisation for Scientific Research) \emph{Learning from Nature} project (WK), the
Generation Challenge Program - Integrated Breeding Platform project 2.2.5 (MM and FvE), the NWO-ALW Technological Top Institute Green Genetics (PF) and the Centre for BioSystems Genomics (RK and JK).
We are grateful to Bjarni Vilhj\'almsson, Arthur Korte, Susanna Atwell and Magnus Nordborg for %providing the data from \cite{atwell_etal_2010} and
valuable discussions on heritability.
We want to thank Doug Speed for his comments on the use of his LDAK-software. Jian Yang is acknowledged for his comments on the calculation of confidence intervals in his GCTA software, and Brian Cullis for feedback on the use of asreml-R. Christian Riedelsheimer, Tobias Schrag and Albrecht Melchinger are acknowledged for providing the SNP-data from Riedelsheimer et al (2012). Finally, we would like to thank two referees for helpful suggestions regarding the terminology.

\section*{Author contributions}

WK and FvE designed the research. WK performed the research.
WK wrote the paper, with input from MM, BE, FvE, PF, MB, JK and RK.
Software was written by WK (R-package heritability) and MB (\verb|scan_GLS|).
Experimental data were obtained by PF (traits LA(S) and LA(H)) and RK (traits BT and LW).

%\section*{References}
% The bibtex filename

%\bibliography{d:/willem/Dropbox/research/statgen3}

%\openup 1em
\bibliographystyle{genetics}
\renewcommand\refname{Literature Cited}

%\bibliography{heritability}

%\openup -1em

\clearpage

%%%%%%%%%%%%%%%%%%%%%%%%%%%%%%%%%%%%%%%%%%%%%%%%%%%%%%%%%%%%%%%%%%%%%%%%%%%%%%%%%%%%%%%%%%%%%%%%%%%%%%%%%%%%%%%%
\appendix
%\begin{appendices}

\section{Marker-based  estimation of heritability} \label{App_marker_based_h2}

%Absence of $G \times E$ is assumed, and we also assume that all additional covariates can be modeled as fixed effects.
We start with a brief summary of existing methodology and the required assumptions, and then define marker-based heritability estimates given genetically identical replicates measured in an unbalanced design.

\subsection*{Marker-based  estimation of heritability given a single observation per genotype} % : using all individuals or genotypic means

Phenotypic variance can be partitioned  as
\begin{equation*}\label{partitioning}
\sigma_{\textrm{Pheno}}^2 = \sigma_G^2 + \sigma_{\textrm{Env}}^2 = \sigma_A^2 + \sigma_D^2 + \sigma_I^2 + \sigma_{\textrm{Env}}^2,
\end{equation*}
where the genetic variance $\sigma_G^2$ is decomposed into additive, dominance and interaction effects (\cite{falconer_mackay_1996}), the dominance effect being absent for inbred lines. % \emph{A. thaliana}
Heritability in the broad sense is defined as $H^2 = \sigma_G^2 / \sigma_{\textrm{Pheno}}^2$, while the narrow-sense heritability $h^2 = \sigma_A^2 / \sigma_{\textrm{Pheno}}^2$ takes into account only the additive genetic effects, which determine breeding values and the selection response.
Defining the residual variance $\sigma_E^2$ to be the sum of the environmental and non-additive genetic variance terms ($\sigma_D^2 + \sigma_I^2 + \sigma_{\textrm{Env}}^2$), we can write
\begin{equation} \label{h2_def}
h^2 = \frac{\sigma_A^2}{\sigma_A^2 + \sigma_E^2}.
\end{equation}
Marker-based estimates of narrow-sense heritability can be obtained using mixed models with random genetic effects. The covariances between these effects are modeled by a genetic relatedness matrix (GRM) estimated from markers, with elements given by \eqref{kinship}. Given a single observation per genotype, the standard infinitesimal model  is%heritability can be estimated using the model
\begin{equation} \label{basicModel}
Y_{i} = \mu + x_{i} \beta + G_i + E_{i} \quad  (i=1,\ldots,n),
\end{equation}
%where $Y_{i}$ is the phenotypic value of genotype $i$,
$\mu$ is the intercept, $G = (G_1,\ldots,G_n)$ has a $N(0,\sigma_A^2 K)$ distribution, and the errors $E_{i}$ have independent normal distributions with variance $\sigma_E^2$.
The optional term  $x_{i} \beta = x_{i}^{(1)} \beta_1  + \ldots + x_{i}^{(k)} \beta_k$ models the effect of $k$ additional covariates. % These can be either continuous or a factor.
Under model \eqref{basicModel}, heritability can be estimated by
\begin{equation} \label{h2estimate}
\hat h^2 = \frac{\hat\sigma_A^2}{\hat\sigma_A^2 + \hat\sigma_E^2},
\end{equation}
where $\hat\sigma_A^2$ and $ \hat\sigma_E^2$ are restricted maximum likelihood (REML) estimates of additive genetic and residual variance (\cite{Yang_HongLee_Goddard_Visscher_2011}, \cite{speed_hemani_johnson_balding_2012}).

It has been noted (e.g. \cite{Hayes_visscher_goddard_2009}) that using model \eqref{basicModel} with genetic relatedness matrix \eqref{kinship} is equivalent to assuming that the effects of the
the standardized marker scores are drawn independently from Gaussian distributions with variance $\sigma_A^2 / p$. %$p^{-1}\sigma_A^2 (f_l (1-f_l))^{-1}$.
Consequently, the model can only account for additive genetic effects, and non-additive effects will get into the residual variance. This is why $\hat h^2$ is an estimate of narrow-sense heritability. The preceding argument also highlights the fact that $\hat h^2$ is a marker based estimate, which %. To avoid bias This
requires the (frequently made) assumptions that every causal locus is tagged by one of the markers, and that linkage-disequilibrium (LD) is constant across the genome (\cite{speed_hemani_johnson_balding_2012}). For the case that LD is not constant over the genome, \cite{speed_hemani_johnson_balding_2012} proposed LD-adjusted kinship matrices.

The parameter $\sigma_A^2$ can only be interpreted as additive genetic variance if $K$ is scaled appropriately. For the kinship matrix used here %by \eqref{kinship}
this is the case, but for different types of kinship matrix (e.g. identity by state), it is necessary to divide each coefficient by $\textrm{tr}(P K P)/(n-1)$, for $P = I_n - \mathbf{1 1'} / n$. Under the condition that $\textrm{tr}(P K P) = n -1$,
\begin{equation} \label{kinship_scaling}
\begin{split}
E \left[\frac{1}{n-1} \sum_{i=1}^n (G_i - \bar G)^2 \right]&= E \left[\frac{1}{n-1} G^t P^t P G \right] = \frac{1}{n-1} E [ G^t P G ]\\
&= \frac{\sigma_A^2}{n-1} \textrm{tr} [P K P] = \sigma_A^2.
\end{split}
\end{equation}

\subsection*{Marker-based  estimation of heritability given genetically identical replicates, for unbalanced designs} % : using all individuals or genotypic means

When for each genotype a number of  genetically identical individuals are observed, \eqref{basicModel} generalizes to
\begin{equation} \label{infModelR_s}
Y_{i,j} = \mu + x_{i,j} \beta + G_i + E_{i,j} \quad (i=1,\ldots,n, \quad j=1,\ldots,r_i),
\end{equation}
where $Y_{i,j}$ is the phenotypic response of replicate $j$ of genotype $i$, $x_{i,j}$ a vector of fixed effects, $G = (G_1,\ldots,G_n)$ has a $N(0,\sigma_A^2 K)$ distribution, and the errors $E_{i,j}$ have independent normal distributions with variance $\sigma_E^2$. The last equation is similar to equation \eqref{infModelR} apart from the additional covariates and the generally unbalanced design.
Analogous to the case of balanced designs, we use the estimator ${\hat h}_r^2$ defined in \eqref{h2repl}, for REML estimates $\hat\sigma_A^2$ and $ \hat\sigma_E^2$ obtained for model \eqref{infModelR_s}.

Heritability estimates based on genotypic means can be constructed as in equations \eqref{infModelM2}-\eqref{h2means}, except that
these genotypic means are no longer equal to the arithmetic averages $\bar Y_i$. We fit the linear model
\begin{equation} \label{firstStage}
Y = X_G \:g + X_C \beta + E,
\end{equation}
where $Y = (Y_{1,1},\ldots,Y_{1,r_1},\ldots,Y_{n,1},\ldots,Y_{n,r_n})^t$ is the vector of phenotypic observations, $r_i$ is the number of replicates of genotype $i$, $X_C$ is the design matrix of the extra fixed effects, and $E = (E_{1,1},\ldots,E_{n,r_n})^t$ is the vector of independent Gaussian errors with variance $\sigma_E^2$. This model is identical to \eqref{infModelR_s},
apart from the factor genotype now being fixed instead of random.
The intercept is included in the incidence matrix $X_G$, which has dimension $N \times n$, for $N = \sum_{i=1}^n r_i$.

We obtain least squares estimates $\hat g = (\hat g_1,\ldots,\hat g_n)^t$ for the genotypic effects in model \eqref{firstStage}, which form the basis for the subsequent analysis.
In mixed-model terminology, this is the best linear unbiased estimator (BLUE) for $g$. In \eqref{firstStage} the only random term is $E$, i.e. we consider an ordinary linear model, as the additional covariates are all assumed to be fixed. The least-squares estimator $(\hat g, \hat \beta)$ is given by
\begin{equation} \label{bluesEquation}
\left(
\begin{array}{l}
\hat g \\
\hat \beta
\end{array}
\right)
=
(X^t X)^{-1} X^t Y =
\left(
\begin{array}{cc}
X_G^t X_G & X_G^t X_C \\
X_C^t X_G & X_C^t X_C
\end{array}
\right)^{-1}
\left(
\begin{array}{l}
X_G^t \\
X_C^t
\end{array}
\right)
Y,
\end{equation}
where $X = [X_G \: X_C]$.
This vector has  covariance matrix
\begin{equation} \label{varG}
\textrm{Var}
\left(
\begin{array}{l}
\hat g \\
\hat \beta
\end{array}
\right)
=
(X^t X)^{-1} \sigma_E^2 =
\left(
\begin{array}{cc}
X_G^t X_G & X_G^t X_C \\
X_C^t X_G & X_C^t X_C
\end{array}
\right)^{-1} \sigma_E^2.
%= R \sigma_E^2.
\end{equation}
Consequently, the covariance matrix of $\hat g$ is
\begin{equation*}
\textrm{Var}(\hat g) =
\left(X_G^t X_G  -  X_G^t X_C (X_C^t X_C)^{-1} X_C^t X_G\right)^{-1} \sigma_E^2  = R \sigma_E^2.
\end{equation*}
The mixed-model \eqref{infModelM2}, used for the estimation of heritability, now naturally generalizes to
\begin{equation} \label{infModelM2_s}
\bar Y_i = \mu + G_i + E_i, \qquad G \sim N(0,\sigma_A^2 K), \quad E \sim N(0,\sigma_E^2 R),
\end{equation}
where, to avoid having twice the letter $g$ in the same equation, we abused the notation by writing $\bar Y_i$ instead of $\hat g_i$.
Hence the matrix $r^{-1} I_n$ in \eqref{infModelM2} is replaced by $R$. Given REML estimates $\hat \sigma_A^2$ and $\hat \sigma_E^2$ for this model, we use ${\hat h}_m^2$ in \eqref{h2means} as heritability estimate.
As in the case of a completely randomized design, ${\hat h}_m^2$ is a so-called two-stage estimator.
The fact that genotype is taken as fixed effect in the first stage \eqref{firstStage} and as random in the subsequent mixed model analysis may appear inelegant, but is common in two-stage analyses, and necessary to avoid shrinking twice.

In case of a completely randomized design with $r_i$ replicates without further covariates, $\hat g_i = \sum_{j=1}^{n_i} Y_{i,j} / n_i$ and
\begin{equation*}
R =(X_G^t X_G)^{-1} = \textrm{diag}(r_1^{-1},\ldots,r_n^{-1}).
\end{equation*}
Note that in contrast to the estimator of broad-sense heritability described below, it is not necessary to define an average number of replicates.
In case of a randomized complete block design with $r$ replicates, $X_C$ is the $nr \times (r-1)$ design matrix for the factor block, and $X_G^t X_C (X_C^t X_C)^{-1} X_C^t X_G$ is the $n \times n$ matrix with elements $(r-1)/n$. Hence, $R$ is not diagonal, even though the design is balanced.
The off-diagonal elements are equal in this case, and small (provided $n$ is large sufficiently). With $n=200$ and $r=3$ for example, the diagonal elements of $R$ are $0.336667$ and the off-diagonal elements are $0.003333$.
For unbalanced designs, some of the off-diagonal elements may be larger and more influential.

\section{Simulations} \label{App_simulations}

Data for genotypes $i=1,\ldots,n$ with replicates $j=1,\ldots,r$ were simulated following the model
\begin{equation} \label{infModel2}
\left\{
\begin{array}{l}
y_{i,j} = \mu +  \sum_{m=1}^q x_{i,m} \alpha_m + g_i + e_{i,j}, \\
g = (g_1,\ldots,g_n)^t \sim N(0,\sigma_a^2 K), \\
e = (e_{1,1},\ldots,e_{n,r})^t \sim N(0,\sigma_e^2 I_{n r}),
\end{array}
\right.
\end{equation}
where $K$ was scaled such that $\textrm{tr}(P K P) = n -1$, for $P = I_n - \mathbf{1 1'} / n$ (see equation \eqref{kinship_scaling}).

We assumed $q$ QTLs located at marker positions, with effect sizes $\alpha_m$ ($m=1,\ldots,q$) and minor allele frequencies $f_m$ ($m=1,\ldots,q$), i.e. genotypes $AA$ and $aa$ occur with frequencies $f_m$ and $(1-f_m)$, $\alpha_m$ being the effect of a single allele.
%Let $n$ still denote the number of genotypes, and let $Z$ be the $(r n) \times n$ matrix that assigns the observations to the genotypes.
Let $x_{i,m} \in \{0,1,2\}$ denote the marker score at QTL $m$ for genotype $i$. In case of \emph{A. thaliana} inbred lines, $x_{i,m} \in \{0,2\}$ and
\begin{equation} \label{h2full}
h^2 = \frac{\sigma_a^2 + 4\sum_{m=1}^q f_m (1-f_m) \alpha_m^2}{\sigma_a^2 + 4\sum_{m=1}^q f_m (1-f_m) \alpha_m^2 + \sigma_e^2} =
\frac{\sigma_A^2}{\sigma_A^2 + \sigma_e^2},
\end{equation}
%For inbred populations ($x_{i,m} \in \{0,1\}$) $c=1$; for outbred populations $c=4$ (correct ?).
where
\begin{equation*}
\sigma_A^2 = \sigma_a^2 + 4\sum_{m=1}^q f_m (1-f_m) \alpha_m^2
\end{equation*}
is the total (additive) genetic variance.
For outbreeding species under Hardy-Weinberg equilibrium, the constant $4$ is to be replaced by a $2$.
In \eqref{h2full} we assume linkage equilibrium between the QTLs.
%The expression in \eqref{h2} is often referred to as a pseudo-heritability; see e.g. \cite{kang_etal_2010}.
%(If we had the total phenotypic variance in the denominator, it would be the narrow-sense heritability ?).

For each simulated trait, QTL-locations were sampled randomly from all available markers whose minor allele frequency exceeds $10\%$. To enforce linkage equilibrium between the sampled QTLs, the sampled locations were discarded and drawn again if the difference between $v_1 = 4\sum_{m=1}^q f_m (1-f_m) \alpha_m^2$ and $v_2 = 4(\alpha_1,\ldots,\alpha_q) S (\alpha_1,\ldots,\alpha_q)'$ was too large, $S$ being the sample covariance matrix of the marker scores at QTL-positions. It was required that $\min(v_1,v_2) / \max(v_1,v_2) > 0.97$. % The QTL-effects were given random signs.

Let
\begin{equation} \label{gamma}
\gamma = \frac{4}{\sigma_A^2} \sum_{m=1}^q f_m (1-f_m) \alpha_m^2
\end{equation}
denote and the proportion of the genetic variance explained by the QTLs. In our main set of simulations we choose $\sigma_e^2 = 1$, $\gamma=0.5$, $q=20$ QTLs.
For the comparison of one- and two-stage association mapping, we choose $\gamma=0.75$, $q=10$.
To achieve the desired level of heritability (for given $\gamma$ and $\sigma_e^2 = 1$), we set $\sigma_A^2 = \frac{h^2 (n-1)}{(1-h^2)n}$ and $\sigma_a^2 = (1-\gamma) \sigma_A^2$, and $\alpha_1,\ldots,\alpha_q$ were chosen equally large, such that $4 q f_m (1-f_m) \alpha_m^2 = \gamma \sigma_A^2$ for each $m$. The QTL-effects were given random signs. In File S4
we repeat the simulations for $\gamma=0.1$ and $q=1$.

For the purpose of genomic prediction, $50$ additional genotypes were drawn randomly from the same subpopulation of the RegMap. These were assigned only genotypic values, defined as the sum of QTL-effects $\sum_{m=1}^q x_{i,m} \alpha_m$ and polygenic effects $g_i$ ($i=201,\ldots,250$). The polygenic effects were simulated such that $(g_1,\ldots,g_{250})^t \sim N(0,\sigma_a^2 K_{total})$, $K_{total}$ being the kinship matrix of the training- and validation set combined.

\section{Genomic prediction with G-BLUP} \label{App_GP}

Genomic prediction with G-BLUP relies on the same models used for marker based estimation of heritability, and can be based on individual plant data (one-stage) or genotypic means (two-stage). We now provide the expressions for one- and two-stage G-BLUP, which are based on models \eqref{infModelR_s} and \eqref{infModelM2_s} in Appendix \ref{App_marker_based_h2}.

\subsection*{One- and two-stage G-BLUP for the training and validation set}

First we consider the G-BLUP for the training-set, for which phenotypic observations are available.
Including the intercept in the design matrix, model \eqref{infModelR_s} can be rewritten as
\begin{equation} \label{infModelR_matrixform}
Y = X\beta + Z G + E,
\end{equation}
where $N$ is the total number of individuals and $Z$ is the $N \times n$ incidence matrix assigning individuals to genotypes.
The BLUP of $G = (G_1,\ldots,G_n)^t$ and the best linear unbiased estimator (BLUE) of $\beta$ are given by
\begin{equation} \label{gblup1}
\hat G = \hat \delta K Z^t (\hat \delta Z K Z^t + I_N)^{-1} (Y - X \hat \beta), \quad \hat \beta = (X^t (\hat \delta Z K Z^t + I_N)^{-1} X)^{-1} X^t (\hat \delta Z K Z^t + I_N)^{-1} Y,
\end{equation}
where $N$ is the total number of individuals, $Z$ is the $N \times n$ incidence matrix assigning individuals to genotypes, and $\hat \delta = \hat \sigma_A^2 / \hat \sigma_E^2$. Numerically $\hat G$ and $\hat \beta$ can be more conveniently obtained by solving the mixed model equations (see e.g. \cite{henderson_1975} or \cite{robinson_1991}).
Similar expressions hold for predictions based on the genotypic means following model \eqref{infModelM2_s}:
\begin{equation} \label{gblup2}
\hat G = \hat \delta K (\hat \delta K + R)^{-1} (\bar Y - 1_n \hat \mu), \quad \hat \mu = (1_n^t (\hat \delta K  + R)^{-1} 1_n)^{-1} 1_n^t (\hat \delta K + R)^{-1} \bar Y,
\end{equation}
where, as in \eqref{infModelM2_s}, $\bar Y$ denotes the vector of genotypic means obtained from preliminary linear model analysis (not necessarily the arithmetic averages).
Since we assume that all covariates have been accounted for already in this preliminary analysis, the only fixed effect in \eqref{gblup2} is an intercept, with design matrix $X=1_n$.

Assuming the one-stage model \eqref{infModelR_matrixform} and $\hat \beta$ given by \eqref{gblup1}, the genetic effects $G_{\textrm{pred}} = (G_{n+1},\ldots,G_{n+m})^t$ of $m$ unobserved (but genotyped) genotypes can be predicted by the conditional means
\begin{equation} \label{Gpred1}
\hat G_{\textrm{pred}}^{(1)} := E[G_{\textrm{pred}} | Y] = \hat \delta K_{\textrm{pred.obs}} Z^t (\hat \delta Z K Z^t + I_N)^{-1} (Y - X \hat \beta), % Y_{\textrm{obs}}
\end{equation}
where $K_{\textrm{pred.obs}}$ is the $m \times n$ matrix of kinship coefficients for the unobserved versus observed genotypes.
Assuming the two-stage model \eqref{infModelM2_s} and $\hat \mu$ given by \eqref{gblup2}, the predictor is given by
\begin{equation} \label{Gpred2}
\hat G_{\textrm{pred}}^{(2)} = \hat \delta K_{\textrm{pred.obs}} (\hat \delta K + R)^{-1} (\bar Y - 1_n \hat \mu).
\end{equation}

\subsection*{Prediction of new observations and cross-validation}

For new observations $Y_{\textrm{pred}}$ at individual plant level, we have the one-stage predictor
\begin{equation} \label{data_prediction}
\hat Y_{\textrm{pred}}^{(1)} = X_{\textrm{pred}} \hat \beta + Z_{\textrm{pred}} \hat G_{\textrm{pred}}^{(1)},
\end{equation}
where $X_{\textrm{pred}}$ and $Z_{\textrm{pred}}$ are the corresponding design matrices.
For predictions with genotypic means, we replace $\hat G_{\textrm{pred}}^{(1)}$ by $\hat G_{\textrm{pred}}^{(2)}$, and $\beta$ by the estimate obtained within the linear model \eqref{firstStage} in the preliminary stage.

\section{Genome-wide association studies} \label{App_GWAS}

In mixed-model based GWAS (\cite{kang_etal_2010}, \cite{lippert_etal_2011}, \cite{Zhou_Stephens_2012},  \cite{lipka_etal_2012})  the phenotype of genotype $i$ is modeled as
\begin{equation} \label{infModelM_gwas}
Y_{i} = \mu +  x_i \gamma + G_i + E_{i},
\end{equation}
where $x_i$ is the marker score, $\gamma$ is the marker effect and the genotypic effects $G = (G_1,\ldots,G_n)$ follow a $N(0,\sigma_A^2 K)$ distribution.
This model assumes a single observation per genotype. If observations on genetically identical individuals are available, the $Y_i$'s can be replaced by genotypic means $\hat g_i$, as in \eqref{bluesEquation}, \eqref{varG} and \eqref{infModelM2_s}. Model \eqref{infModelM_gwas} then generalizes to
\begin{equation} \label{infModelM_gwas_2}
\hat g_{i} = \mu +  x_i \gamma + G_i + E_{i}, \qquad G \sim N(0,\sigma_A^2 K), \quad E \sim N(0,\sigma_E^2 R).
\end{equation}
This amounts to a two-stage approach: first the genotypic means are calculated using \eqref{bluesEquation}, and next association mapping is performed following \eqref{infModelM_gwas_2}. %{infModelM_gwas_2}
In practice GWAS is often performed on the arithmetic averages $\bar Y_i$, which implicitly assumes a balanced completely randomized design, without any missing values or replicate effects.
Here we use the more general model \eqref{infModelM_gwas_2}. Apart from the additional marker effect, this is the same model we used in \eqref{infModelM2_s} to construct the heritability estimate ${\hat h}_m^2$.

Alternatively, association mapping can be based on the one-stage model
\begin{equation} \label{infModelR_gwas}
Y_{i,j} = \mu +  x_i \gamma + (X_C)_{i,j} \beta +  G_i + E_{i,j},
\end{equation}
where the term $(X_C)_{i,j} \beta$ models additional covariates, as in \eqref{firstStage}. In the two-stage GWAS, this information is accounted for in the genotypic means $\hat g_{i}$.
In models \eqref{infModelM_gwas_2} and \eqref{infModelR_gwas} we test the hypothesis $\gamma=0$ using the F-test, conditional on estimates of the variance components obtained from a model without markers (\cite{kang_etal_2010}). When marker effects are small, these estimates are a good approximation of the exact estimates, obtained when genetic- and residual variance are re-estimated for each marker. % \cite{kang_etal_2010}.

\section{Phenotypic data} \label{App_pheno}

\textbf{Collection of phenotypic data}
For the two flowering traits from \cite{atwell_etal_2010} (LDV and LD), details can be found in the original publication. In the case of LD, all plants that had not flowered by the end of the experiment were given a phenotypic value of $200$ days. The leaf area trait (LA(S)/LA(H)) was measured in two separate experiments on the Swedish RegMap and the HapMap, using the same phenotyping platform.
The plants were imaged top down for projected leaf area every day. The images were taken using near infrared light (790nm) so as not to influence the photoperiod during the night measurements.
In our final experiment, Bolting time (BT) and leaf width (LW) were measured for the HapMap.
BT was noted as the number of days after vernalization on which the plant started to bolt. LW was measured on photographs taken from the longest leaf two weeks after flowering.

\textbf{Plant growth conditions}
In the leaf-area experiments, seeds from the Arabidopsis Swedish RegMap (298 accessions) and Arabidopsis HapMap (350 accessions) were stratified at $4\,^{\circ}\mathrm{C}$ for 4 days, and sown on Rockwool blocks which had been covered with black foamed PVC sheet to prevent algal growth and provide a uniform background for automated image analysis. The growth conditions were a light intensity of 200 $\mu \: mol \: m^{-2} \: s^{-1}$, 10h SD, $20/18\,^{\circ}\mathrm{C}$ day/night cycle, 70\% RH.
In the experiment measuring BT and LW on the HapMap, seeds were sown on filter paper with demi water and stratified at $4\,^{\circ}\mathrm{C}$ in dark conditions for 5 days. Following stratification, seeds were transferred to a culture room (16h LD, $24\,^{\circ}\mathrm{C}$) to induce seed germination for 42h. Germinating seeds were then transplanted to wet Rockwool blocks of 4 x 4 cm in a climate chamber with a light intensity of 125 $\mu \: mol \: m^{-2} \: s^{-1}$ 16h LD, $20/18\,^{\circ}\mathrm{C}$ day/ night cycle, 70\% RH. All plants were watered every morning for 5 min at 9am with 1/1000 Hyponex solution (Hyponex, Osaka, Japan). 19 days after germination, all plants were vernalized for 6 weeks in a cold room (12h L, $4\,^{\circ}\mathrm{C}$). After the 6-weeks vernalization period plants were transferred back to the same climate chamber in the same order, but given more space to grow.

\textbf{Genotypic means}
All experiments were laid out as randomized complete block designs, apart from the final HapMap experiment (BT and LW), where the same randomization was used within each replicate.
In all experiments we included a replicate (complete block) effect. The numbers of replicates in the different experiments were $6$ (LD and LDV), $4$ (LA(S)/LA(H)) and $3$ (BT and LW). Due to non-germinating seeds and dead plants, some accessions had lower numbers of replicates. For the leaf area experiments (LA(S)/LA(H)) we additionally included a row and column effect to model the within image position of each plant ($x=1,2,3$ and $y=1,2,3,4$). These factors correct for technical artifacts, which are known to be consistent across replicates.

%\end{appendices}

%%%%%%%%%%%%%%%%%%%%%%%%%%%%%%%%%%%%%%%%%%%%%%%%%%%%%%%%%%%%%%%%%%%%%%%%%%%%%%%%%%%%%%%%%%%%%%%%%%%%%%%%%%%%%%%%%%

\section*{Figures}

% figure produced with the script h2_sim_PLOTS_h2_levels_NEW.r

%\label{atwell_h2}

%%%%%%%%%%%%%%%%%%%%%%%%%%% GENOMIC SELECTION %%%%%%%%%%%%%%%%%%%%%%%%%%%%%%%%%%%%

%%%%%%%%%%%%%%%%%%%%%%% validation set

\clearpage

%%%%%%%%%%%%%%%%%%%%%%%%%%%%%%%%%
% gwas, simulations

\clearpage

\section*{Tables}

\begin{table}[!ht]
\begin{tabular}{|l|l|}
  \hline
  abbreviation & trait \\
  \hline
  LDV & Days to flowering time under Long Day and Vernalization \\
  LD & Days to flowering time under Long Day \\
  \hline
  LA(S) & Leaf Area 13 days after sowing (Swedish RegMap)\\
  \hline
  LA(H) & Leaf Area 13 days after sowing (HapMap)\\
  \hline
  BT & Bolting Time (HapMap) \\
  LW & Leaf Width (HapMap) \\
  \hline
\end{tabular}
\caption{
{\bf Abbreviations of the trait-names.} The first two traits were taken from (Atwell \emph{et al.} 2010). Horizontal lines separate traits measured in different experiments.
}
\label{trait_abbreviations}
\end{table}

\begin{table}[ht]
\centering
\begin{tabular}{|c|r|r|r|r|}
  \hline
population & $r$ & $h^2=0.2$ & $h^2=0.5$ & $h^2=0.8$ \\
  \hline
structured RegMap & 1 & 0.0702 (1.00) & 0.0762 (1.00) & 0.0497 (1.00) \\
  structured RegMap & 2 & 0.0428 (0.85) & 0.0406 (0.71) & 0.0187 (0.49) \\
  structured RegMap & 3 & 0.0337 (0.80) & 0.0323 (0.66) & 0.0153 (0.46) \\
  structured RegMap & 4 & 0.0289 (0.77) & 0.0284 (0.64) & 0.0139 (0.46) \\ \hline
  HapMap & 1 & 0.1137 (1.00) & 0.1310 (1.00) & 0.1040 (1.00) \\
  HapMap & 2 & 0.0356 (0.46) & 0.0292 (0.31) & 0.0139 (0.19) \\
  HapMap & 3 & 0.0250 (0.40) & 0.0226 (0.29) & 0.0115 (0.18) \\
  HapMap & 4 & 0.0204 (0.37) & 0.0199 (0.29) & 0.0106 (0.19) \\ \hline
  Swedish RegMap & 1 & 0.0781 (1.00) & 0.0827 (1.00) & 0.0574 (1.00) \\
  Swedish RegMap & 2 & 0.0391 (0.73) & 0.0344 (0.58) & 0.0160 (0.36) \\
  Swedish RegMap & 3 & 0.0291 (0.67) & 0.0269 (0.54) & 0.0132 (0.34) \\
  Swedish RegMap & 4 & 0.0244 (0.64) & 0.0237 (0.52) & 0.0121 (0.33) \\ \hline
  French RegMap & 1 & 0.1141 (1.00) & 0.1135 (1.00) & 0.0768 (1.00) \\
  French RegMap & 2 & 0.0497 (0.66) & 0.0416 (0.52) & 0.0192 (0.33) \\
  French RegMap & 3 & 0.0360 (0.60) & 0.0322 (0.49) & 0.0158 (0.31) \\
  French RegMap & 4 & 0.0297 (0.58) & 0.0283 (0.48) & 0.0145 (0.31) \\ \hline
  Van Heerwaarden et al. (2012) & 1 & 0.0680 (1.00) & 0.0725 (1.00) & 0.0502 (1.00) \\
  Van Heerwaarden et al. (2012) & 2 & 0.0329 (0.71) & 0.0290 (0.56) & 0.0136 (0.37) \\
  Van Heerwaarden et al. (2012) & 3 & 0.0244 (0.65) & 0.0227 (0.53) & 0.0113 (0.35) \\
  Van Heerwaarden et al. (2012) & 4 & 0.0204 (0.62) & 0.0200 (0.52) & 0.0103 (0.36) \\ \hline
    Riedelsheimer et al. (2012) & 1 & 0.0698 (1.00) & 0.0679 (1.00) & 0.0399 (1.00) \\
  Riedelsheimer et al. (2012) & 2 & 0.0374 (0.81) & 0.0346 (0.74) & 0.0168 (0.60) \\
  Riedelsheimer et al. (2012) & 3 & 0.0286 (0.77) & 0.0280 (0.72) & 0.0141 (0.59) \\
  Riedelsheimer et al. (2012) & 4 & 0.0244 (0.76) & 0.0249 (0.72) & 0.0129 (0.60) \\ \hline
  Zhao et al. (2011) & 1 & 0.0642 (1.00) & 0.0625 (1.00) & 0.0329 (1.00) \\
  Zhao et al. (2011) & 2 & 0.0399 (0.87) & 0.0351 (0.79) & 0.0156 (0.65) \\
  Zhao et al. (2011) & 3 & 0.0311 (0.83) & 0.0280 (0.76) & 0.0128 (0.62) \\
  Zhao et al. (2011) & 4 & 0.0264 (0.81) & 0.0246 (0.74) & 0.0116 (0.62) \\
   \hline
\end{tabular}
%\begin{flushleft}%Table caption
\caption{{\bf
Asymptotic distribution of marker-based heritability estimators: standard deviations of the estimators based on replicates, and between brackets the ratios of the standard deviations of the estimators based on replicates over those based on means.} Asymptotic variances were obtained using the expressions in equation \eqref{as_var_formula}. Four populations of \emph{A. thaliana} were considered (containing respectively $250,350,305$ and $204$ accessions), two populations of \emph{Z. mays} (Van Heerwaarden \emph{et al.} 2012, $400$ accessions; Riedelsheimer \emph{et al.} 2012, 280 accessions) and one population of \emph{O. sativa} (Zhao et al. 2011, $413$ accessions). The second column gives the number of replicates $(r)$.}
%\end{flushleft}
\label{asymptotic_variance_table}
\end{table}

%\definecolor{lightgray}{gray}{0.9}
%\definecolor{kugray}{RGB}{224,224,224}

% latex table generated in R 3.0.1 by xtable 1.7-1 package
% Fri Jan 10 11:00:40 2014
\begin{table}[!ht]
\centering
\begin{tabular}{|l|r|r|r|}
  \hline
 & bias & standard error & relative standard error \\
  \hline
\multicolumn{4}{|l|}{\textbf{Structured RegMap}} \\
%\multicolumn{4}{|>{\columncolor[gray]{.8}}l|}{Structured regmap} \\
%
%\multicolumn{4}{\columncolor[grey]}|l|}{Structured regmap} \\
  \hline
%\multicolumn{4}{|>{\columncolor[kugray]{.8}}c|}{$h^2 = 0.2$} \\
\multicolumn{4}{|c|}{$h^2 = 0.2$} \\
  \hline
%broad-sense ($H^2$) & -0.00314 & 0.06079 & 1.00000 \\
%  individual level ($h_r^2$) & -0.00268 & 0.05306 & 0.87284 \\
%  means ($h_m^2$) & 0.00475 & 0.08670 & 1.42612 \\
broad-sense ($H^2$) & -0.00314 & 0.06079 & 0.70120 \\
  individual level ($h_r^2$) & -0.00268 & 0.05306 & 0.61204 \\
  means ($h_m^2$) & 0.00475 & 0.08670 & 1.00000 \\
  \hline
%\multicolumn{4}{|>{\columncolor[gray]{.8}}c|}{$h^2 = 0.5$} \\
\multicolumn{4}{|c|}{$h^2 = 0.5$} \\
  \hline
%broad-sense ($H^2$) & -0.00938 & 0.07193 & 1.00000 \\
%  individual level ($h_r^2$) & -0.00400 & 0.05093 & 0.70816 \\
%  means ($h_m^2$) & -0.00164 & 0.12761 & 1.77416 \\
broad-sense ($H^2$) & -0.00938 & 0.07193 & 0.56365 \\
  individual level ($h_r^2$) & -0.00400 & 0.05093 & 0.39915 \\
  means ($h_m^2$) & -0.00164 & 0.12761 & 1.00000 \\
   \hline
%\multicolumn{4}{|>{\columncolor[gray]{.8}}c|}{$h^2 = 0.8$} \\
\multicolumn{4}{|c|}{$h^2 = 0.8$} \\
  \hline
%broad-sense ($H^2$) & -0.00825 & 0.04309 & 1.00000 \\
%  individual level ($h_r^2$) & -0.00138 & 0.02473 & 0.57395 \\
%  means ($h_m^2$) & -0.00388 & 0.11940 & 2.77111 \\
broad-sense ($H^2$) & -0.00825 & 0.04309 & 0.36087 \\
  individual level ($h_r^2$) & -0.00138 & 0.02473 & 0.20712 \\
  means ($h_m^2$) & -0.00388 & 0.11940 & 1.00000 \\
   \hline
\multicolumn{4}{|l|}{\textbf{HapMap}} \\
%\multicolumn{4}{|>{\columncolor[gray]{.8}}l|}{Hapmap} \\
  \hline
%\multicolumn{4}{|>{\columncolor[gray]{.8}}c|}{$h^2 = 0.2$} \\
\multicolumn{4}{|c|}{$h^2 = 0.2$} \\
\hline
%broad-sense ($H^2$) & -0.00128 & 0.04663 & 1.00000 \\
%  individual level ($h_r^2$) & -0.00154 & 0.04642 & 0.99550 \\
%  means ($h_m^2$) & 0.05328 & 0.24033 & 5.15354 \\
broad-sense ($H^2$) & -0.00128 & 0.04663 & 0.19404 \\
  individual level ($h_r^2$) & -0.00154 & 0.04642 & 0.19317 \\
  means ($h_m^2$) & 0.05328 & 0.24033 & 1.00000 \\
  \hline
%\multicolumn{4}{|>{\columncolor[gray]{.8}}c|}{$h^2 = 0.5$} \\
\multicolumn{4}{|c|}{$h^2 = 0.5$} \\
   \hline
%broad-sense ($H^2$) & -0.00288 & 0.04280 & 1.00000 \\
%  individual level ($h_r^2$) & -0.00295 & 0.04120 & 0.96281 \\
%  means ($h_m^2$) & 0.04400 & 0.31115 & 7.27064 \\
broad-sense ($H^2$) & -0.00288 & 0.04280 & 0.13754 \\
  individual level ($h_r^2$) & -0.00295 & 0.04120 & 0.13242 \\
  means ($h_m^2$) & 0.04400 & 0.31115 & 1.00000 \\
  \hline
%\multicolumn{4}{|>{\columncolor[gray]{.8}}c|}{$h^2 = 0.8$} \\
\multicolumn{4}{|c|}{$h^2 = 0.8$} \\
   \hline
%broad-sense ($H^2$) & -0.00282 & 0.02385 & 1.00000 \\
%  individual level ($h_r^2$) & -0.00231 & 0.02214 & 0.92841 \\
%  means ($h_m^2$) & -0.06137 & 0.28863 & 12.10168 \\
broad-sense ($H^2$) & -0.00282 & 0.02385 & 0.08263 \\
  individual level ($h_r^2$) & -0.00231 & 0.02214 & 0.07672 \\
  means ($h_m^2$) & -0.06137 & 0.28863 & 1.00000 \\
   \hline
\end{tabular}
\caption{{ \bf Comparison of the marker-based estimators heritability estimators $\hat h_r^2$ and $\hat h_m^2$ for simulated data.} We simulated 5000 traits, for random samples of 200 accessions drawn from the structured RegMap and HapMap. 20 unlinked QTLs were simulated, which explained 50 percent of the genetic variance. The simulated heritability was $0.2$, $0.5$ and $0.8$. Standard errors are given relative to those of
$\hat h_m^2$.} % %the broad sense heritability estimator ($H^2$)
\label{sd_h2_sim_pop1_AND_pop3_regmap_new_gamma5_5000traits_200acc_20QTLs_3rep_ALL}
\end{table}

% latex table generated in R 3.0.1 by xtable 1.7-1 package
% Fri Jan 10 11:00:40 2014
\begin{table}[!ht]
\centering
\begin{tabular}{|l|r|r|}
  \hline
 & coverage & interval width \\
  \hline
\multicolumn{3}{|l|}{\textbf{Structured RegMap}} \\
%\multicolumn{3}{|>{\columncolor[gray]{.8}}l|}{Structured regmap} \\
  \hline
\multicolumn{3}{|c|}{$h^2 = 0.2$} \\
  \hline
broad-sense & 0.864 & 0.178 \\
  individual level (standard) & 0.926 & 0.201 \\
  individual level (log-transformed) & 0.957 & 0.203 \\
  means (standard) & 0.912 & 0.314 \\
  means (log-transformed) & 0.965 & 0.322 \\
   \hline
\multicolumn{3}{|c|}{$h^2 = 0.5$} \\
  \hline
broad-sense & 0.735 & 0.160 \\
  individual level (standard) & 0.948 & 0.196 \\
  individual level (log-transformed) & 0.956 & 0.194 \\
  means (standard) & 0.909 & 0.460 \\
  means (log-transformed) & 0.961 & 0.433 \\
   \hline
\multicolumn{3}{|c|}{$h^2 = 0.8$} \\
  \hline
broad-sense & 0.671 & 0.086 \\
  individual level (standard) & 0.941 & 0.094 \\
  individual level (log-transformed) & 0.941 & 0.094 \\
  means (standard) & 0.923 & 0.422 \\
  means (log-transformed) & 0.946 & 0.492 \\
   \hline
%\multicolumn{3}{|>{\columncolor[gray]{.8}}l|}{Hapmap} \\
\multicolumn{3}{|l|}{\textbf{HapMap}} \\
  \hline
\multicolumn{3}{|c|}{$h^2 = 0.2$} \\
  \hline
broad-sense & 0.948 & 0.178 \\
  individual level (standard) & 0.945 & 0.181 \\
  individual level (log-transformed) & 0.969 & 0.182 \\
  means (standard) & 0.835 & 0.538 \\
  means (log-transformed) & 0.916 & 0.678 \\
   \hline
\multicolumn{3}{|c|}{$h^2 = 0.5$} \\
  \hline
broad-sense & 0.935 & 0.160 \\
  individual level (standard) & 0.948 & 0.164 \\
  individual level (log-transformed) & 0.952 & 0.163 \\
  means (standard) & 0.861 & 0.806 \\
  means (log-transformed) & 0.984 & 0.832 \\
   \hline
\multicolumn{3}{|c|}{$h^2 = 0.8$} \\
  \hline
broad-sense & 0.924 & 0.084 \\
  individual level (standard) & 0.948 & 0.085 \\
  individual level (log-transformed) & 0.950 & 0.085 \\
  means (standard) & 0.892 & 0.882 \\
  means (log-transformed) & 0.914 & 0.915 \\   \hline
\end{tabular}
\caption{{ \bf Marker-based estimation of heritability: width and coverage confidence intervals obtained from
the individual plant data and the genotypic means.} Results for broad sense heritability intervals are reported for comparison. We simulated 5000 traits, for random samples of 200 accessions drawn from the structured RegMap (top) and HapMap (bottom). 20 unlinked QTLs were simulated, which explained 50 percent of the genetic variance. The simulated heritability was $0.2$, $0.5$ and $0.8$.}
\label{conf_h2_sim_pop1_and_pop3_regmap_new_gamma5_5000traits_200acc_20QTLs_3rep_ALL}
\end{table}
%

%%%%%%%%%%%%%%%%%%%%%%%%%%%%%%%%%%%%%%%%%%%%%%%%%%%%%%%%%%%%%%%%%%%%%%%%

\begin{table}[!ht]
\begin{tabular}{|r|l|l|l|}
  \hline
trait & ${\hat h}_r^2$ & ${\hat h}_m^2$ & ${\hat H}^2$\\
  \hline
LDV & 0.801 (0.756,0.840) & 0.510 (0.303,0.714) & 0.858 (0.826,0.886) \\
  LD & 0.933 (0.917,0.947) & 1.000 (0.000,1.000) & 0.966 (0.957,0.973) \\
  \hline
LA(S) & 0.209 (0.141,0.298) & 0.088 (0.028,0.244) & 0.235 (0.167,0.306) \\
  \hline
LA(H) & 0.378 (0.315,0.444) & 0.339 (0.134,0.631) & 0.388 (0.327,0.451) \\
   \hline
BT & 0.941 (0.929,0.950) & 1.000 (0.000,1.000) & 0.956 (0.947,0.963) \\
  LW & 0.553 (0.491,0.614) & 0.155 (0.028,0.538) & 0.530 (0.468,0.589) \\
   \hline
\end{tabular}
%\begin{flushleft}%Table caption
\caption{{\bf Heritability estimates and confidence intervals for two flowering traits from (Atwell \emph{et al.} 2010)% \cite{atwell_etal_2010}
(LDV and LD), and $4$ traits from new experiments.} Three estimators were used: the marker-based estimator using individual plant data (${\hat h}_r^2$), the marker-based estimator using genotypic means (${\hat h}_m^2$), and the ANOVA-based estimator of broad-sense heritability (${\hat H}^2$).
Horizontal lines separate traits measured in different experiments.
Trait abbreviations are given in Table \ref{trait_abbreviations}.}
\label{h2_values}
\end{table}

\begin{table}[!ht]
\centering
\begin{tabular}{|l|r|r|r|r|r|r|}
  \hline
interval & ${\hat h}_r^2$ & ${\hat h}_m^2$ & $r$ (replicates) & $r$ (means) & $r$ (replicates) & $r$ (means) \\
&&& Training set & Training set & Validation set & Validation set \\
  \hline
$[0,0.1)$ & 0.00 \% & 2.58 \% &  & 0.890 &  & 0.289  \\
  $[0.1,0.3)$ & 0.00 \% & 8.34 \% &  & 0.937 &  & 0.373 \\
  $[0.3,0.5)$ & 0.00 \% & 12.34 \% &  & 0.954 &  & 0.409 \\
  $[0.5,0.7)$ & 0.04 \% & 15.90 \% & 0.942 & 0.959 & 0.208 & 0.423 \\
  $[0.7,0.9)$ & 99.96 \% & 15.62 \% & 0.961 & 0.961  & 0.431 & 0.443 \\
  $[0.9,1]$ & 0.00 \% & 45.22 \% &  & 0.961 &  & 0.448 \\
\hline
  [0,1] & 100 \% & 100 \% & 0.961 & 0.956 & 0.431 & 0.428 \\
   \hline
\end{tabular}
\caption{{ \bf Prediction accuracy ($r$) of G-BLUP for $5000$ simulated traits, for the HapMap population.} Each trait was simulated for a randomly drawn training (200 accessions) and validation set (50 accessions). Genetic effects were predicted using G-BLUP, based on either a mixed model for the individual plants (replicates) or for the genotypic means. The second and third column contain the percentage of the $5000$ traits for which the corresponding heritability estimates (${\hat h}_r^2$ and ${\hat h}_m^2$) were contained in the intervals in the first column. The remaining columns show the correlation ($r$) between simulated and predicted genetic effects,  averaged over these traits. The simulated heritability was $0.8$; 20 QTLs were simulated, which explained 50 percent of the genetic variance.
}
\label{accuracy_table}
\end{table}

\clearpage

%\newpage

\section*{File S1: confidence intervals for broad-sense heritability}

Confidence intervals for the broad-sense heritability estimates obtain from the ANOVA mean sums of squares are traditionally obtained from the ratio $F = MS(G) / MS(E)$ and the quantiles of the F-distribution with the corresponding degrees of freedom. Given $n$ genotypes with $r_1,\ldots,r_n$ replicates, the intervals are given by
\begin{equation*}
\frac{F / F_{\textrm{df}1,\textrm{df}2,0.95} - 1}{F / F_{\textrm{df}1,\textrm{df}2,0.95} + \bar r - 1} < H^2 < \frac{F / F_{\textrm{df}1,\textrm{df}2,0.05} - 1}{F / F_{\textrm{df}1,\textrm{df}2,0.05} + \bar r - 1},
\end{equation*}
where $\textrm{df}1 = n-1$, $\textrm{df}2=\sum (r_i-1)$ %(BAS: or $n(\bar r-1)$ ?)
and $\bar r = (n-1)^{-1}  (\sum r_i  - (\sum r_i^2)/(\sum r_i))$. In case of a balanced design with $r_i=r$ replicates, this reduces to $\bar r = r$ and $\textrm{df}2 = n(r-1)$. See \cite{lynch_walsh_1998} (p.563) or \cite{Singh_Ceccarelli_Hamblin_1993}.

\newpage
%%%%%%%%%%%%%%%%%%%%%%%%%%%%%%%%%%%%%%%%%%%%%%%%%%%%%%%%%%%%%%%%%%%%%%%%%
% TEXT 2

\section*{File S2: analysis of flowering traits of \cite{atwell_etal_2010}}

Our broad-sense heritability estimates differ from those reported in Supplementary table 7 of \cite{atwell_etal_2010}, for the following three reasons. %: $0.94$ for LDV and $0.99$ for LD
%There are two reasons for this difference. %s can be explained by the two
First, the broad-sense heritability estimates in \cite{atwell_etal_2010} were calculated using the formula
\begin{equation} \label{naive_h2_estimate}
\frac{MS(G)}{MS(G) + MS(E)}.
\end{equation}
Although this quantity may be an adequate criterion to compare heritabilities of traits within the same experiment (as long as they have the same number of replicates), this is a biased estimator of broad-sense-heritability. Since the expectation of $MS(G)$ is $r \sigma_G^2 + \sigma_E^2$, $MS(G) / (MS(G) + MS(E))$ will tend to overestimate heritability. The usual estimator defined in the materials and methods section is also biased, but this bias is usually small, and (in contrast to \eqref{naive_h2_estimate}) tends to zero when the number of genotypes increases (\cite{nyquist_baker_1991}, \cite{Singh_Ceccarelli_Hamblin_1993}).

Second, broad-sense heritability estimates in \cite{atwell_etal_2010} were based on more accessions:  $189$ for LDV and $186$ for LD. To allow a direct comparison with mixed model analysis we restricted our analysis to genotyped accessions, excluding $21$ accessions for LDV and for $19$ LD. This had little impact on heritability estimates.

Third, the analysis of variance in \cite{atwell_etal_2010} did not include a replicate effect. In our analysis, the mean sums of squares for replicates removes some environmental variance, therefore giving higher estimates than in an analysis without a replicate effect. This however did not compensate for the use of \eqref{naive_h2_estimate}; hence our heritability estimates are lower than those reported in \cite{atwell_etal_2010}.

\newpage
%%%%%%%%%%%%%%%%%%%%%%%%%%%%%%%%%%%%%%%%%%%%%%%%%%%%%%%%%%%%%%%%%%%%%%%%%
% TEXT 3

\section*{File S3: the likelihood is constant for a kinship matrix with compound symmetry structure}

Mixed model based estimation of  heritability using genotypic means may become problematic when the sample size is small and the kinship matrix is close to compound symmetry, i.e. the structure where all off-diagonal elements are equal.
%This is case for the Arabidopsis hapmap: Figure S4 shows the log-likelihood profile for one of the simulated traits.
%In human genetics, panels of unrelated individuals also have a structure that is close to compound symmetry, most kinship coefficients being close to zero. However, the numbers of genetically different individuals are usually %orders of magnitude larger than in \emph{A. thaliana}.
Here we show that in the case the kinship matrix is exactly compound symmetry, the likelihood is constant in $\eta = \sigma_E^2 / \sigma_A^2$.
We use the notation $\eta$ to avoid confusion with $\delta = \sigma_A^2 / \sigma_E^2$, used in our results on genomic prediction.
We write $1_n$ for the $n \times 1$ column vector of ones, and $I_n$ for the $n$-dimensional identity matrix. Finally, let $J_n$ be the $n \times n$ matrix of ones.

Suppose that $K=I_n + a J_n$, for some $a>0$.
The key observation is that the covariance matrix of the data can be written as
\begin{equation*}
\Sigma = \sigma_A^2 K + \sigma_E^2 I_n = \tilde \sigma_A^2 \tilde K + \tilde \sigma_E^2 I_n = \tilde \sigma_A^2 (\tilde K + \tilde \eta I_n),
\end{equation*}
where $\tilde \sigma_A^2 = \sigma_A^2$, $\tilde \sigma_E^2 = \sigma_A^2 + \sigma_E^2$, $\tilde K = a J_n = a 1_n 1_n^t$ and $\tilde \eta = (\sigma_A^2 + \sigma_E^2)/ \sigma_A^2 > 0$.
We can then directly apply the results in section 3 of \cite{lippert_etal_2011_supp}, with (in their notation) $k=1$, $d=1$ and $X=1_n$ (we only include an intercept, and no marker effect),
and replacing $\sigma_A^2$, $\sigma_E^2$, $\eta$ and $K$ by respectively $\tilde \sigma_A^2$, $\tilde \sigma_E^2$, $\tilde \eta$ and $\tilde K$.
In particular, we have the spectral decomposition % $\tilde K = U S U^t$,  % following
\begin{equation*}
\begin{split}
\tilde K & = U S U^t = [ U_1, U_2 ] \left[
\begin{array}{cc}
S_1 & 0 \\
0 &S_2
\end{array}
\right] [ U_1, U_2 ]^t \\
& =\left( \begin{array}{cccc}
n^{-\frac{1}{2}} & 0 & \hdots & 0 \\
\vdots & \vdots & & \vdots \\
n^{-\frac{1}{2}} & 0 & \hdots & 0
\end{array} \right)
\left( \begin{array}{cccc}
n a & 0 & \hdots & 0 \\
0 & 0 & \hdots & 0 \\
\vdots & & \ddots  & \vdots \\
0 & \hdots& \hdots& 0
\end{array} \right)
\left( \begin{array}{ccc}
n^{-\frac{1}{2}} & \hdots & n^{-\frac{1}{2}} \\
0 & \hdots & 0 \\
0 & \hdots & 0
\end{array} \right),
\end{split}
\end{equation*}
i.e. the only non-zero eigenvalue of $\tilde K$ is $an$, with eigenvector $(n^{-\frac{1}{2}},\ldots,n^{-\frac{1}{2}})$.

For this choice of $X$ and $\tilde K$, the expressions for the (RE)ML estimates of $\beta$ and $\tilde \sigma_A^2$ given in sections 3.2 and 4 of \cite{lippert_etal_2011_supp} greatly simplify:
$\hat \beta = \bar y$ and the REML-estimate of $\tilde \sigma_A^2$ is $ \sum_{i=1}^n (y_i - \bar y)^2 / (\tilde \eta (n-1))$. The extra terms
\begin{equation*}
\frac{1}{2} \left(d \log(2\pi \tilde \sigma_A^2) + \log|X^t X| - \log|X^t (\tilde K + \tilde \eta I_n)^{-1} X| \right)
\end{equation*}
in the REML-log-likelihood (see the first equation in section 4 of \cite{lippert_etal_2011_supp}), now equal
\begin{equation*}
\frac{1}{2} \left(d \log(2\pi \tilde \sigma_A^2) + \log n - \log\left(\frac{n}{na + \tilde \eta}\right) \right).
\end{equation*}
Combining this with their equation (3.7), it follows that the REML-log-likelihood is constant in $\tilde \eta = (\sigma_A^2 + \sigma_E^2)/ \sigma_A^2 > 0$, and hence also constant in $\eta = \tilde \eta - 1 = \sigma_E^2/ \sigma_A^2$.

\newpage
%%%%%%%%%%%%%%%%%%%%%%%%%%%%%%%%%%%%%%%%%%%%%%%%%%%%%%%%%%%%%%%%%%%%%%%%%
% TEXT 4

\section*{File S4: Simulation results for a different genetic architecture.}

\begin{table}[!ht]
\centering
\caption{{ \bf Comparison of the marker-based estimators heritability estimators $h_r^2$ and $h_m^2$ for simulated data.} We simulated 5000 traits, for random samples of 200 accessions drawn from the Structured regmap and Hapmap. A single QTL was simulated, which explained 90 percent of the genetic variance. The simulated heritability was $0.2$, $0.5$ and $0.8$. Standard errors are given relative to those of the broad sense heritability estimator ($H^2$).}
\begin{tabular}{|l|r|r|r|}
  \hline
 & bias & standard error & relative standard error \\
  \hline
\multicolumn{4}{|>{\columncolor[gray]{.8}}l|}{Structured regmap} \\
%\multicolumn{4}{\columncolor[grey]}|l|}{Structured regmap} \\
  \hline
%\multicolumn{4}{|>{\columncolor[kugray]{.8}}c|}{$h^2 = 0.2$} \\
\multicolumn{4}{|c|}{$h^2 = 0.2$} \\
  \hline
broad-sense ($H^2$) & -0.00127 & 0.04787 & 1.00000 \\
  replicates ($h_r^2$) & -0.00066 & 0.05102 & 1.06585 \\
  means ($h_m^2$) & 0.00782 & 0.08626 & 1.80191 \\
  \hline
%\multicolumn{4}{|>{\columncolor[gray]{.8}}c|}{$h^2 = 0.5$} \\
\multicolumn{4}{|c|}{$h^2 = 0.5$} \\
  \hline
broad-sense ($H^2$) & -0.00279 & 0.04500 & 1.00000 \\
  replicates ($h_r^2$) & -0.00571 & 0.07001 & 1.55569 \\
  means ($h_m^2$) & 0.01295 & 0.16461 & 3.65791 \\
   \hline
%\multicolumn{4}{|>{\columncolor[gray]{.8}}c|}{$h^2 = 0.8$} \\
\multicolumn{4}{|c|}{$h^2 = 0.8$} \\
  \hline
broad-sense ($H^2$) & -0.00257 & 0.02458 & 1.00000 \\
  replicates ($h_r^2$) & -0.01163 & 0.05404 & 2.19850 \\
  means ($h_m^2$) & 0.00337 & 0.20855 & 8.48496 \\
   \hline
\multicolumn{4}{|>{\columncolor[gray]{.8}}l|}{Hapmap} \\
  \hline
%\multicolumn{4}{|>{\columncolor[gray]{.8}}c|}{$h^2 = 0.2$} \\
\multicolumn{4}{|c|}{$h^2 = 0.2$} \\
\hline
broad-sense ($H^2$) & -0.00110 & 0.04344 & 1.00000 \\
  replicates ($h_r^2$) & -0.00098 & 0.04320 & 0.99453 \\
  means ($h_m^2$) & 0.06629 & 0.26168 & 6.02448 \\
  \hline
%\multicolumn{4}{|>{\columncolor[gray]{.8}}c|}{$h^2 = 0.5$} \\
\multicolumn{4}{|c|}{$h^2 = 0.5$} \\
   \hline
broad-sense ($H^2$) & -0.00123 & 0.03437 & 1.00000 \\
  replicates ($h_r^2$) & -0.00187 & 0.03736 & 1.08695 \\
  means ($h_m^2$) & 0.03062 & 0.33527 & 9.75477 \\
  \hline
%\multicolumn{4}{|>{\columncolor[gray]{.8}}c|}{$h^2 = 0.8$} \\
\multicolumn{4}{|c|}{$h^2 = 0.8$} \\
   \hline
  broad-sense ($H^2$) & -0.00027 & 0.01633 & 1.00000 \\
  replicates ($h_r^2$) & -0.00106 & 0.02029 & 1.24235 \\
  means ($h_m^2$) & -0.07852 & 0.33486 & 20.50621 \\
   \hline
\end{tabular}
\label{sd_h2_sim_pop1_AND_pop3_regmap_new_gamma9_5000traits_200acc_1QTLs_3rep_ALL}
\end{table}

\begin{table}[!ht]
\centering
\caption{{ \bf Marker-based estimation of heritability: width and coverage confidence intervals obtained from
the individual plant data and the genotypic means.} Results for broad sense heritability intervals are reported for comparison. We simulated 5000 traits, for random samples of 200 accessions drawn from the structured regmap (top) and Hapmap (bottom). A single QTL was simulated, which explained 90 percent of the genetic variance. The simulated heritability was $0.2$, $0.5$ and $0.8$.}
\begin{tabular}{|l|r|r|}
  \hline
 & coverage & interval width \\
  \hline
\multicolumn{3}{|>{\columncolor[gray]{.8}}l|}{Structured regmap} \\
  \hline
\multicolumn{3}{|c|}{$h^2 = 0.2$} \\
  \hline
broad-sense & 0.940 & 0.178 \\
  replicates (standard) & 0.945 & 0.201 \\
  replicates (log-transformed) & 0.962 & 0.202 \\
  means (standard) & 0.911 & 0.315 \\
  means (log-transformed) & 0.960 & 0.321 \\
   \hline
\multicolumn{3}{|c|}{$h^2 = 0.5$} \\
  \hline
broad-sense & 0.926 & 0.160 \\
  replicates (standard) & 0.837 & 0.194 \\
  replicates (log-transformed) & 0.847 & 0.192 \\
  means (standard) & 0.814 & 0.446 \\
  means (log-transformed) & 0.886 & 0.427 \\
   \hline
\multicolumn{3}{|c|}{$h^2 = 0.8$} \\
  \hline
broad-sense & 0.914 & 0.084 \\
  replicates (standard) & 0.674 & 0.097 \\
  replicates (log-transformed) & 0.666 & 0.097 \\
  means (standard) & 0.714 & 0.437 \\
  means (log-transformed) & 0.840 & 0.547 \\
   \hline
\multicolumn{3}{|>{\columncolor[gray]{.8}}l|}{Hapmap} \\
  \hline
\multicolumn{3}{|c|}{$h^2 = 0.2$} \\
  \hline
broad-sense & 0.961 & 0.178 \\
  replicates (standard) & 0.961 & 0.181 \\
  replicates (log-transformed) & 0.972 & 0.182 \\
  means (standard) & 0.807 & 0.537 \\
  means (log-transformed) & 0.899 & 0.675 \\
   \hline
\multicolumn{3}{|c|}{$h^2 = 0.5$} \\
  \hline
broad-sense & 0.979 & 0.160 \\
  replicates (standard) & 0.971 & 0.164 \\
  replicates (log-transformed) & 0.975 & 0.163 \\
  means (standard) & 0.800 & 0.766 \\
  means (log-transformed) & 0.967 & 0.819 \\
   \hline
\multicolumn{3}{|c|}{$h^2 = 0.8$} \\
  \hline
broad-sense & 0.990 & 0.084 \\
  replicates (standard) & 0.963 & 0.085 \\
  replicates (log-transformed) & 0.964 & 0.085 \\
  means (standard) & 0.820 & 0.840 \\
  means (log-transformed) & 0.849 & 0.903 \\
  \hline
\end{tabular}
\label{conf_h2_sim_pop1_and_pop3_regmap_new_gamma9_5000traits_200acc_1QTLs_3rep_ALL}
\end{table}

% figure produced with the script h2_sim_PLOTS_h2_levels_NEW.r
\begin{figure}[!ht]
\setlength{\fboxsep}{1pt}%
\setlength{\fboxrule}{1pt}%
\fbox{\includegraphics[width=16cm,height=9.5cm]{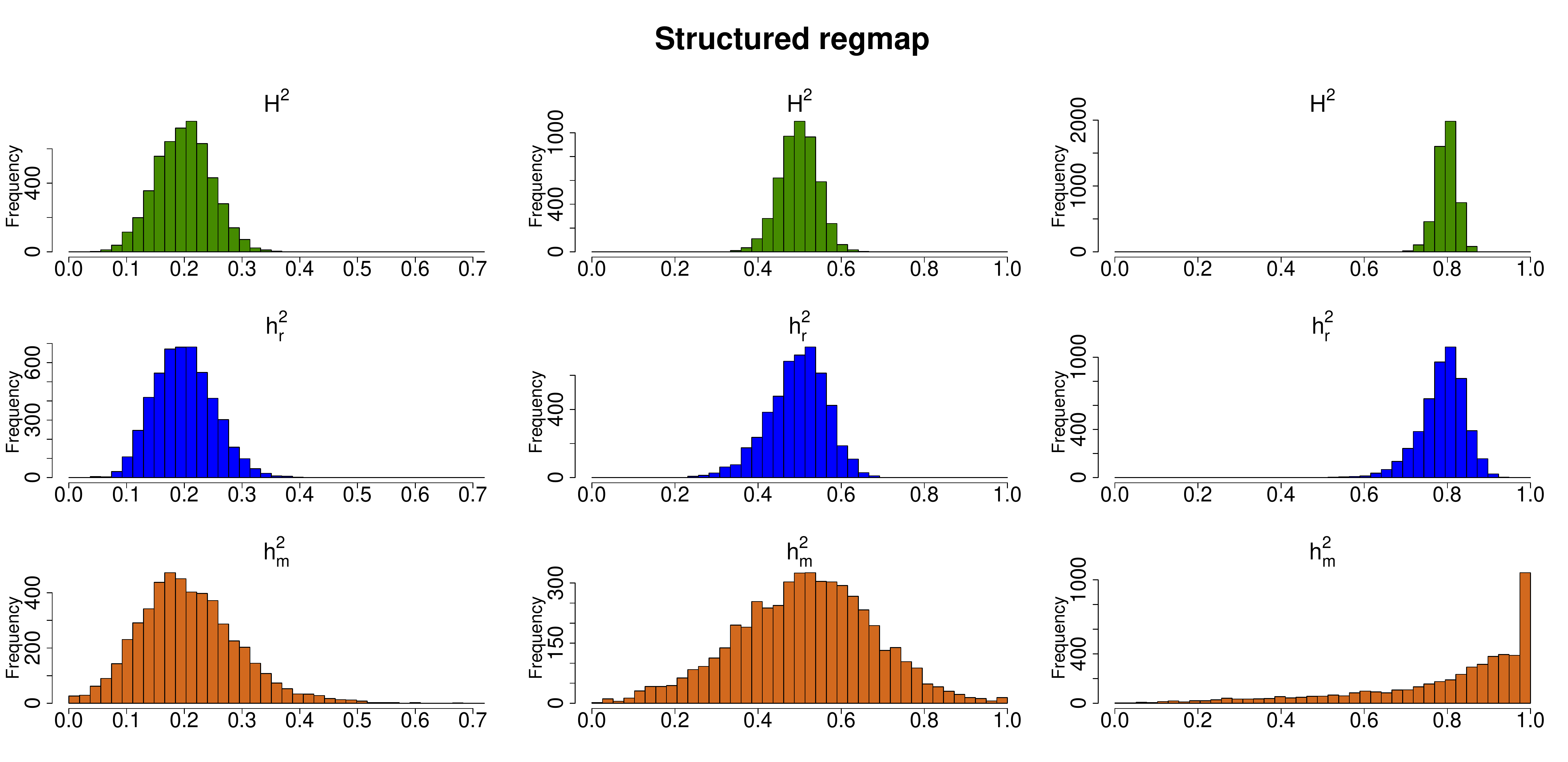}}
%\vspace{15pt}
\fbox{\includegraphics[width=16cm,height=9.5cm]{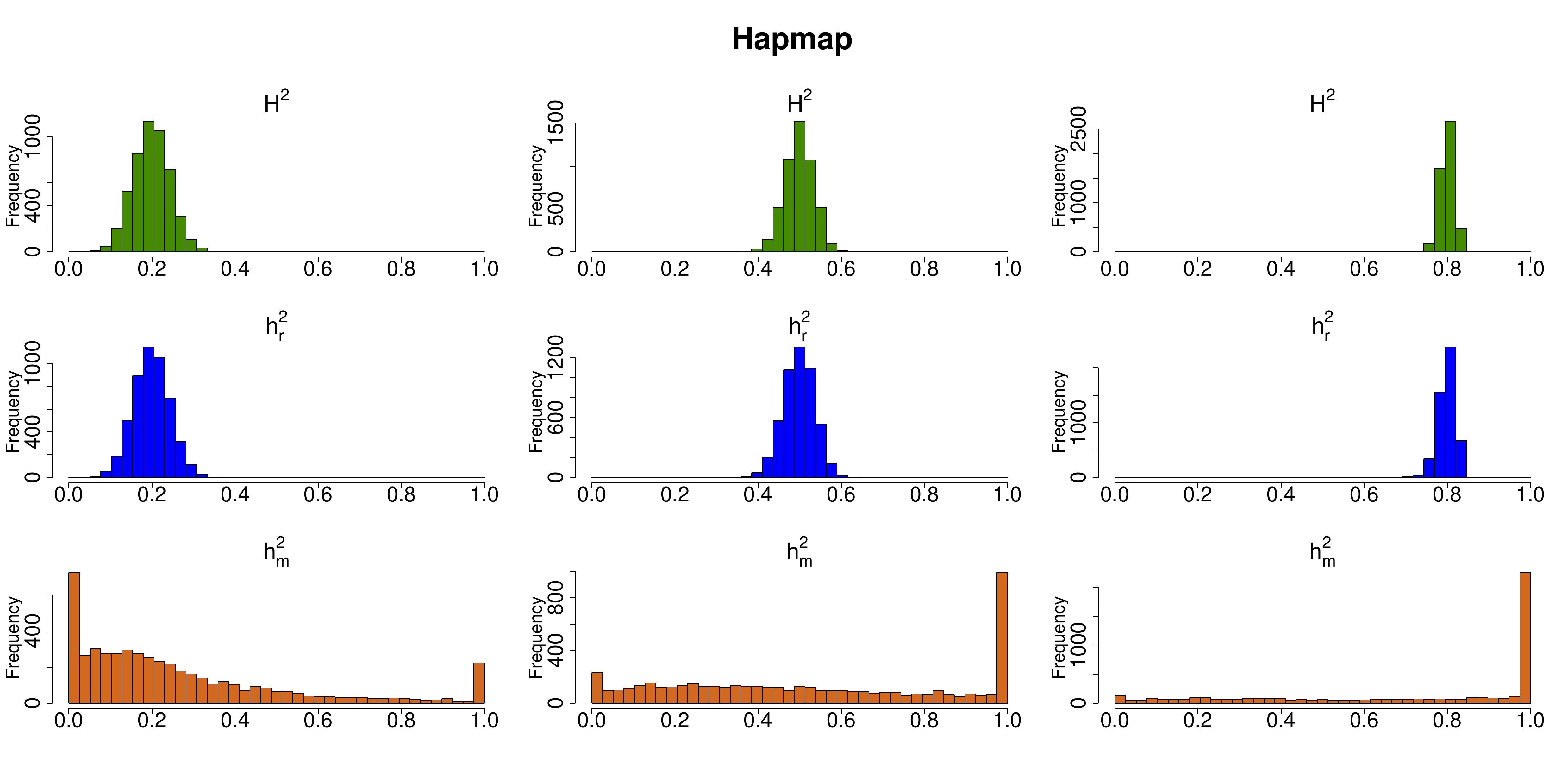}}
\caption{{\bf Heritability estimates for 5000 simulated traits for random samples of 200 accessions drawn from the Structured regmap (top panel) and the Hapmap (bottom panel).} 1 QTL was simulated, which explained $90\%$ of the genetic variance. The simulated heritability was 0.2 (left column), 0.5 (middle column) and 0.8 (right column). Within each panel, the first row shows the ANOVA-based estimates of broad-sense heritability, the second row the mixed model based estimates based on the individual data, and the third row the mixed model based estimates based on genotypic means.}%\label{boxplot3}
%\label{h2_sim histograms_structured_and_hapmap_b}
%\label{h2_sim histograms_structured_b}
\end{figure}

%\section
%\subsection*{s2}

\clearpage
\newpage
%%%%%%%%%%%%%%%%%%%%%%%%%%%%%%%%%%%%%%%%%%%%%%%%%%%%%%%%%%%%%%%%%%%%%%%%%
% TEXT 5

\section*{File S5: prediction error variance in the training- and validation set.}

We assume a balanced and completely random design, with $n$ genotypes and $r$ replicates.
Given the model $y_{i,j} = \mu + G_i + E_{i,j}$, the best linear unbiased predictor (BLUP) of $G = (G_1,\ldots,G_n)^t$ and the best linear unbiased estimator (BLUE) of $\mu$ are given by
\begin{equation} \label{gblup1}
\hat G = \delta K Z^t (\delta Z K Z^t + I_N)^{-1} (y - \hat \mu 1_N), \quad \hat \mu = \frac{1_N^t (\delta Z K Z^t + I_N)^{-1} y}{1_N^t (\delta Z K Z^t + I_N)^{-1} 1_N},
\end{equation}
where $\delta = \sigma_A^2 / \sigma_E^2$ is the shrinkage parameter, $N$ is the total number of individuals and $Z$ is the $N \times n$ incidence matrix assigning individuals to genotypes. See e.g. \cite{henderson_1975} or \cite{robinson_1991}, or equation (23) in the present work (Appendix C). The parameter $\delta = h^2 / (1- h^2)$ is a function of the heritability, and determines the extent to which the phenotypic data $y$ are 'shrunk' towards zero. When the heritability is high, $\delta$ is large, and there is little shrinkage, i.e. $\hat G$ will be close to the observed phenotypic observations $y$. For low heritability, $\delta$ is small, and $y$ will be shrunk towards the vector of zeros. When BLUPs are based on the genotypic means the same expressions hold, with $N=n$ and $Z = I_n$, and $\hat G = \delta_r K (\delta_r K + I_n)^{-1} ((\bar y_1,\ldots,\bar y_n)^t - \hat \mu 1_n)$. Since the noise level is reduced from $\sigma_E^2$ to $r^{-1} \sigma_E^2$, the shrinkage parameter $\delta$ becomes $\sigma_A^2 / (r^{-1}\sigma_E^2)$.

The preceding expressions assume the shrinkage parameter to be known, while it is usually estimated from the data.
As a consequence, the standard error of $\hat \mu$ and prediction error variance of $\hat G$ obtained by setting $\delta=\hat \delta = \hat h^2 / (1- \hat h^2)$ in \eqref{gblup1} are larger than what would be obtained when $\delta$ is known (\cite{kackar_harville_1984}, \cite{kenward_roger_1997}).
Before we give examples of too much or too little shrinkage (section \ref{misspec}), we first give expressions for the prediction error variance for the training and validation set, for the case when heritability is known ($\hat \delta = \delta$). These can be derived as a special case of the more general expressions in e.g. \cite{henderson_1975} or \cite{robinson_1991}.

%In section \ref{misspec} we consider the prediction error variance in case $\hat \delta \neq \delta$, and there is .

\subsection*{Prediction error variance when $\delta=\hat \delta$}

First we consider the genetic effects $G= (G_1,\ldots,G_n)^t$ of the genotypes in the training sample.
If we assume that $G \sim N(0,\sigma_A^2 K)$ (i.e. in equation (21) in the main text (Appendix B), $\gamma$ and the QTL-effects $\alpha_m$ are zero), the
%prediction error variance are the diagonal elements of
prediction error variance is given by the diagonal elements of
\begin{equation}\label{gblup1mse}
E (\hat G - G) (\hat G - G)^t = (Z^t Z + {\delta}^{-1} K^{-1} - J_n)^{-1},
\end{equation}
where $Z$ is the $N \times n$ incidence matrix assigning plants to genotypes, and $J_n$ is the $n \times n$ matrix with identical elements $1/n$. In case the phenotypic data consists of genotypic means, $N=n$.
For efficient computation, see \cite{gilmour_etal_2004} \cite{welham_etal_2004}.
%Estimates of the prediction error variance are obtained by plugging in the estimate $\delta = \hat \delta$.
%Usually, these

The genetic effects $G_{\textrm{pred}} = (G_{n+1},\ldots,G_{n+m})^t$ of $m$ unobserved (but genotyped) genotypes can be predicted with the conditional mean
\begin{equation} \label{gblup2sup}
\hat G_{\textrm{pred}} := E[G_{\textrm{pred}} | y] = \hat \delta K_{\textrm{pred.obs}} Z^t (\hat \delta Z K Z^t + I_N)^{-1} (y - \hat \mu 1_N), % y_{\textrm{obs}}
\end{equation}
where $K_{\textrm{pred.obs}}$ is the $m \times n$ matrix of kinship coefficients for the unobserved versus observed genotypes.
To give expressions for the prediction error variance $E(\hat G_{\textrm{pred}} - G_{\textrm{pred}})_{i'}^2$  ($i' = 1,\ldots,m$) we assume again that $\gamma=0$, all genetic signal being polygenic.
Writing $K_{\textrm{pred.pred}}$ for the $m \times m$ kinship matrix of the unobserved genotypes, it is assumed that the kinship matrix is the $(n+m) \times (n+m)$ block matrix with $K$ and $K_{\textrm{pred.pred}}$ on the diagonal and off-diagonal blocks $K_{\textrm{pred.obs}}$ and $K_{\textrm{pred.obs}}^t$. Then the conditional distribution of $G_{\textrm{pred}} | G$ is
\begin{equation*}
G_{\textrm{pred}} | G \sim N\left(  K_{\textrm{pred.obs}} K^{-1} G, \sigma_A^2 \left(K_{\textrm{pred.pred}} - K_{\textrm{pred.obs}} K^{-1} K_{\textrm{pred.obs}}^t \right) \right).
\end{equation*}
Since $\hat G_{\textrm{pred}} = K_{\textrm{pred.obs}} K^{-1} \hat G$ (by comparing \eqref{gblup1} and \eqref{gblup2sup}), it follows that
\begin{equation*}
\begin{split}
& (\hat G_{\textrm{pred}} - G_{\textrm{pred}}) | (\hat G - G) = K_{\textrm{pred.obs}} K^{-1} (\hat G - G) - Y, \\
& \qquad \textrm{where} \quad Y \sim N\left(0, \sigma_A^2 (K_{\textrm{pred.pred}} - K_{\textrm{pred.obs}} K^{-1} K_{\textrm{pred.obs}}^t)\right).
\end{split}
\end{equation*}

Consequently, the prediction error variances $E(\hat G_{\textrm{pred}} - G_{\textrm{pred}})_i^2$ are the diagonal elements of
\begin{equation}\label{gblup2mse}
\begin{split}
& E (\hat G_{\textrm{pred}} - G_{\textrm{pred}}) (\hat G_{\textrm{pred}} - G_{\textrm{pred}})^t
= E \left[E (\hat G_{\textrm{pred}} - G_{\textrm{pred}}) (\hat G_{\textrm{pred}} - G_{\textrm{pred}})^t \mid (\hat G - G) \right] \\
& \quad =  (K_{\textrm{pred.obs}} K^{-1}) \left[ E (\hat G - G) (\hat G - G)^t \right] K^{-1} K_{\textrm{pred.obs}}^t \\
& \qquad \quad + \sigma_A^2 (K_{\textrm{pred.pred}} - K_{\textrm{pred.obs}} K^{-1} K_{\textrm{pred.obs}}^t).
\end{split}
\end{equation}
Hence, the prediction error variance for the validation set contains a term depending on $\delta^{-1} = \sigma_E^2 / \sigma_A^2$ (see \eqref{gblup1mse}), as well as a term which depends only on the genetic variance $\sigma_A$.

\subsection*{Prediction error variance with incorrect shrinkage ($\delta \neq \hat \delta$)} \label{misspec}

For the case that the amount of shrinkage is not chosen correctly ($ \hat \delta \neq \delta = \sigma_A^2 / (r^{-1}\sigma_E^2$)), we  now give an expression for the prediction error variance for the training set based on genotypic means, under the additional assumption that $\mu$ is known to be zero.
%The BLUP for $G$ then simplifies to
%\begin{equation} \label{gblup1simplified}
%\hat G = \hat \delta K Z^t (\hat \delta Z K Z^t + I_N)^{-1} y,
%\end{equation}
%where we recall that we still assume a balanced and completely random design and that $y_{i,j} = G_i + E_{i,j}$, with $E_{i,j} \sim N(0,\sigma_E^2)$ and $G = (G_1,\ldots,G_n)^t \sim N(0,\sigma_A^2)$.
The BLUP for $G$ then simplifies to
\begin{equation} \label{gblup2simplified}
\hat G = \hat \delta K (\hat \delta K  + I_n)^{-1} \bar y,
\end{equation}
where we recall that we still assume a balanced and completely random design. Hence $\bar y_{i} = G_i + \bar E_{i}$, with $\bar E_{i} \sim N(0,r^{-1} \sigma_E^2)$ and $G = (G_1,\ldots,G_n)^t \sim N(0,\sigma_A^2 K)$.
Since $\bar y = (\bar y_1,\ldots,\bar y_n)^t  \sim N(0,\sigma_A^2 K + r^{-1} \sigma_E^2 I_n) = N(0,\sigma_E^2 (\delta K + r^{-1} I_n)$, the variance-covariance matrix of $\hat G - G$ equals
\begin{equation*}
\begin{split}
\textrm{Var}(\hat G - G) = \sigma_A^2 K - 2 \hat \delta K ( \hat \delta K + I_n)^{-1} \sigma_A^2 K
+  \hat \delta K( \hat \delta K + I_n)^{-1} (\delta K + r^{-1} I_n) ( \hat \delta K + I_n)^{-1} \hat \delta K \sigma_E^2,
\end{split}
\end{equation*}
where we used that (by the independence of $G$ and $E$)
\begin{equation*}
\textrm{Cov}(G,\hat G) = \textrm{Cov}(G,\hat \delta K (\hat  \delta K + I_n)^{-1} G) =  \hat \delta K (\hat  \delta K + I_n)^{-1} \sigma_A^2 K
\end{equation*}
and that (using $\bar y \sim  N(0,\sigma_E^2 (\delta K + r^{-1} I_n)$ and the symmetry of $K$ and $I_n$)
\begin{equation*}
\hat G = \hat \delta K (\hat  \delta K + I_n)^{-1} \bar y \sim N(0, \hat \delta K (\hat  \delta K + I_n)^{-1} (\delta K + r^{-1}I_n) ( \hat \delta K + I_n)^{-1} \hat \delta K \sigma_E^2).
\end{equation*}

In particular, when $\hat \delta = \infty$  (i.e. $\hat h^2 = 1$), there is no shrinkage, and $\hat G = \bar y$. The prediction error variance is then completely determined by the residual variance, since $\hat G - G = \bar y - G = \bar E$, and
\begin{equation*}
E (\hat G - G) (\hat G - G)^t = r^{-1} \sigma_E^2 I_n.
\end{equation*}
On the other hand, when $\hat \delta = 0$  (i.e. $\hat h^2 = 0$), there is 'total' shrinkage towards zero, i.e. $\hat G = 0$, and
\begin{equation*}
E (\hat G - G) (\hat G - G)^t = E (G G^t) = \sigma_A^2 K.
\end{equation*}
This explains the asymmetry in the observed accuracy in our simulations, in particular when $h^2=0.5$: when the number of replicates $r$ is sufficiently large, overestimating the heritability will have less impact on the prediction error variance (and hence accuracy) than underestimating it.

\newpage
%%%%%%%%%%%%%%%%%%%%%%%%%%%%%%%%%%%%%%%%%%%%%%%%%%%%%%%%%%%%%%%%%%%%%%%%%
% Figure 1

\section*{Figure S1: histograms of the off-diagonal kinship coefficients, for 4 sub-populations of the regmap.}

\renewcommand{\figurename}{Figure S1}
 \renewcommand{\thefigure}{}

% figure produced with the script VISUALISE_POPULATION_STRUCTURE.r
\begin{figure}[!ht]
\begin{center}
\includegraphics[width=6in]{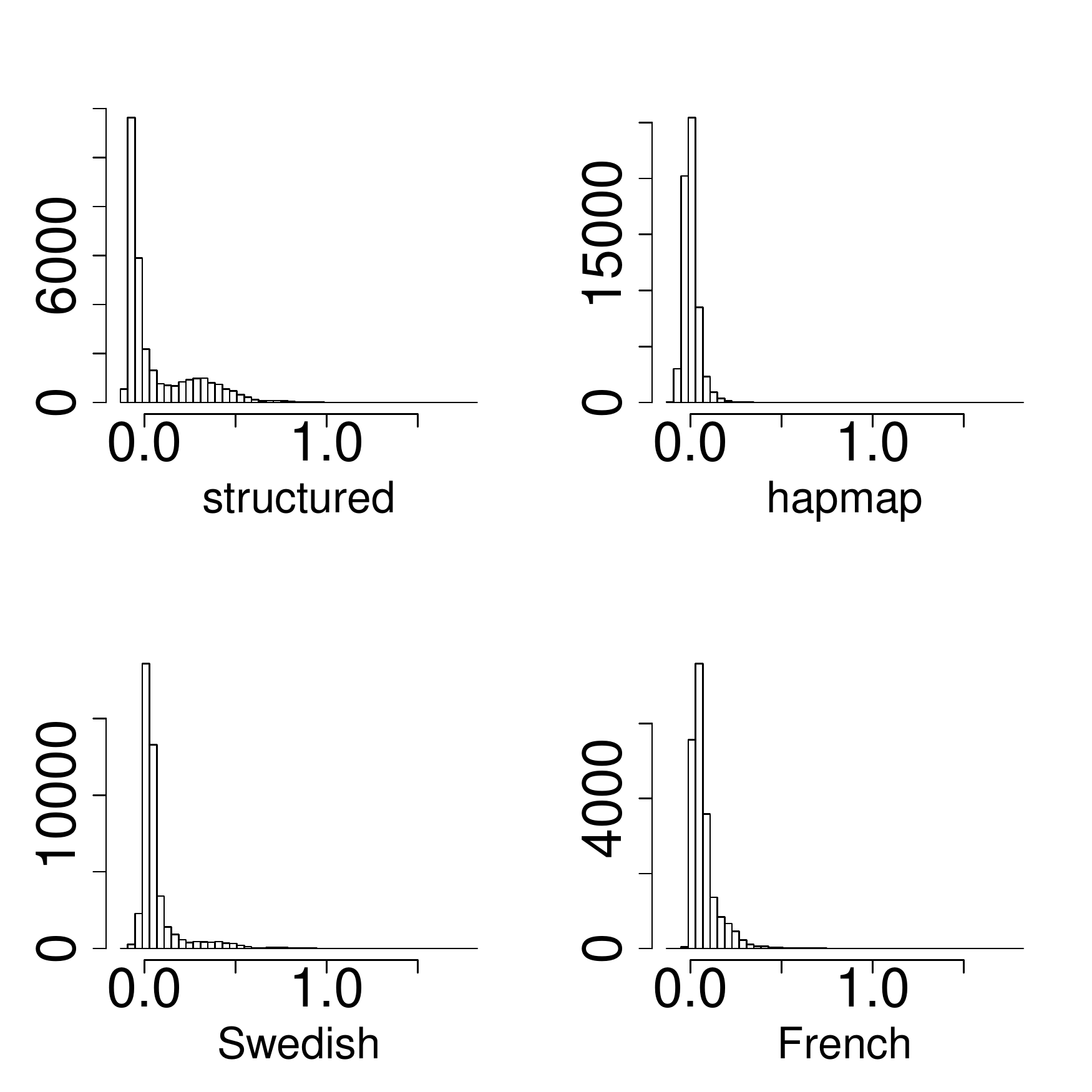}
\end{center}
\caption{
{\bf Off-diagonal coefficients of the genetic relatedness matrix (equation (1) in the main text), for 4 sub-populations of the regmap.}
}
\label{GRM_histograms}
\end{figure}

\newpage
%%%%%%%%%%%%%%%%%%%%%%%%%%%%%%%%%%%%%%%%%%%%%%%%%%%%%%%%%%%%%%%%%%%%%%%%%
% Figure 2

\section*{Figure S2: histograms of the off-diagonal identity-by-state coefficients, for 4 sub-populations of the regmap.}

\renewcommand{\figurename}{Figure S2}
 \renewcommand{\thefigure}{}

% figure produced with the script VISUALISE_POPULATION_STRUCTURE.r
%
\begin{figure}[!ht]
\begin{center}
\includegraphics[width=6in]{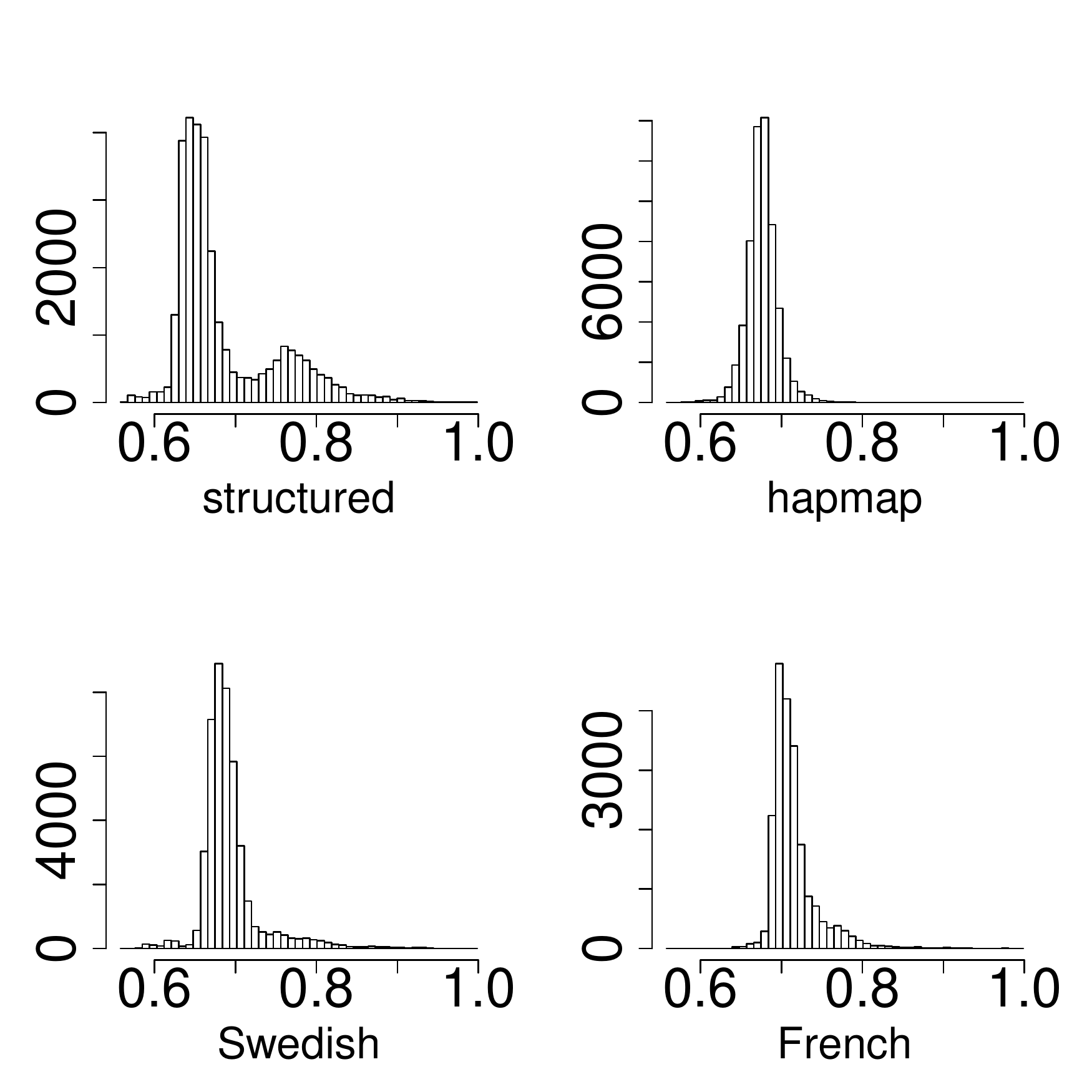}
\end{center}
\caption{
{\bf Off-diagonal identity-by-state kinship coefficients, for 4 sub-populations of the regmap.}
}
\label{IBS_histograms}
\end{figure}

\newpage
%%%%%%%%%%%%%%%%%%%%%%%%%%%%%%%%%%%%%%%%%%%%%%%%%%%%%%%%%%%%%%%%%%%%%%%%%
% Figure 3

\section*{Figure S3}

\renewcommand{\figurename}{Figure S3}
 \renewcommand{\thefigure}{}

% figure produced with the script h2_sim_PLOTS_h2_levels_NEW.r
\begin{figure}[!ht]
\setlength{\fboxsep}{1pt}%
\setlength{\fboxrule}{1pt}%
\fbox{\includegraphics[width=16cm,height=9.5cm]{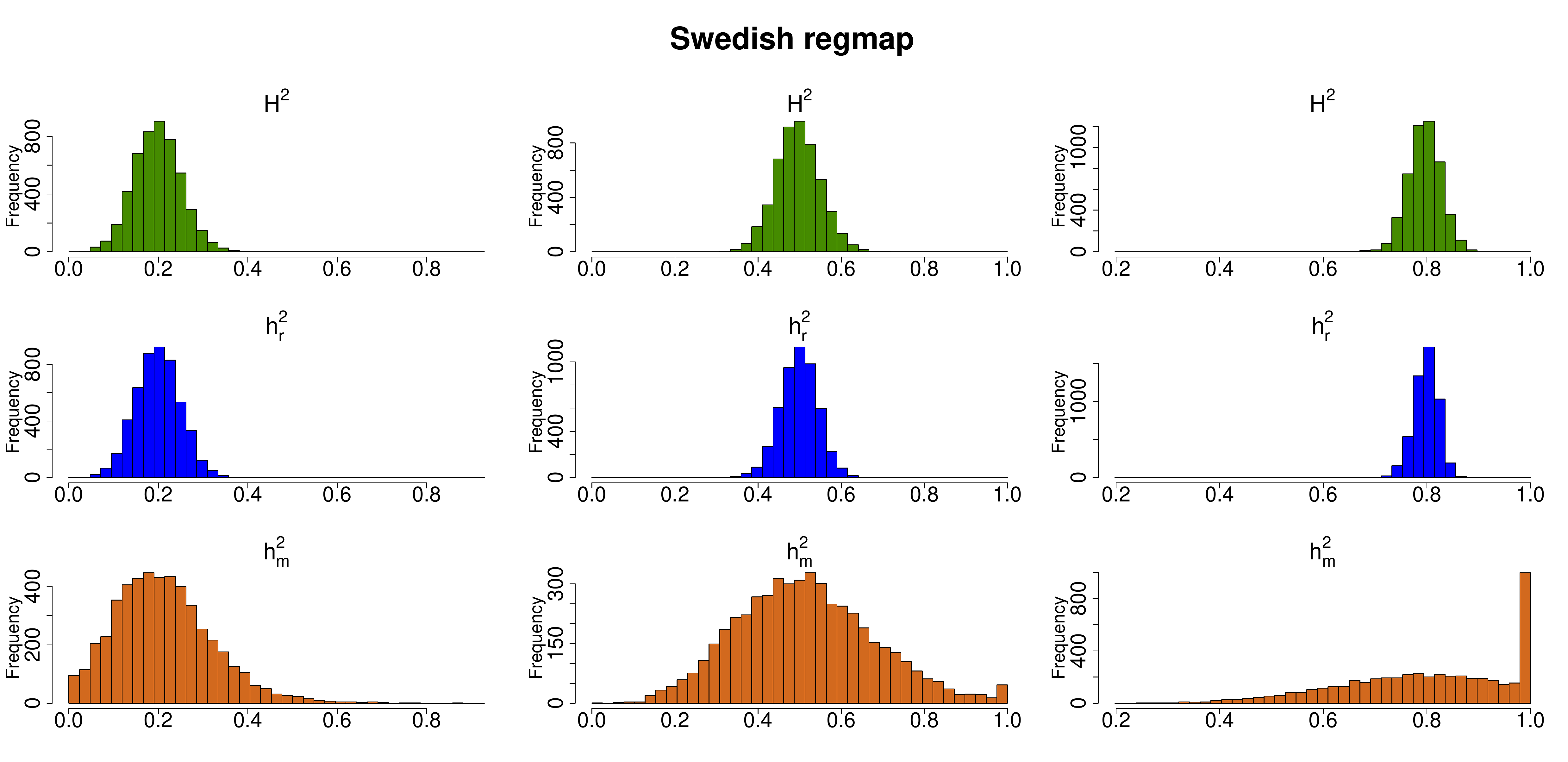}}
%\vspace{15pt}
\fbox{\includegraphics[width=16cm,height=9.5cm]{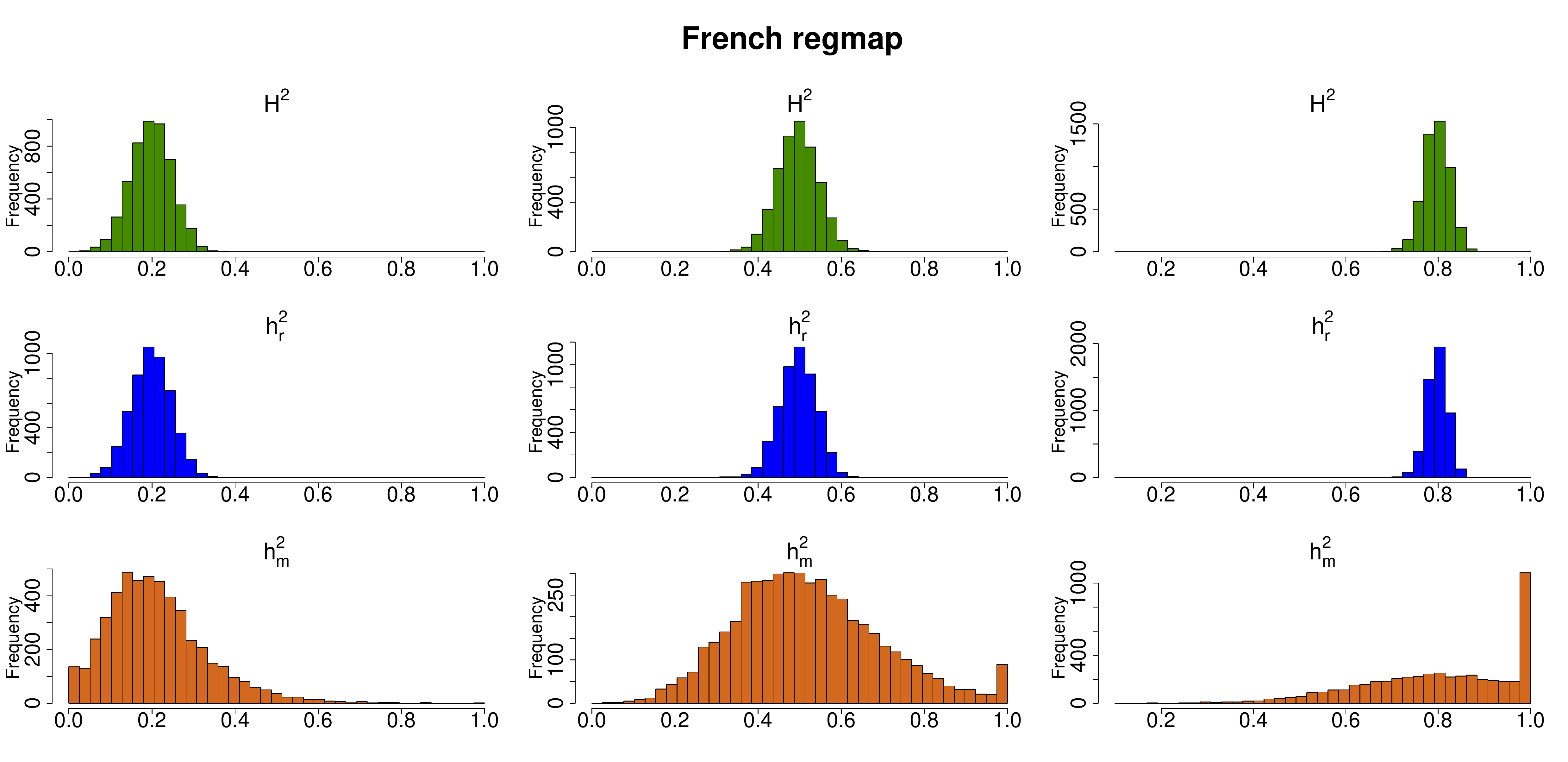}}
\caption{{\bf Heritability estimates for 5000 simulated traits for random samples of 200 accessions drawn from the Swedish regmap (top panel) and the French regmap (bottom panel).} 20 QTLs were simulated, which explained half of the genetic variance. The simulated heritability was 0.2 (left column), 0.5 (middle column) and 0.8 (right column). Within each panel, the first row shows the ANOVA-based estimates of broad-sense heritability, the second row the mixed model based estimates based on the individual data, and the third row the mixed model based estimates based on genotypic means.}%\label{boxplot3}
%\label{h2_sim histograms_swedish_french}
%\label{h2_sim histograms_structured}
\end{figure}
%\section
%\subsection*{s2}

\newpage
%%%%%%%%%%%%%%%%%%%%%%%%%%%%%%%%%%%%%%%%%%%%%%%%%%%%%%%%%%%%%%%%%%%%%%%%%
% Figure 4

\section*{Figure S4: Monotone likelihood}

\renewcommand{\figurename}{Figure S4}
 \renewcommand{\thefigure}{}

% figure produced with the script plot_REML_likelihood.r
\begin{figure}[!ht]
\centering
\begin{minipage}{0.45\linewidth}
\includegraphics[width=7cm,height=7cm]{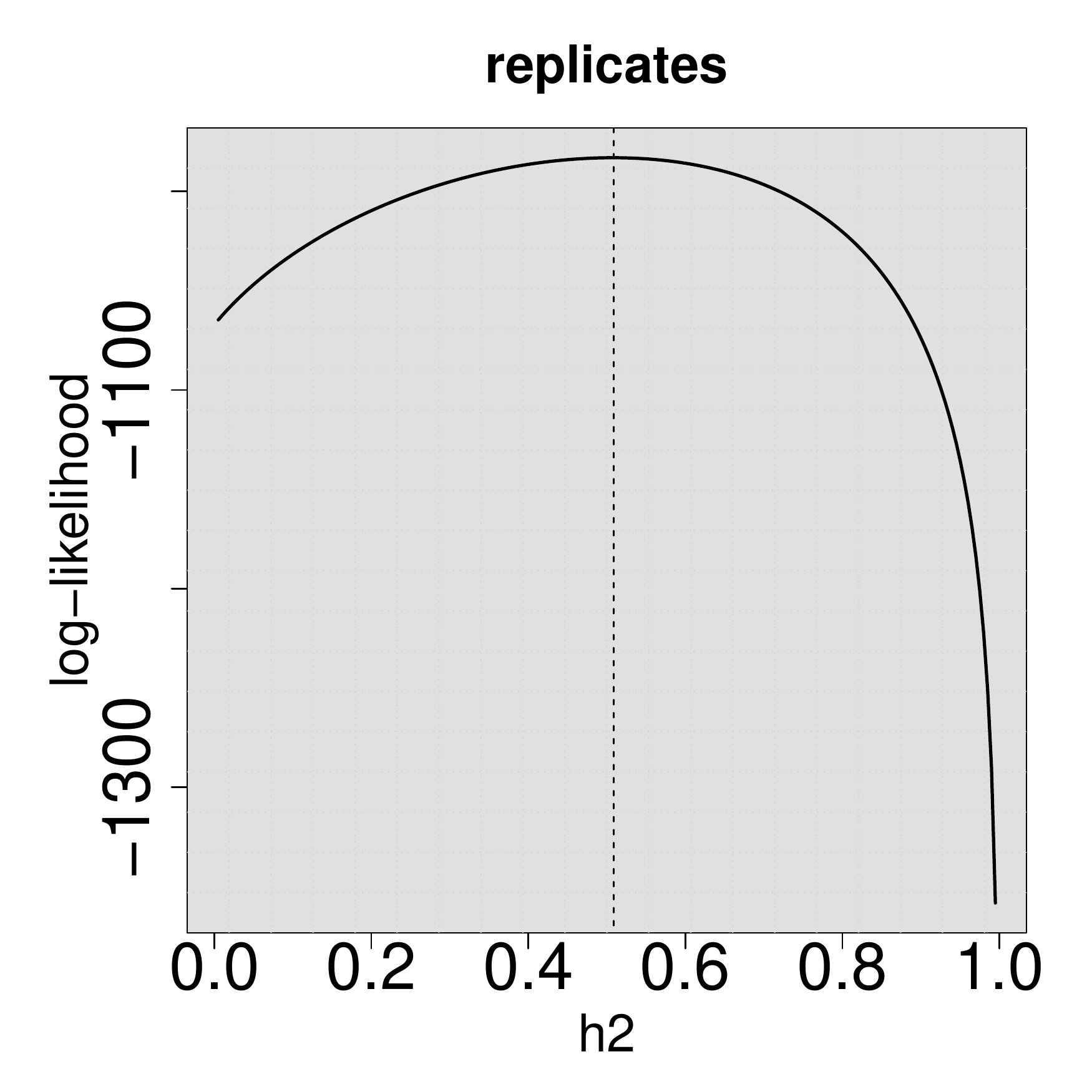}
\end{minipage} \quad
\begin{minipage}{0.45\linewidth}
\includegraphics[width=7cm,height=7cm]{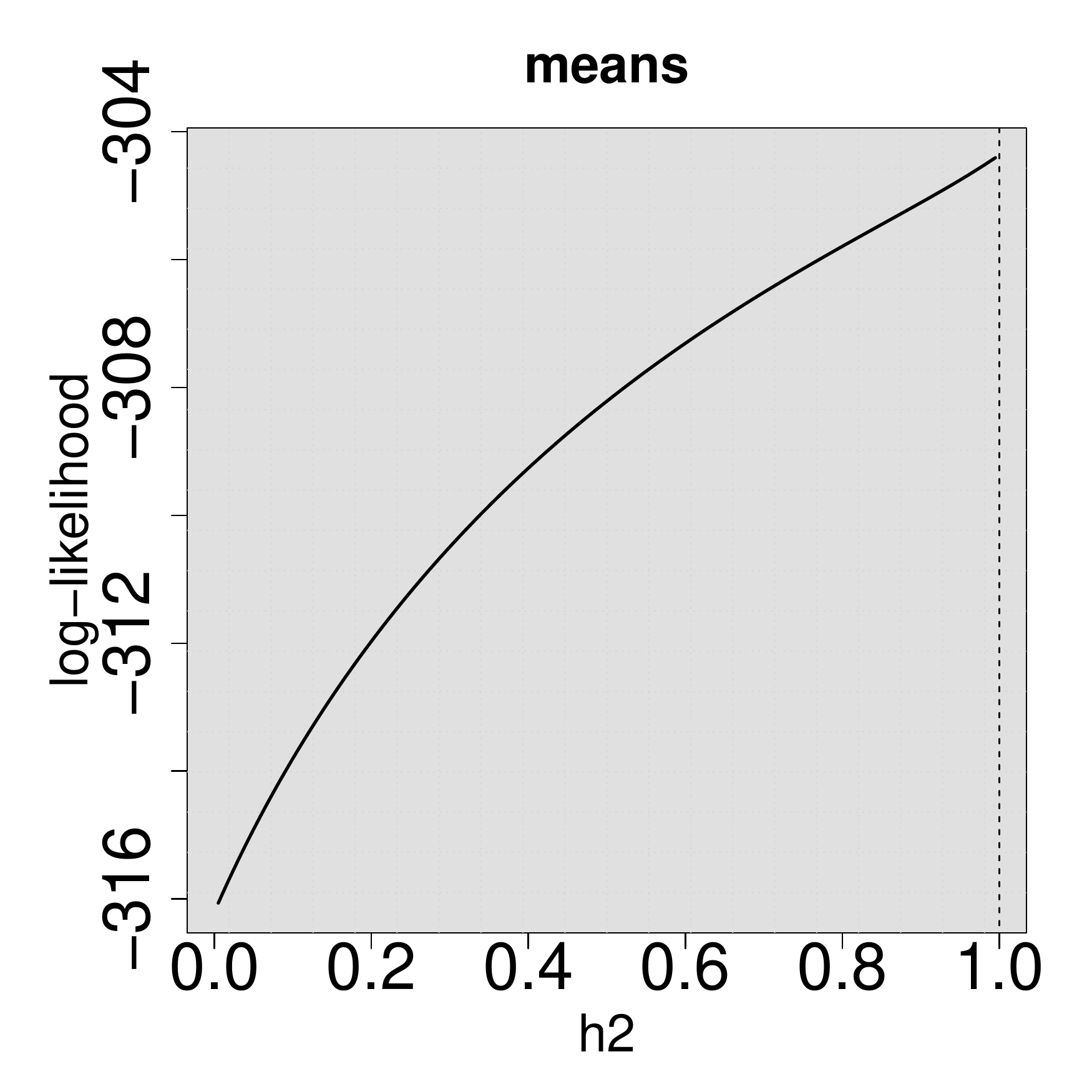}
\end{minipage}
%\begin{center}
%\includegraphics[width=16cm,height=9cm]{plots/log_lik_profile_replicates.pdf}
%\end{center}
%\caption{
%{\bf Log-likelihood of the heritability (h2) for 1 simulated trait with heritability 0.5, based on the individuals observations and the genotypic means (right).}
%}
\caption{{\bf Log-likelihood as function of the heritability, for one of the 5000 simulated traits from Figure 1 (in the main text), for accessions drawn from the HapMap and a simulated heritability of 0.5} Here we choose one of the 882 traits ($17.6 \%$) for which the heritability estimate based on means ($\hat h_m^2$) was larger than $0.99$. For these traits, the heritability estimate based on replicates ($\hat h_r^2$) was on average $0.502$. For the trait shown here, $\hat h_m^2 = 0.5087$ (left) and $\hat h_m^2 = 0.9999$ (right).}
\label{profile_lik}
\end{figure}

\newpage
%%%%%%%%%%%%%%%%%%%%%%%%%%%%%%%%%%%%%%%%%%%%%%%%%%%%%%%%%%%%%%%%%%%%%%%%%
% Figure 5

\section*{Figure S5}

\renewcommand{\figurename}{Figure S5}
 \renewcommand{\thefigure}{}

% figure produced with the script CONFIDENCE_INTERVAL_PLOTS_VERTICAL_ALL_LDAK.r
\begin{figure}[!ht]
\begin{center}
\includegraphics[height=6in]{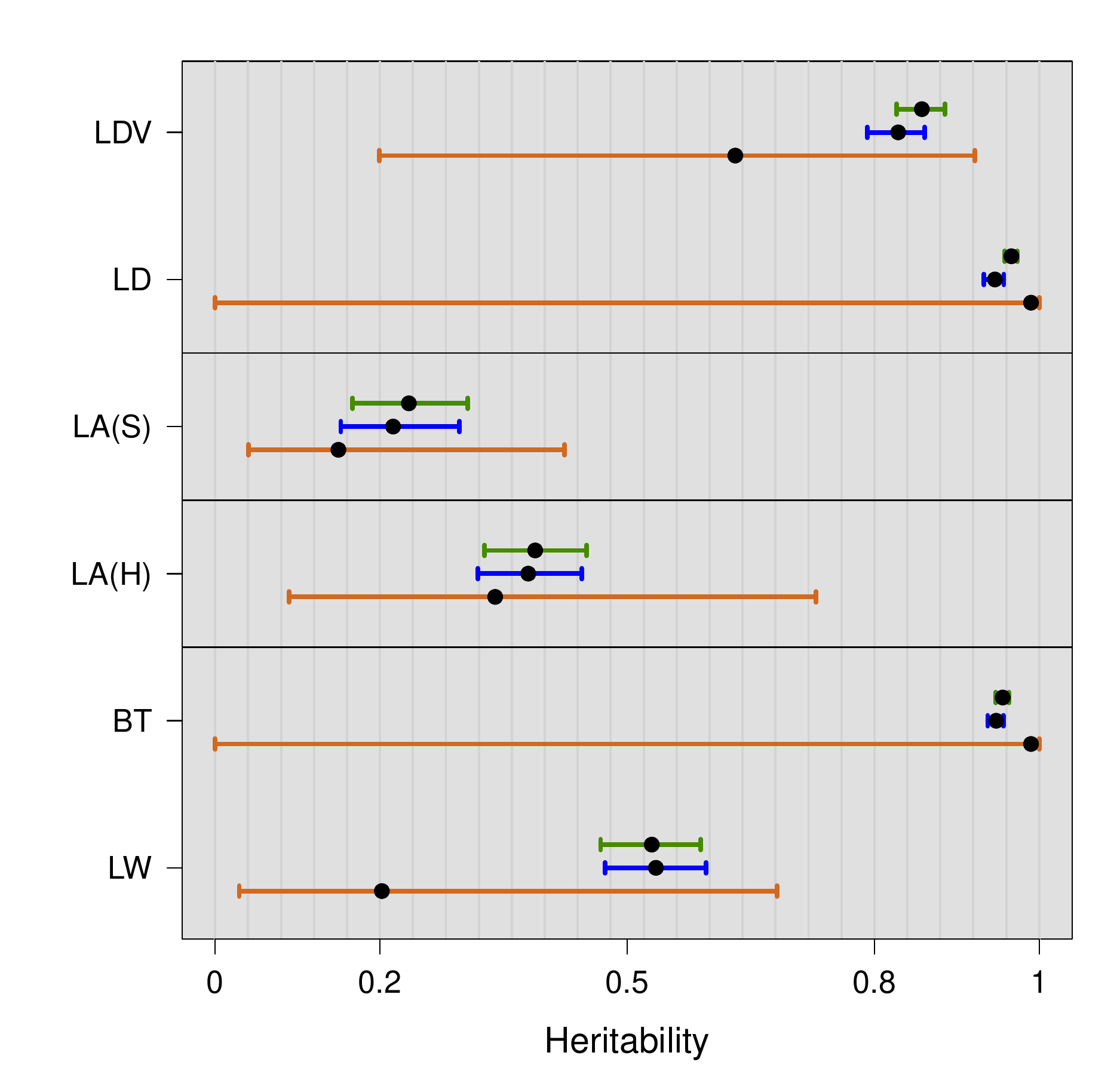}
\end{center}
%\caption{
%{\bf Heritability estimates and confidence intervals for two flowering traits from \cite{atwell_etal_2010} (LDV and LD), and $4$ traits from new experiments.} Trait abbreviations are given in Table 1 of the main text. Three estimators were used: broad-sense heritability (${\hat H}^2$, green),
%mixed model based on replicates (${\hat h}_r^2$, blue) and mixed model based on genotypic means (${\hat h}_m^2$, brown). Traits from the $4$ experiments are separated by the black horizontal lines.
%}
%\label{all_h2}
\caption{
{\bf Heritability estimates and confidence intervals for two flowering traits from \cite{atwell_etal_2010} (LDV and LD), and $4$ traits from new experiments.}  Three estimators were used: the ANOVA-based estimator of broad-sense heritability (${\hat H}^2$, green),
the marker-based estimator using individual plant data (${\hat h}_r^2$, blue) and the marker-based estimator using genotypic means (${\hat h}_m^2$, brown). Traits from different experiments are separated by the black horizontal lines. Trait abbreviations are given in Table 1 of the main text.
The LD-adjusted kinship matrix was computed using version 2.0 of the LDAK-software \cite{speed_hemani_johnson_balding_2012}.%, available at \verb|http://dougspeed.com/ldak/|.
We used sections of 1000 SNPs, with a buffer of 200. The maximum distance considered for LD was 250kb; the 'halflife' parameter (modeling LD-decay) was set to 20kb.
}
\end{figure}

%\protect\UseVerb
% \protect{\verb|http://dougspeed.com/ldak/|}
% \protect\UseVerb{http://dougspeed.com/ldak/}

%\newpage
%\maketitle

\newpage
%%%%%%%%%%%%%%%%%%%%%%%%%%%%%%%%%%%%%%%%%%%%%%%%%%%%%%%%%%%%%%%%%%%%%%%%%
% Figure 6

\section*{Figure S6}

\renewcommand{\figurename}{Figure S6}
 \renewcommand{\thefigure}{}
% plot made using the script GWAS_correlations_in_simulations.r

\begin{figure}[!ht]
\includegraphics[width=15cm,height=15cm]{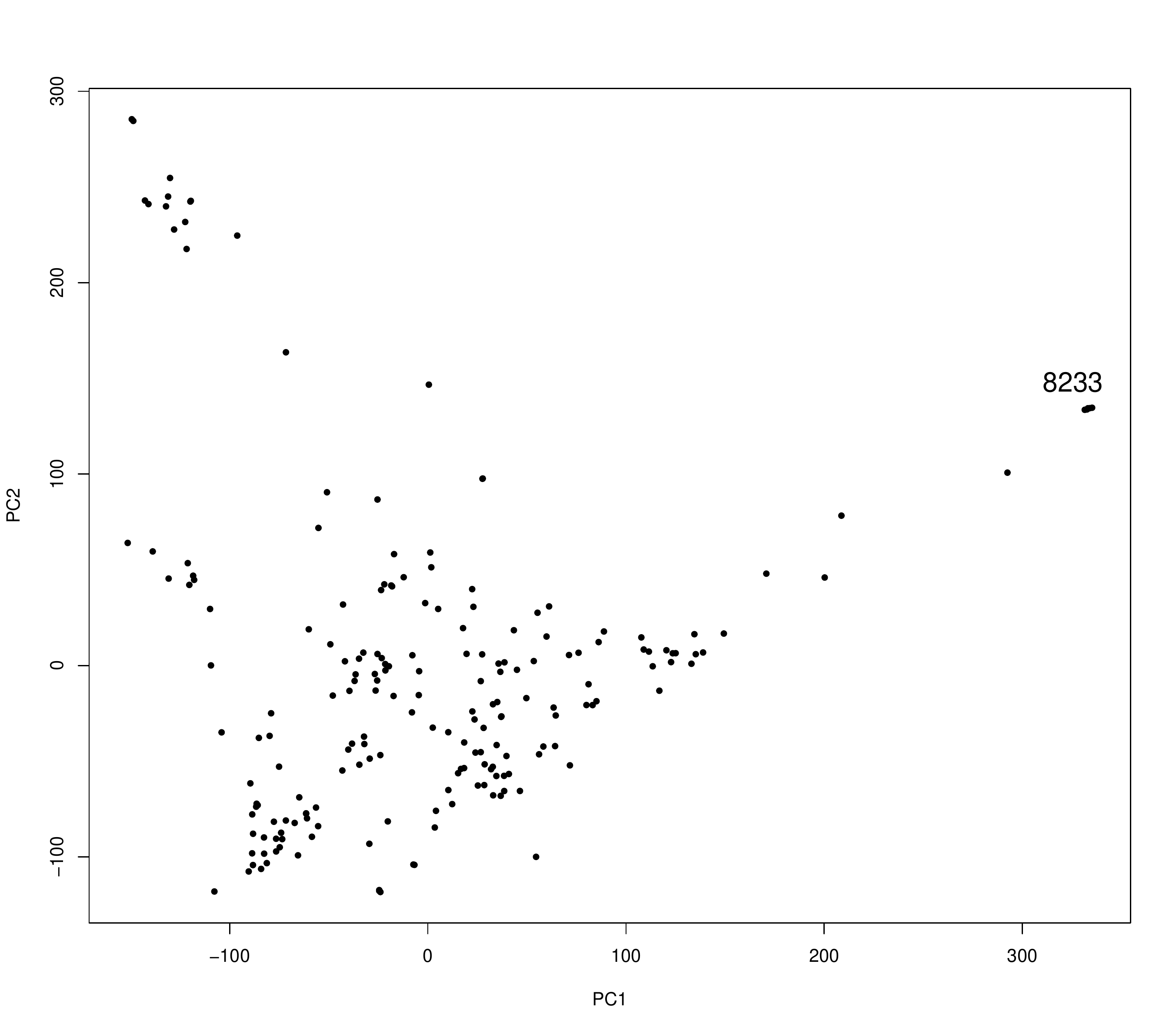}
\caption{{ \bf The first two principal component of the genetic markers for the panel of \cite{atwell_etal_2010}, restricted for the $168$ accessions for which the trait LDV was measured.}
On the very right are the accessions with ecotype ID's $8233$, $7526$ and $7515$.}
%\label{GWAS_correlations_pop1_and_pop3}
\end{figure}

\newpage
%%%%%%%%%%%%%%%%%%%%%%%%%%%%%%%%%%%%%%%%%%%%%%%%%%%%%%%%%%%%%%%%%%%%%%%%%
% Figure 7

\section*{Figure S7}

\renewcommand{\figurename}{Figure S7}
 \renewcommand{\thefigure}{}

\begin{figure}[!ht]
\includegraphics[width=15cm,height=10cm]{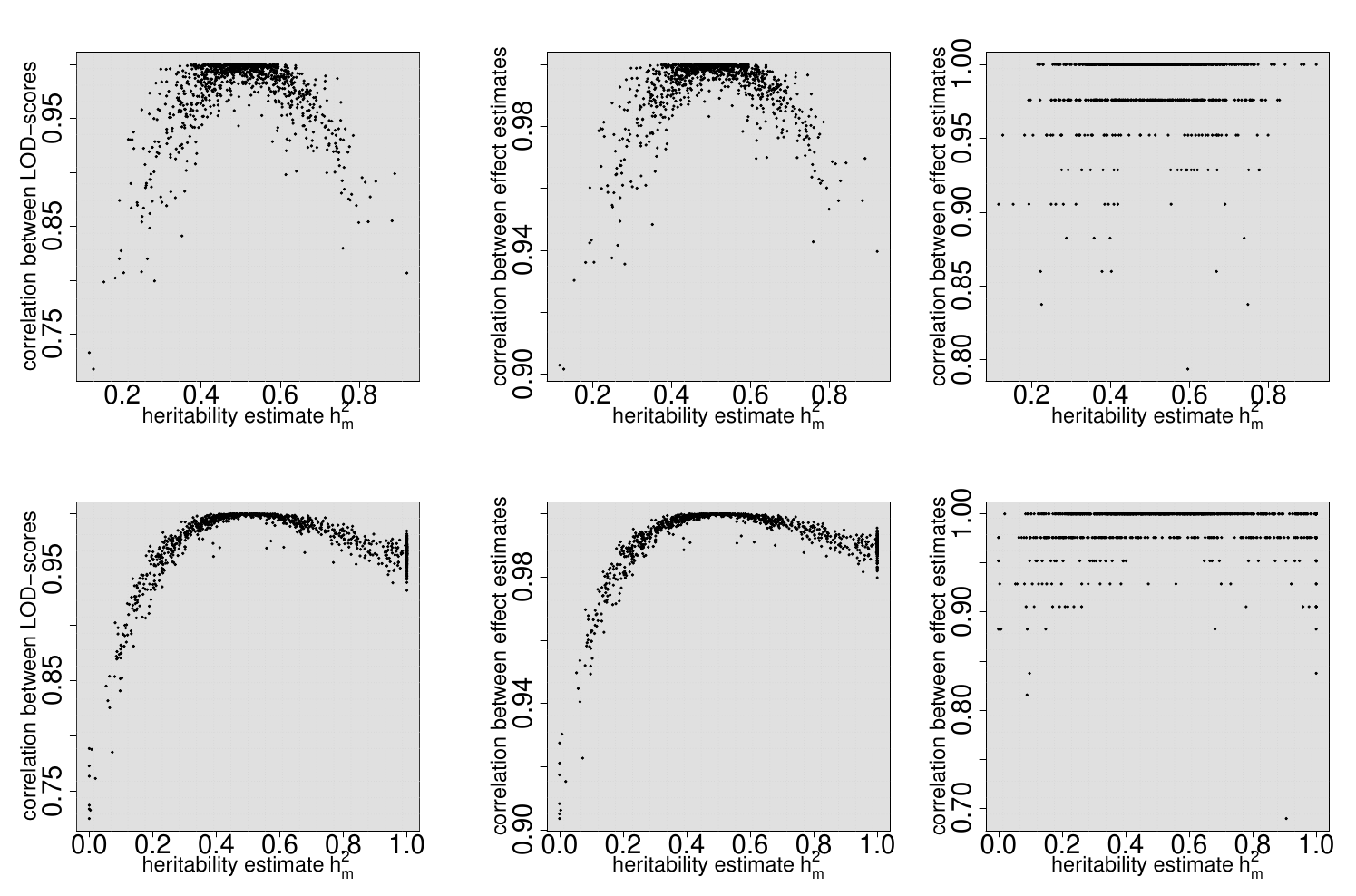}
\caption{{ \bf Rank correlation (Spearman $\rho^2$) between effect-size estimates obtained with a one- and two-stage approach, versus the heritability estimates obtained in the two-stage approach (${\hat h_m}^2$).}
         1000 traits were simulated for the Structured RegMap (first row) and the HapMap (second row), with a simulated heritability of $0.5$. 10 QTLs were simulated, which explained $75 \%$ of the genetic variance. Left column: rank correlation between LOD-scores of all SNPs. Middle column: rank correlation between effect-size estimates for all SNPs. Right column: rank correlation between effect-size estimates for the $10$ simulated QTLs.}
%\label{GWAS_correlations_pop1_and_pop3}
\end{figure}

\newpage
%%%%%%%%%%%%%%%%%%%%%%%%%%%%%%%%%%%%%%%%%%%%%%%%%%%%%%%%%%%%%%%%%%%%%%%%%
% Table 1

\section*{Table S1: ecotype-IDs of the structured regmap.}

Accession information was taken from the Bergelson lab \newline (\verb|http://bergelson.uchicago.edu/Members/mhorton/resources/snps/accession_coordinates.xls|). The following table contains geographic information for $242$ of the $250$ accessions from our structured regmap. Geographic information was not available for eight accessions: 6909, 8428, 5712, 6143, 5708, 5730, 5829 and 8254.
Accessions were selected based on the variance of the off-diagonal kinship coefficients of each row: the accessions corresponding to the rows with the $250$ highest variances were chosen.
%The set mainly consists of  American ($103$), British ($40$) and Swedish ($83$) accessions (Table \ref{accession_table}).

\vspace{1cm}
% table produced with the script VISUALISE_POPULATION_STRUCTURE.r
%\begin{table}[!ht]
%\centering
%\caption{
%\bf{Geographic origin of the accession of the structured regmap.}}
\begin{tabular}{|l|r|}
  \hline
country & number of accessions \\
  \hline
Czech Republic & 7 \\
  Finland & 2 \\
  France & 2 \\
  Ireland & 2 \\
  Sweden & 83 \\
  Tajikistan & 3 \\
  United Kingdom & 40 \\
  United States of America & 103 \\
  unknown & 8 \\
   \hline
\end{tabular}
%\begin{flushleft}%Table caption
%The structured regmap is a subset of the regmap \cite{horton_etal_2012}, chosen on the basis of on the variance of the off-diagonal (EIGENSTRAT) kinship coefficients of each row. Accessions were ranked according to the %variance of their kinship coefficients; accessions corresponding to the 250 highest variances were chosen.
%\end{flushleft}
%\label{accession_table}
%\end{table}

\vspace{1cm}

% table produced with the script VISUALISE_POPULATION_STRUCTURE.r
\begin{longtable}{|c|c|c|c|c|c|}%[!ht]
  \hline
ecotype-id & country & latitude & longitude & population & native-name \\
  \hline
1716 & United States of America & 42.405 & -85.398 & Americas & KBS-Mac-8 \\
  1718 & United States of America & 42.405 & -85.398 & Americas & KBS-Mac-15 \\
  1719 & United States of America & 42.405 & -85.398 & Americas & KBS-Mac-16 \\
  1722 & United States of America & 42.405 & -85.398 & Americas & KBS-Mac-23 \\
  1724 & United States of America & 42.405 & -85.398 & Americas & KBS-Mac-28 \\
  1726 & United States of America & 42.405 & -85.398 & Americas & KBS-Mac-33 \\
  1729 & United States of America & 42.405 & -85.398 & Americas & KBS-Mac-41 \\
  1730 & United States of America & 42.405 & -85.398 & Americas & KBS-Mac-43 \\
  1733 & United States of America & 42.405 & -85.398 & Americas & KBS-Mac-53 \\
  1736 & United States of America & 42.405 & -85.398 & Americas & KBS-Mac-58 \\
  1738 & United States of America & 42.405 & -85.398 & Americas & KBS-Mac-64 \\
  1740 & United States of America & 42.405 & -85.398 & Americas & KBS-Mac-72 \\
  1743 & United States of America & 42.405 & -85.398 & Americas & KBS-Mac-75 \\
  1744 & United States of America & 42.405 & -85.398 & Americas & KBS-Mac-76 \\
  1745 & United States of America & 42.405 & -85.398 & Americas & KBS-Mac-78 \\
  1749 & United States of America & 42.405 & -85.398 & Americas & KBS-Mac-88 \\
  1750 & United States of America & 42.405 & -85.398 & Americas & KBS-Mac-89 \\
  1751 & United States of America & 42.405 & -85.398 & Americas & KBS-Mac-91 \\
  1752 & United States of America & 42.405 & -85.398 & Americas & KBS-Mac-95 \\
  1753 & United States of America & 42.405 & -85.398 & Americas & KBS-Mac-96 \\
  1782 & United States of America & 42.184 & -86.358 & Americas & Ker-38 \\
  1829 & United States of America & 42.051 & -86.509 & Americas & Mdn-1 \\
  1850 & United States of America & 43.595 & -86.2657 & Americas & MNF-Pot-8 \\
  1853 & United States of America & 43.595 & -86.2657 & Americas & MNF-Pot-21 \\
  1858 & United States of America & 43.595 & -86.2657 & Americas & MNF-Pot-47 \\
  1864 & United States of America & 43.595 & -86.2657 & Americas & MNF-Pot-60 \\
  1871 & United States of America & 43.595 & -86.2657 & Americas & MNF-Pot-76 \\
  1872 & United States of America & 43.595 & -86.2657 & Americas & MNF-Pot-75 \\
  1873 & United States of America & 43.595 & -86.2657 & Americas & MNF-Pot-79 \\
  1874 & United States of America & 43.595 & -86.2657 & Americas & MNF-Pot-80 \\
  1938 & United States of America & 43.5251 & -86.1843 & Americas & MNF-Che-41 \\
  1941 & United States of America & 43.5251 & -86.1843 & Americas & MNF-Che-45 \\
  1948 & United States of America & 43.5251 & -86.1843 & Americas & MNF-Che-58 \\
  1960 & United States of America & 43.5187 & -86.1739 & Americas & MNF-Jac-22 \\
  1963 & United States of America & 43.5187 & -86.1739 & Americas & MNF-Jac-26 \\
  2057 & United States of America & 42.166 & -86.412 & Americas & Map-42 \\
  2148 & United States of America & 42.148 & -86.431 & Americas & Paw-1 \\
  2150 & United States of America & 42.148 & -86.431 & Americas & Paw-3 \\
  2157 & United States of America & 42.148 & -86.431 & Americas & Paw-11 \\
  2160 & United States of America & 42.148 & -86.431 & Americas & Paw-14 \\
  2180 & United States of America & 42.148 & -86.431 & Americas & Paw-40 \\
  2201 & United States of America & 43.7623 & -86.3929 & Americas & Pent-22 \\
  2204 & United States of America & 43.7623 & -86.3929 & Americas & Pent-30 \\
  2214 & United States of America & 43.7623 & -86.3929 & Americas & Pent-49 \\
  2274 & United States of America & 43.665 & -86.496 & Americas & SLSP-30 \\
  2280 & United States of America & 43.665 & -86.496 & Americas & SLSP-58 \\
  2294 & United States of America & 42.03 & -86.514 & Americas & Ste-9 \\
  2300 & United States of America & 42.03 & -86.514 & Americas & Ste-15 \\
  6927 & United States of America & 41.2816 & -86.621 & Americas & Kno-10 \\
  6983 & United States of America & 37.45 & -119.35 & Americas & Yo-0 \\
  7033 & United States of America & 41.3599 & -122.755 & Americas & Buckhorn Pass \\
  7515 & United States of America & 41.5609 & -86.4251 & Americas & RRS-10 \\
  7523 & United States of America & 42.0945 & -86.3253 & Americas & Pna-17 \\
  7524 & United States of America & 42.036 & -86.511 & Americas & Rmx-A02 \\
  7525 & United States of America & 42.036 & -86.511 & Americas & Rmx-A180 \\
  7526 & United States of America & 42.0945 & -86.3253 & Americas & Pna-10 \\
  7566 & United States of America & 42.093 & -86.359 & Americas & 627ME-13Y1 \\
  7578 & United States of America & 42.0945 & -86.3253 & Americas & 627PNA-1Y1 \\
  7580 & United States of America & 42.0945 & -86.3253 & Americas & 627PNA-2B3 \\
  7584 & United States of America & 42.0945 & -86.3253 & Americas & 627PNA-3M4 \\
  7787 & United States of America & 41.273 & -86.625 & Americas & KNO2.77 \\
  7837 & United States of America & 42.093 & -86.359 & Americas & ME3.41 \\
  7847 & United States of America & 42.093 & -86.359 & Americas & ME3.51 \\
  7867 & United States of America & 42.093 & -86.359 & Americas & ME4.20 \\
  8122 & United States of America & 42.036 & -86.511 & Americas & RMX3.11 \\
  8233 & United States of America & 41.1876 & -87.1923 & Americas & Dem-4 \\
  8557 & United States of America & 42.093 & -86.359 & Americas & 328ME032 \\
  8612 & United States of America & 42.093 & -86.359 & Americas & 11ME1.34 \\
  8616 & United States of America & 42.093 & -86.359 & Americas & 11ME1.41 \\
  8619 & United States of America & 42.093 & -86.359 & Americas & 11ME1.44 \\
  8629 & United States of America & 42.093 & -86.359 & Americas & 11ME2.10 \\
  8673 & United States of America & 42.0945 & -86.3253 & Americas & 328PNA032 \\
  8724 & United States of America & 42.0945 & -86.3253 & Americas & 11PNA1.15 \\
  8725 & United States of America & 42.0945 & -86.3253 & Americas & 11PNA1.4 \\
  8727 & United States of America & 42.0945 & -86.3253 & Americas & 11PNA1.6 \\
  8730 & United States of America & 42.0945 & -86.3253 & Americas & 11PNA1.9 \\
  8760 & United States of America & 42.0945 & -86.3253 & Americas & 11PNA3.19 \\
  8770 & United States of America & 42.0945 & -86.3253 & Americas & 11PNA3.65 \\
  8777 & United States of America & 42.0945 & -86.3253 & Americas & 11PNA3.75 \\
  8787 & United States of America & 42.0945 & -86.3253 & Americas & 11PNA3.86 \\
  8796 & United States of America & 42.0945 & -86.3253 & Americas & 11PNA4.101 \\
  8824 & United States of America & 42.0945 & -86.3253 & Americas & 11PNA4.129 \\
  8954 & United States of America & 42.036 & -86.511 & Americas & RMX413.1 \\
  8961 & United States of America & 42.036 & -86.511 & Americas & RMX413.16 \\
  8965 & United States of America & 42.036 & -86.511 & Americas & RMX413.2 \\
  8966 & United States of America & 42.036 & -86.511 & Americas & RMX413.20 \\
  8967 & United States of America & 42.036 & -86.511 & Americas & RMX413.21 \\
  8969 & United States of America & 42.036 & -86.511 & Americas & RMX413.24 \\
  8970 & United States of America & 42.036 & -86.511 & Americas & RMX413.25 \\
  8973 & United States of America & 42.036 & -86.511 & Americas & RMX413.29 \\
  8975 & United States of America & 42.036 & -86.511 & Americas & RMX413.30 \\
  8976 & United States of America & 42.036 & -86.511 & Americas & RMX413.31 \\
  8977 & United States of America & 42.036 & -86.511 & Americas & RMX413.32 \\
  8992 & United States of America & 42.036 & -86.511 & Americas & RMX413.48 \\
  8996 & United States of America & 42.036 & -86.511 & Americas & RMX413.51 \\
  9001 & United States of America & 42.036 & -86.511 & Americas & RMX413.57 \\
  9004 & United States of America & 42.036 & -86.511 & Americas & RMX413.6 \\
  9006 & United States of America & 42.036 & -86.511 & Americas & RMX413.62 \\
  9007 & United States of America & 42.036 & -86.511 & Americas & RMX413.63 \\
  9012 & United States of America & 42.036 & -86.511 & Americas & RMX413.7 \\
  9041 & United States of America & 42.039 & -86.5154 & Americas & RMXF413.10 \\
  9045 & United States of America & 42.039 & -86.5154 & Americas & RMXF413.15 \\
  9053 & United States of America & 42.039 & -86.5154 & Americas & RMXF413.6 \\
  5991 & Czech Republic & 49.4112 & 16.2815 & Austria-Hungary & DraIV 6-20 \\
  5997 & Czech Republic & 49.4112 & 16.2815 & Austria-Hungary & DraIV 6-27 \\
  6427 & Czech Republic & 49.3853 & 16.2544 & Austria-Hungary & ZdrI 2-1 \\
  6435 & Czech Republic & 49.3853 & 16.2544 & Austria-Hungary & ZdrI 2-10 \\
  6444 & Czech Republic & 49.3853 & 16.2544 & Austria-Hungary & ZdrI 2-20 \\
  6903 & Czech Republic & 49.4013 & 16.2326 & Austria-Hungary & Bor-4 \\
  7461 & Czech Republic & 49 & 15 & Austria-Hungary & H55 \\
  4632 & United Kingdom & 50.4 & -4.7 & British-Isles & UKSW06-025 \\
  4675 & United Kingdom & 50.4 & -4.7 & British-Isles & UKSW06-070 \\
  4862 & United Kingdom & 50.3 & -4.9 & British-Isles & UKSW06-262 \\
  5106 & United Kingdom & 51.3 & 0.5 & British-Isles & UKSE06-254 \\
  5133 & United Kingdom & 52.2 & -1.7 & British-Isles & UKSE06-302 \\
  5207 & United Kingdom & 51.3 & 0.4 & British-Isles & UKSE06-429 \\
  5232 & United Kingdom & 51.2 & 0.4 & British-Isles & UKSE06-466 \\
  5292 & United Kingdom & 51.3 & 1.1 & British-Isles & UKSE06-556 \\
  5331 & United Kingdom & 51.1 & 0.4 & British-Isles & UKSE06-618 \\
  5341 & United Kingdom & 51.1 & 0.4 & British-Isles & UKSE06-628 \\
  5380 & United Kingdom & 54.4 & -3 & British-Isles & UKNW06-059 \\
  5381 & United Kingdom & 54.4 & -3 & British-Isles & UKNW06-060 \\
  5385 & United Kingdom & 54.4 & -3 & British-Isles & UKNW06-078 \\
  5469 & United Kingdom & 54.4 & -3 & British-Isles & UKNW06-210 \\
  5582 & United Kingdom & 54.7 & -3.4 & British-Isles & UKNW06-410 \\
  5678 & United Kingdom & 54.6 & -3.1 & British-Isles & UKNW99-025 \\
  5709 & United Kingdom & 54.6 & -2.6 & British-Isles & UKID2 \\
  5719 & Ireland & 54.1335 & -6.1667 & British-Isles & Bur-0 \\
  5720 & United Kingdom & 53.3 & -1.6 & British-Isles & Cal-2 \\
  5723 & United Kingdom & 51.3 & 1 & British-Isles & Chr-1 \\
  5724 & United Kingdom & 51.4 & 0.1 & British-Isles & UKID17 \\
  5731 & United Kingdom & 54.9 & -2.9 & British-Isles & Crl-1 \\
  5732 & United Kingdom & 54.9 & -2.9 & British-Isles & UKID25 \\
  5737 & United Kingdom & 51.3 & 0.1 & British-Isles & UKID32 \\
  5758 & United Kingdom & 50.3 & -5.2 & British-Isles & UKID53 \\
  5774 & United Kingdom & 51.1 & 0.6 & British-Isles & Sis-1 \\
  5780 & United Kingdom & 53.1 & -1 & British-Isles & UKID75 \\
  5788 & United Kingdom & 50.8 & -2 & British-Isles & UKID83 \\
  5792 & United Kingdom & 50.8 & -0.7 & British-Isles & UKID87 \\
  5793 & United Kingdom & 51.3 & 0.6 & British-Isles & UKID88 \\
  5798 & United Kingdom & 53.1 & -3.3 & British-Isles & UKID93 \\
  5804 & United Kingdom & 50.8 & -1.1 & British-Isles & UKID100 \\
  5807 & United Kingdom & 51.8 & -0.5 & British-Isles & UKID103 \\
  6905 & Ireland & 54.1 & -6.2 & British-Isles & Bur-0 \\
  6923 & United Kingdom & 51.4083 & -0.6383 & British-Isles & HR-10 \\
  6924 & United Kingdom & 51.4083 & -0.6383 & British-Isles & HR-5 \\
  6944 & United Kingdom & 51.4083 & -0.6383 & British-Isles & NFA-8 \\
  7064 & United Kingdom & 51.3 & 1.1 & British-Isles & Cnt-1 \\
  7109 & United Kingdom & 51.3 & 0.5 & British-Isles & Ema-1 \\
  7483 & United Kingdom & 51.2878 & 0.0565 & British-Isles & PHW-14 \\
  9490 & United Kingdom & 55.9218 & -3.17108 & British-Isles & 02B6 \\
  9504 & United Kingdom & 55.8877 & -3.16377 & British-Isles & 12A1 \\
  6929 & Tajikistan & 38.48 & 68.49 & Eastern-Range & Kondara \\
  6962 & Tajikistan & 38.35 & 68.48 & Eastern-Range & Shahdara \\
  7168 & Tajikistan & 38.48 & 68.49 & Eastern-Range & Hodja-Obi-Garm \\
  1247 & Sweden & 59.4333 & 17.0167 & Fennoscandia & Tos-31-374 \\
  1254 & Sweden & 59.4333 & 17.0167 & Fennoscandia & Tos-82-387 \\
  1409 & Sweden & 62.8 & 18.2 & Fennoscandia & Röd-38 \\
  1416 & Sweden & 62.8 & 18.2 & Fennoscandia & Röd-45 \\
  1435 & Sweden & 62.8 & 18.2 & Fennoscandia & Röd-17-319 \\
  1552 & Sweden & 63.0833 & 18.3667 & Fennoscandia & Sku-30 \\
  5835 & Sweden & 63.324 & 18.484 & Fennoscandia & Bil-3 \\
  5856 & Sweden & 63.0167 & 17.4914 & Fennoscandia & Dör-10 \\
  5860 & Sweden & 62.6814 & 18.0165 & Fennoscandia & Dra-3 \\
  6009 & Sweden & 62.877 & 18.177 & Fennoscandia & Eden-1 \\
  6010 & Sweden & 62.877 & 18.177 & Fennoscandia & Eden-5 \\
  6011 & Sweden & 62.877 & 18.177 & Fennoscandia & Eden-6 \\
  6012 & Sweden & 62.877 & 18.177 & Fennoscandia & Eden-7 \\
  6013 & Sweden & 62.877 & 18.177 & Fennoscandia & Eden-9 \\
  6016 & Sweden & 62.9 & 18.4 & Fennoscandia & Eds-1 \\
  6017 & Sweden & 62.9 & 18.4 & Fennoscandia & Eds-9 \\
  6024 & Sweden & 55.7509 & 13.3712 & Fennoscandia & Fly2-2 \\
  6025 & Sweden & 62.6437 & 17.7339 & Fennoscandia & Gro-3 \\
  6030 & Sweden & 62.806 & 18.1896 & Fennoscandia & Grön-5 \\
  6043 & Sweden & 62.801 & 18.079 & Fennoscandia & Löv-1 \\
  6046 & Sweden & 62.801 & 18.079 & Fennoscandia & Löv-5 \\
  6064 & Sweden & 62.9513 & 18.2763 & Fennoscandia & Nyl-2 \\
  6069 & Sweden & 62.9513 & 18.2763 & Fennoscandia & Nyl-7 \\
  6071 & Sweden & 62.9308 & 18.3448 & Fennoscandia & Omn-5 \\
  6077 & Sweden & 55.6942 & 13.4504 & Fennoscandia & Rev-3 \\
  6091 & Sweden & 55.6525 & 13.215 & Fennoscandia & T1010 \\
  6104 & Sweden & 55.7 & 13.2 & Fennoscandia & T1160 \\
  6118 & Sweden & 55.7 & 13.2 & Fennoscandia & T610 \\
  6154 & Sweden & 62.6422 & 17.7406 & Fennoscandia & TAA 04 \\
  6163 & Sweden & 62.6425 & 17.7356 & Fennoscandia & TAA 14 \\
  6166 & Sweden & 62.6425 & 17.7372 & Fennoscandia & TAA 17 \\
  6169 & Sweden & 62.8714 & 18.3447 & Fennoscandia & TÅD 01 \\
  6170 & Sweden & 62.8717 & 18.3442 & Fennoscandia & TÅD 02 \\
  6171 & Sweden & 62.8717 & 18.3444 & Fennoscandia & TÅD 03 \\
  6172 & Sweden & 62.8717 & 18.3436 & Fennoscandia & TÅD 04 \\
  6173 & Sweden & 62.8717 & 18.3419 & Fennoscandia & TÅD 05 \\
  6174 & Sweden & 62.8719 & 18.3422 & Fennoscandia & TÅD 06 \\
  6177 & Sweden & 62.6322 & 17.69 & Fennoscandia & TÄL 03 \\
  6180 & Sweden & 62.6322 & 17.6906 & Fennoscandia & TÄL 07 \\
  6184 & Sweden & 62.8892 & 18.4522 & Fennoscandia & TBÖ 01 \\
  6209 & Sweden & 62.8836 & 18.1842 & Fennoscandia & TEDEN 02 \\
  6210 & Sweden & 62.8839 & 18.1836 & Fennoscandia & TEDEN 03 \\
  6212 & Sweden & 63.0175 & 18.3239 & Fennoscandia & TFÄ 02 \\
  6214 & Sweden & 63.0175 & 18.3281 & Fennoscandia & TFÄ 04 \\
  6215 & Sweden & 63.0172 & 18.3281 & Fennoscandia & TFÄ 05 \\
  6216 & Sweden & 63.0167 & 18.3283 & Fennoscandia & TFÄ 06 \\
  6217 & Sweden & 63.0169 & 18.3283 & Fennoscandia & TFÄ 07 \\
  6218 & Sweden & 63.0172 & 18.3283 & Fennoscandia & TFÄ 08 \\
  6220 & Sweden & 62.806 & 18.1896 & Fennoscandia & TGR 01 \\
  6221 & Sweden & 62.806 & 18.1896 & Fennoscandia & TGR 02 \\
  6226 & Sweden & 62.7994 & 17.9033 & Fennoscandia & THÖ 08 \\
  6231 & Sweden & 62.96 & 18.2844 & Fennoscandia & TNY 04 \\
  6235 & Sweden & 62.9611 & 18.3589 & Fennoscandia & TOM 01 \\
  6236 & Sweden & 62.9617 & 18.36 & Fennoscandia & TOM 02 \\
  6237 & Sweden & 62.9619 & 18.35 & Fennoscandia & TOM 03 \\
  6238 & Sweden & 62.9619 & 18.35 & Fennoscandia & TOM 04 \\
  6240 & Sweden & 62.9622 & 18.35 & Fennoscandia & TOM 06 \\
  6241 & Sweden & 62.9614 & 18.3608 & Fennoscandia & TOM 07 \\
  6244 & Sweden & 62.9169 & 18.4728 & Fennoscandia & TRÄ 01 \\
  6900 & Sweden & 63.324 & 18.484 & Fennoscandia & Bil-5 \\
  6901 & Sweden & 63.324 & 18.484 & Fennoscandia & Bil-7 \\
  6913 & Sweden & 62.877 & 18.177 & Fennoscandia & Eden-2 \\
  6917 & Sweden & 63.0165 & 18.3174 & Fennoscandia & Fäb-2 \\
  6918 & Sweden & 63.0165 & 18.3174 & Fennoscandia & Fäb-4 \\
  6968 & Finland & 60 & 23.5 & Fennoscandia & Tamm-2 \\
  6969 & Finland & 60 & 23.5 & Fennoscandia & Tamm-27 \\
  8218 & Sweden & 62.877 & 18.177 & Fennoscandia & Eden-4 \\
  8227 & Sweden & 62.7989 & 17.9103 & Fennoscandia & THÖ 03 \\
  8335 & Sweden & 55.71 & 13.2 & Fennoscandia & Lund \\
  8376 & Sweden & 62.69 & 18 & Fennoscandia & Sanna-2 \\
  9321 & Sweden & 62.8622 & 18.336 & Fennoscandia & Ådal 1 \\
  9323 & Sweden & 62.8622 & 18.336 & Fennoscandia & Ådal 3 \\
  9332 & Sweden & 62.8698 & 18.381 & Fennoscandia & Bar 1 \\
  9354 & Sweden & 62.8762 & 18.1746 & Fennoscandia & Eden 15 \\
  9355 & Sweden & 62.8762 & 18.1746 & Fennoscandia & Eden 16 \\
  9356 & Sweden & 62.8762 & 18.1746 & Fennoscandia & Eden 17 \\
  9363 & Sweden & 62.9147 & 18.4045 & Fennoscandia & EdJ 2 \\
  9371 & Sweden & 63.016 & 18.3175 & Fennoscandia & FäL 1 \\
  9378 & Sweden & 63.0165 & 18.3174 & Fennoscandia & FäU 4 \\
  9386 & Sweden & 62.806 & 18.1896 & Fennoscandia & Grön 12 \\
  9388 & Sweden & 62.806 & 18.1896 & Fennoscandia & Grön 14 \\
  9427 & Sweden & 62.8815 & 18.4055 & Fennoscandia & Näs 2 \\
  9433 & Sweden & 62.9513 & 18.2763 & Fennoscandia & Nyl 13 \\
  9434 & Sweden & 62.8959 & 18.3659 & Fennoscandia & Öde 2 \\
  9471 & Sweden & 56.0648 & 13.9707 & Fennoscandia & UllA 1 \\
  171 & France & 47.3833 & 5.31667 & France & MIB-20 \\
  228 & France & 47.3833 & 5.31667 & France & MIB-9 \\
   \hline
%\end{tabular}
\end{longtable}

\newpage
%%%%%%%%%%%%%%%%%%%%%%%%%%%%%%%%%%%%%%%%%%%%%%%%%%%%%%%%%%%%%%%%%%%%%%%%%
% Table 2

\section*{Table S2}

\renewcommand{\tablename}{Table S2}
 \renewcommand{\thetable}{}

%
% latex table generated in R 3.0.1 by xtable 1.7-1 package
% Fri Jan 10 11:00:49 2014
\begin{table}[!ht]
\centering
\caption{{ \bf Comparison of the marker-based estimators heritability estimators $h_r^2$ and $h_m^2$ for simulated data.} We simulated 5000 traits, for random samples of 200 accessions drawn from Swedish and French regmap. 20 unlinked QTLs were simulated, which explained 50 percent of the genetic variance. The simulated heritability was $0.2$, $0.5$ and $0.8$. Standard errors are given relative to those of the broad sense heritability estimator ($H^2$).}
\begin{tabular}{|l|r|r|r|}
  \hline
 & bias & standard error & relative standard error \\
  \hline
\multicolumn{4}{|>{\columncolor[gray]{.8}}l|}{Swedish regmap} \\
  \hline
\multicolumn{4}{|c|}{$h^2 = 0.2$} \\
  \hline
broad-sense ($H^2$) & -0.00162 & 0.05227 & 1.00000 \\
  replicates ($h_r^2$) & -0.00109 & 0.04991 & 0.95498 \\
  means ($h_m^2$) & 0.01302 & 0.11018 & 2.10798 \\
   \hline
\multicolumn{4}{|c|}{$h^2 = 0.5$} \\
  \hline
broad-sense ($H^2$) & -0.00403 & 0.05373 & 1.00000 \\
  replicates ($h_r^2$) & -0.00173 & 0.04506 & 0.83860 \\
  means ($h_m^2$) & 0.01494 & 0.16662 & 3.10123 \\
   \hline
\multicolumn{4}{|c|}{$h^2 = 0.8$} \\
  \hline
broad-sense ($H^2$) & -0.00458 & 0.03130 & 1.00000 \\
  replicates ($h_r^2$) & -0.00180 & 0.02319 & 0.74095 \\
  means ($h_m^2$) & -0.00104 & 0.16227 & 5.18435 \\
   \hline
\multicolumn{4}{|>{\columncolor[gray]{.8}}l|}{French regmap} \\
  \hline
\multicolumn{4}{|c|}{$h^2 = 0.2$} \\
  \hline
broad-sense ($H^2$) & -0.00183 & 0.04958 & 1.00000 \\
  replicates ($h_r^2$) & -0.00196 & 0.04780 & 0.96421 \\
  means ($h_m^2$) & 0.01306 & 0.12049 & 2.43043 \\
   \hline
\multicolumn{4}{|c|}{$h^2 = 0.5$} \\
  \hline
broad-sense ($H^2$) & -0.00396 & 0.04930 & 1.00000 \\
  replicates ($h_r^2$) & -0.00396 & 0.04409 & 0.89431 \\
  means ($h_m^2$) & 0.01952 & 0.17547 & 3.55941 \\
   \hline
\multicolumn{4}{|c|}{$h^2 = 0.8$} \\
  \hline
broad-sense ($H^2$) & -0.00341 & 0.02808 & 1.00000 \\
  replicates ($h_r^2$) & -0.00236 & 0.02246 & 0.79988 \\
  means ($h_m^2$) & 0.00202 & 0.16461 & 5.86175 \\
   \hline
\end{tabular}
\end{table}

\newpage
%%%%%%%%%%%%%%%%%%%%%%%%%%%%%%%%%%%%%%%%%%%%%%%%%%%%%%%%%%%%%%%%%%%%%%%%%
% Table 3

\section*{Table S3}

\renewcommand{\tablename}{Table S3:}
 \renewcommand{\thetable}{}

% latex table generated in R 3.0.1 by xtable 1.7-1 package
% Fri Jan 10 11:00:49 2014
\begin{table}[!ht]
\caption{{ \bf Marker-based estimation of heritability: width and coverage confidence intervals obtained from
the individual plant data and the genotypic means.} Results for broad sense heritability intervals are reported for comparison. We simulated 5000 traits, for random samples of 200 accessions drawn from the Swedish regmap (top) and French (bottom). 20 unlinked QTLs were simulated, which explained 50 percent of the genetic variance. The simulated heritability was $0.2$, $0.5$ and $0.8$.}
\centering
\begin{tabular}{|l|r|r|}
  \hline
 & coverage & interval width \\
  \hline
\multicolumn{3}{|>{\columncolor[gray]{.8}}l|}{Swedish regmap} \\
  \hline
\multicolumn{3}{|c|}{$h^2 = 0.2$} \\
  \hline
broad-sense & 0.912 & 0.178 \\
  replicates (standard) & 0.933 & 0.188 \\
  replicates (log-transformed) & 0.961 & 0.189 \\
  means (standard) & 0.899 & 0.381 \\
  means (log-transformed) & 0.968 & 0.405 \\
   \hline
\multicolumn{3}{|c|}{$h^2 = 0.5$} \\
  \hline
broad-sense & 0.864 & 0.160 \\
  replicates (standard) & 0.946 & 0.176 \\
  replicates (log-transformed) & 0.950 & 0.175 \\
  means (standard) & 0.921 & 0.594 \\
  means (log-transformed) & 0.970 & 0.560 \\
   \hline
\multicolumn{3}{|c|}{$h^2 = 0.8$} \\
  \hline
broad-sense & 0.823 & 0.085 \\
  replicates (standard) & 0.945 & 0.088 \\
  replicates (log-transformed) & 0.947 & 0.088 \\
  means (standard) & 0.960 & 0.635 \\
  means (log-transformed) & 0.938 & 0.748 \\
   \hline
\multicolumn{3}{|>{\columncolor[gray]{.8}}l|}{French regmap} \\
  \hline
\multicolumn{3}{|c|}{$h^2 = 0.2$} \\
  \hline
broad-sense & 0.929 & 0.178 \\
  replicates (standard) & 0.941 & 0.184 \\
  replicates (log-transformed) & 0.962 & 0.185 \\
  means (standard) & 0.898 & 0.396 \\
  means (log-transformed) & 0.960 & 0.431 \\
   \hline
\multicolumn{3}{|c|}{$h^2 = 0.5$} \\
  \hline
broad-sense & 0.898 & 0.160 \\
  replicates (standard) & 0.953 & 0.173 \\
  replicates (log-transformed) & 0.956 & 0.171 \\
  means (standard) & 0.927 & 0.619 \\
  means (log-transformed) & 0.976 & 0.585 \\
   \hline
\multicolumn{3}{|c|}{$h^2 = 0.8$} \\
  \hline
broad-sense & 0.866 & 0.084 \\
  replicates (standard) & 0.947 & 0.088 \\
  replicates (log-transformed) & 0.947 & 0.088 \\
  means (standard) & 0.966 & 0.652 \\
  means (log-transformed) & 0.948 & 0.767 \\
   \hline
\end{tabular}
%\caption{{ \bf Width and coverage of confidence intervals for broad sense heritability ($H^2$) and narrow sense heritability (replicates,  means), based on 5000 simulated traits, for random samples of 200 accessions drawn from %the Swedish regmap (top) and French (bottom).} 20 unlinked QTLs were simulated, which explained 50 percent of the genetic variance. The simulated heritability was $0.2$, $0.5$ and $0.8$.}
%\label{conf_h2_sim_pop2_regmap_new_gamma5_5000traits_200acc_20QTLs_3rep_h20.2}
\end{table}

\newpage
%%%%%%%%%%%%%%%%%%%%%%%%%%%%%%%%%%%%%%%%%%%%%%%%%%%%%%%%%%%%%%%%%%%%%%%%%
% Table 4

\section*{Table S4}

\renewcommand{\tablename}{Table S4:}
 \renewcommand{\thetable}{}

% table produced using the script CONFIDENCE_INTERVALS_PLOTS_LDAK.r
% The table is written to the file /tables/atwell_h2_table_LDAK.tex
% .. and copied to this file, to adapt the caption and labels 'manually'
\begin{table}[!ht]
%\centering
\caption{
\bf{Heritability estimates and confidence intervals,} for two flowering traits from \cite{atwell_etal_2010} and four
traits measured in new experiments
(trait abbreviations given in Table 1 of the main text).
}
\begin{tabular}{|r|l|l|l|}
  \hline
trait & replicates & means & broad-sense\\
  \hline
LDV & 0.829 (0.791,0.861) & 0.631 (0.199,0.922) & 0.858 (0.827,0.885) \\
  LD & 0.946 (0.933,0.957) & 1.000 (0.000,1.000) & 0.966 (0.958,0.973) \\
  \hline
LA(S) & 0.216 (0.153,0.297) & 0.150 (0.040,0.424) & 0.235 (0.167,0.306) \\
  \hline
LA(H) & 0.380 (0.319,0.445) & 0.340 (0.090,0.729) & 0.388 (0.327,0.451) \\
  \hline
BT & 0.948 (0.937,0.956) & 1.000 (0.000,1.000) & 0.956 (0.947,0.963) \\
  LW & 0.535 (0.473,0.596) & 0.202 (0.029,0.682) & 0.530 (0.468,0.589) \\
   \hline
\end{tabular}
\begin{flushleft}%Table caption
Three estimators were used: mixed model based on replicates (${\hat h}_r^2$), mixed model based on genotypic means (${\hat h}_m^2$), and the usual ANOVA-based broad-sense heritability estimator (${\hat H}^2$).
An LD-adjusted kinship matrix was used in the mixed model for ${\hat h}_r^2$ and ${\hat h}_m^2$.
\end{flushleft}
%\label{h2_values_LDAK}
\end{table}

The LD-adjusted kinship matrix was computed using version 2.0 of the LDAK-software \cite{speed_hemani_johnson_balding_2012}, available at \verb|http://dougspeed.com/ldak/|.
We used sections of 1000 SNPs, with a buffer of 200. The maximum distance considered for LD was 250kb; the 'halflife' parameter (modeling LD-decay) was set to 20kb.
%\maketitle

\newpage
%%%%%%%%%%%%%%%%%%%%%%%%%%%%%%%%%%%%%%%%%%%%%%%%%%%%%%%%%%%%%%%%%%%%%%%%%
% Table 5

\section*{Table S5}

\renewcommand{\tablename}{Table S5:}
 \renewcommand{\thetable}{}

%\subsection*{s2}
In the following tables, the second and third column contain the percentage of the $5000$ traits for which the corresponding heritability estimates (${\hat h}_r^2$ and ${\hat h}_m^2$) were contained in the intervals in the first column. The remaining columns show the correlation ($r$) between simulated and predicted genetic effects,  averaged over these traits. 20 QTLs were simulated, which explained 50 percent of the genetic variance.
Each trait was simulated for a randomly drawn training (200 accessions) and validation set (50 accessions). Genetic effects were predicted using G-BLUP, based on either a mixed model for the individual plants (replicates) or for the genotypic means.

% latex table generated in R 3.0.1 by xtable 1.7-1 package
% Fri Apr 11 11:06:41 2014
\begin{table}[!ht]
\renewcommand{\tablename}{Table S5(a)}
\centering
\caption{{ \bf Prediction accuracy ($r$) of G-BLUP for $5000$ simulated traits, for the structured regmap population, and a simulated heritability of $0.2$.}}
\begin{tabular}{|l|r|r|r|r|r|r|}
  \hline
interval & ${\hat h}_r^2$ & ${\hat h}_m^2$ & $r$ (replicates) & $r$ (means) & $r$ (replicates) & $r$ (means) \\
&&& Training set & Training set & Validation set & Validation set \\
  \hline
$[0,0.1)$ & 3.08 \% & 9.88 \% & 0.637 & 0.654 & 0.216 & 0.218 \\
  $[0.1,0.3)$ & 93.96 \% & 76.38 \% & 0.770 & 0.770 & 0.280 & 0.279 \\
  $[0.3,0.5)$ & 2.96 \% & 13.52 \% & 0.816 & 0.803 & 0.325 & 0.313 \\
  $[0.5,0.7)$ & 0 \% & 0.22 \% &  & 0.782 &  & 0.287 \\
  $[0.7,0.9)$ & 0 \% & 0 \% &  &  &  &  \\
  $[0.9,1]$ & 0 \% & 0 \% &  &  &  &  \\
  $[0,1]$ & 100 \% & 100 \% & 0.767 & 0.763 & 0.279 & 0.278 \\
   \hline
\end{tabular}
\end{table}
% latex table generated in R 3.0.1 by xtable 1.7-1 package
% Fri Apr 11 11:06:45 2014
\begin{table}[!ht]
\renewcommand{\tablename}{Table S5(b)}
\centering
%\caption{{ \bf Correlation (r2) between true and predicted genetic effects. 5000 traits were simulated, for random samples of 200 accessions drawn from the structured regmap. For each simulated trait, 50 additional genotypes were %drawn from the structured regmap, whose genetic effects were predicted.} 20 QTLs were simulated, which explained 50 percent of the genetic variance.}
\caption{{ \bf Prediction accuracy ($r$) of G-BLUP for $5000$ simulated traits, for the structured regmap population and a simulated heritability of $0.5$.}
%Each trait was simulated for a randomly drawn training (200 accessions) and validation set (50 accessions). Genetic effects were predicted using G-BLUP, based on either a mixed model for the individual plants (replicates) or for the genotypic means. The second and third column contain the percentage of the $5000$ traits for which the corresponding heritability estimates (${\hat h}_r^2$ and ${\hat h}_m^2$) were contained in the intervals in the first column. The remaining columns show the correlation ($r$) between simulated and predicted genetic effects,  averaged over these traits. 20 QTLs were simulated, which explained 50 percent of the genetic variance.
}
\begin{tabular}{|l|r|r|r|r|r|r|}
  \hline
interval & ${\hat h}_r^2$ & ${\hat h}_m^2$ & $r$ (replicates) & $r$ (means) & $r$ (replicates) & $r$ (means) \\
&&& Training set & Training set & Validation set & Validation set \\
  \hline
$[0,0.1)$ & 0 \% & 0.04 \% &  & 0.709 &  & 0.328 \\
  $[0.1,0.3)$ & 0 \% & 5.24 \% &  & 0.836 &  & 0.269 \\
  $[0.3,0.5)$ & 51.42 \% & 46.54 \% & 0.886 & 0.887 & 0.302 & 0.300 \\
  $[0.5,0.7)$ & 48.58 \% & 42.12 \% & 0.905 & 0.903 & 0.333 & 0.337 \\
  $[0.7,0.9)$ & 0 \% & 5.84 \% &  & 0.905 &  & 0.343 \\
  $[0.9,1]$ & 0 \% & 0.22 \% &  & 0.888 &  & 0.386 \\
  $[0,1]$ & 100 \% & 100 \% & 0.895 & 0.892 & 0.317 & 0.317 \\
   \hline
\end{tabular}
\end{table}
% latex table generated in R 3.0.1 by xtable 1.7-1 package
% Fri Apr 11 11:06:49 2014
\begin{table}[!ht]
\renewcommand{\tablename}{Table S5(c)}
\centering
%\caption{{ \bf Correlation (r2) between true and predicted genetic effects. 5000 traits were simulated, for random samples of 200 accessions drawn from the structured regmap. For each simulated trait, 50 additional genotypes were %drawn from the structured regmap, whose genetic effects were predicted.} 20 QTLs were simulated, which explained 50 percent of the genetic variance.}
\caption{{ \bf Prediction accuracy ($r$) of G-BLUP for $5000$ simulated traits, for the structured regmap population and a simulated heritability of $0.8$.}
%Each trait was simulated for a randomly drawn training (200 accessions) and validation set (50 accessions). Genetic effects were predicted using G-BLUP, based on either a mixed model for the individual plants (replicates) or for the genotypic means. The second and third column contain the percentage of the $5000$ traits for which the corresponding heritability estimates (${\hat h}_r^2$ and ${\hat h}_m^2$) were contained in the intervals in the first column. The remaining columns show the correlation ($r$) between simulated and predicted genetic effects,  averaged over these traits. 20 QTLs were simulated, which explained 50 percent of the genetic variance.
}
\begin{tabular}{|l|r|r|r|r|r|r|}
  \hline
interval & ${\hat h}_r^2$ & ${\hat h}_m^2$ & $r$ (replicates) & $r$ (means) & $r$ (replicates) & $r$ (means) \\
&&& Training set & Training set & Validation set & Validation set \\
  \hline
$[0,0.1)$ & 0 \% & 0 \% &  &  &  &  \\
  $[0.1,0.3)$ & 0 \% & 0.02 \% &  & 0.877 &  & 0.299 \\
  $[0.3,0.5)$ & 0 \% & 1.42 \% &  & 0.930 &  & 0.283 \\
  $[0.5,0.7)$ & 0.04 \% & 19.26 \% & 0.953 & 0.955 & 0.400 & 0.318 \\
  $[0.7,0.9)$ & 99.96 \% & 59.26 \% & 0.964 & 0.964 & 0.343 & 0.344 \\
  $[0.9,1]$ & 0 \% & 20.04 \% &  & 0.965 &  & 0.365 \\
  $[0,1]$ & 100 \% & 100 \% & 0.964 & 0.962 & 0.343 & 0.343 \\
   \hline
\end{tabular}
\end{table}
% latex table generated in R 3.0.1 by xtable 1.7-1 package
% Fri Apr 11 11:06:54 2014
\begin{table}[!ht]
\renewcommand{\tablename}{Table S5(d)}
\centering
%\caption{{ \bf Correlation (r2) between true and predicted genetic effects. 5000 traits were simulated, for random samples of 200 accessions drawn from the hapmap. For each simulated trait, 50 additional genotypes were drawn from %the hapmap, whose genetic effects were predicted.} 20 QTLs were simulated, which explained 50 percent of the genetic variance.}
\caption{{ \bf Prediction accuracy ($r$) of G-BLUP for $5000$ simulated traits, for the HapMap population and a simulated heritability of $0.2$.}
%Each trait was simulated for a randomly drawn training (200 accessions) and validation set (50 accessions). Genetic effects were predicted using G-BLUP, based on either a mixed model for the individual plants (replicates) or for the genotypic means. The second and third column contain the percentage of the $5000$ traits for which the corresponding heritability estimates (${\hat h}_r^2$ and ${\hat h}_m^2$) were contained in the intervals in the first column. The remaining columns show the correlation ($r$) between simulated and predicted genetic effects,  averaged over these traits. 20 QTLs were simulated, which explained 50 percent of the genetic variance.
}
\begin{tabular}{|l|r|r|r|r|r|r|}
  \hline
interval & ${\hat h}_r^2$ & ${\hat h}_m^2$ & $r$ (replicates) & $r$ (means) & $r$ (replicates) & $r$ (means) \\
&&& Training set & Training set & Validation set & Validation set \\
  \hline
$[0,0.1)$ & 1.74 \% & 28.64 \% & 0.616 & 0.632 & 0.259 & 0.273 \\
  $[0.1,0.3)$ & 96.7 \% & 40.7 \% & 0.673 & 0.674 & 0.341 & 0.348 \\
  $[0.3,0.5)$ & 1.56 \% & 17.8 \% & 0.711 & 0.684 & 0.364 & 0.382 \\
  $[0.5,0.7)$ & 0 \% & 5.8 \% &  & 0.681 &  & 0.366 \\
  $[0.7,0.9)$ & 0 \% & 2.84 \% &  & 0.675 &  & 0.370 \\
  $[0.9,1]$ & 0 \% & 4.22 \% &  & 0.669 &  & 0.357 \\
  $[0,1]$ & 100 \% & 100 \% & 0.672 & 0.664 & 0.340 & 0.335 \\
   \hline
\end{tabular}
\end{table}
% latex table generated in R 3.0.1 by xtable 1.7-1 package
% Fri Apr 11 11:06:58 2014
\begin{table}[!ht]
\renewcommand{\tablename}{Table S5(e)}
\centering
%\caption{{ \bf Correlation (r2) between true and predicted genetic effects. 5000 traits were simulated, for random samples of 200 accessions drawn from the hapmap. For each simulated trait, 50 additional genotypes were drawn from %the hapmap, whose genetic effects were predicted.} 20 QTLs were simulated, which explained 50 percent of the genetic variance.}
\caption{{ \bf Prediction accuracy ($r$) of G-BLUP for $5000$ simulated traits, for the HapMap population and a simulated heritability of $0.5$.}
%Each trait was simulated for a randomly drawn training (200 accessions) and validation set (50 accessions). Genetic effects were predicted using G-BLUP, based on either a mixed model for the individual plants (replicates) or for the genotypic means. The second and third column contain the percentage of the $5000$ traits for which the corresponding heritability estimates (${\hat h}_r^2$ and ${\hat h}_m^2$) were contained in the intervals in the first column. The remaining columns show the correlation ($r$) between simulated and predicted genetic effects,  averaged over these traits. 20 QTLs were simulated, which explained 50 percent of the genetic variance.
}
\begin{tabular}{|l|r|r|r|r|r|r|}
  \hline
interval & ${\hat h}_r^2$ & ${\hat h}_m^2$ &$r$ (replicates) & $r$ (means) & $r$ (replicates) & $r$ (means) \\
&&& Training set & Training set & Validation set & Validation set \\
  \hline
$[0,0.1)$ & 0 \% & 6 \% &  & 0.811 &  & 0.285 \\
  $[0.1,0.3)$ & 0 \% & 21.02 \% &  & 0.851 &  & 0.366 \\
  $[0.3,0.5)$ & 51.78 \% & 22.56 \% & 0.862 & 0.867 & 0.395 & 0.413 \\
  $[0.5,0.7)$ & 48.22 \% & 17.5 \% & 0.877 & 0.871 & 0.416 & 0.428 \\
  $[0.7,0.9)$ & 0 \% & 10.86 \% &  & 0.873 &  & 0.426 \\
  $[0.9,1]$ & 0 \% & 22.06 \% &  & 0.871 &  & 0.422 \\
  $[0,1]$ & 100 \% & 100 \% & 0.869 & 0.863 & 0.405 & 0.401 \\
   \hline
\end{tabular}
\end{table}
% latex table generated in R 3.0.1 by xtable 1.7-1 package
% Fri Apr 11 11:07:02 2014
\begin{table}[!ht]
\renewcommand{\tablename}{Table S5(f) (given as Table 6 in the main text)}
\centering
%\caption{{ \bf Correlation (r2) between true and predicted genetic effects. 5000 traits were simulated, for random samples of 200 accessions drawn from the hapmap. For each simulated trait, 50 additional genotypes were drawn from %the hapmap, whose genetic effects were predicted.} 20 QTLs were simulated, which explained 50 percent of the genetic variance.}
\caption{{ \bf Prediction accuracy ($r$) of G-BLUP for $5000$ simulated traits, for the HapMap population and a simulated heritability of $0.8$.}
%Each trait was simulated for a randomly drawn training (200 accessions) and validation set (50 accessions). Genetic effects were predicted using G-BLUP, based on either a mixed model for the individual plants (replicates) or for the genotypic means. The second and third column contain the percentage of the $5000$ traits for which the corresponding heritability estimates (${\hat h}_r^2$ and ${\hat h}_m^2$) were contained in the intervals in the first column. The remaining columns show the correlation ($r$) between simulated and predicted genetic effects,  averaged over these traits. 20 QTLs were simulated, which explained 50 percent of the genetic variance.
}
\begin{tabular}{|l|r|r|r|r|r|r|}
  \hline
interval & ${\hat h}_r^2$ & ${\hat h}_m^2$ & $r$ (replicates) & $r$ (means) & $r$ (replicates) & $r$ (means) \\
&&& Training set & Training set & Validation set & Validation set \\
  \hline
$[0,0.1)$ & 0 \% & 2.58 \% &  & 0.890 &  & 0.289 \\
  $[0.1,0.3)$ & 0 \% & 8.34 \% &  & 0.937 &  & 0.373 \\
  $[0.3,0.5)$ & 0 \% & 12.34 \% &  & 0.954 &  & 0.409 \\
  $[0.5,0.7)$ & 0.04 \% & 15.9 \% & 0.942 & 0.959 & 0.208 & 0.423 \\
  $[0.7,0.9)$ & 99.96 \% & 15.62 \% & 0.961 & 0.961 & 0.431 & 0.443 \\
  $[0.9,1]$ & 0 \% & 45.22 \% &  & 0.961 &  & 0.448 \\
  $[0,1]$ & 100 \% & 100 \% & 0.961 & 0.956 & 0.431 & 0.428 \\
   \hline
\end{tabular}
\end{table}

\end{document}